\documentclass[3p]{elsarticle}

\usepackage{mymacros}
\usepackage{enumitem}
\usepackage{lipsum}
\usepackage{amssymb}
\usepackage{amsmath}
\usepackage{stmaryrd}
\usepackage{comment}
\usepackage{float}
\usepackage{xcolor}
\usepackage{bussproofs}
\usepackage{hyperref}

\setlist{nolistsep}
\newtheorem{definition}{Definition}
\newtheorem{theorem}{Theorem}
\newtheorem{proposition}{Proposition}

\newtheorem{example}{Example}

\usepackage{setspace}

\begin{document}

\title{Measuring Robustness in Cyber-Physical Systems under    Sensor Attacks\tnoteref{hscc23}}

\tnotetext[hscc23]{This is an extended and revised version of a preliminary paper appeared in the proceedings of the \emph{26th ACM International Conference on Hybrid Systems: Computation and Control} (HSCC'23)~\cite{hscc2023}.}

\author[a]{Jian Xiang}
\ead{jian.xiang@charlotte.edu}
\author[b]{Ruggero Lanotte}
\ead{ruggero.lanotte@uninsubria.it}
\author[b]{Simone Tini}
\ead{simone.tini@uninsubria.it}
\author[c]{Stephen Chong}
\ead{chong@seas.harvard.edu} 
\author[d]{Massimo Merro}
\ead{massimo.merro@univr.it}

\affiliation[a]{organization={UNC Charlotte}, addressline={
9201 University City Blvd},city={Charlotte},postcode={NC 28223}, country={US}}

\affiliation[b]{organization={University of Insubria}, addressline={Via Valleggio 11},city={Como},postcode={22100}, country={Italy}}

\affiliation[c]{organization={Harvard University}, addressline={150 Western Ave},city={Boston},postcode={MA 02134}, country={US}}

\affiliation[d]{organization={University of Verona}, addressline={Strada le Grazie 15},city={Verona},postcode={37134}, country={Italy}}

\newif{\ifanonymous}
\anonymousfalse

\ifanonymous
\else






\fi


\begin{abstract}
  
  This paper contributes a formal framework for quantitative analysis of bounded sensor attacks on cyber-physical systems, using the formalism of differential dynamic logic. Given a precondition and postcondition of a system, we formalize two quantitative safety notions, quantitative forward and backward safety, which respectively express  (1) how strong the strongest postcondition of the system is with respect to the specified postcondition, and 
  (2) how strong the specified precondition is with respect to the weakest precondition of the system needed to ensure the specified postcondition holds. 
  We introduce two notions, forward and backward robustness, to characterize the robustness of a system against sensor attacks as the loss of safety. To reason about robustness, we introduce two simulation distances, forward and backward simulation distances, which are defined based on the behavioral distances between the original system and the system with compromised sensors. Forward and backward distances, respectively, characterize upper bounds of the degree of forward and backward safety loss caused by the sensor attacks. 
%
 {We verify the two simulation distances by expressing them as modalities, i.e., formulas of differential dynamic logic, and develop an ad-hoc proof system to reason with such formulas. We showcase our formal notions and reasoning techniques on two non-trivial case studies: an autonomous vehicle that needs to avoid collision and a water tank system.} 

%

\end{abstract}






\maketitle
`


\setlength{\belowcaptionskip}{-10pt}
\setlength{\abovecaptionskip}{2pt}

\setlength{\abovedisplayskip}{3pt}
\setlength{\belowdisplayskip}{3pt}

\setlist[itemize,1]{leftmargin=\dimexpr 26pt-.15in}




\newcommand{\CPS}{CPS\xspace} 
\newcommand{\CPSs}{CPSs\xspace}

\newcommand{\FHsimulation}[5]{\ensuremath{#1 \sqsubseteq^{\scriptsize\rm{F}}_{{#3,#4,#5}} #2}\xspace}

\newcommand{\BHsimulation}[5]{\ensuremath{#1 \sqsubseteq^{\scriptsize\rm{B}}_{{#3,#4,#5}} #2}\xspace}

\newcommand{\Frobustn}{forward robustness\xspace}
\newcommand{\Brobustn}{backward robustness\xspace}
\newcommand{\FrobustN}{Forward robustness\xspace}
\newcommand{\BrobustN}{Backward robustness\xspace}

\section{Introduction}

Cyber-Physical Systems (\CPSs) are integrations of networking and distributed computing systems with physical processes, where feedback loops allow physical processes to affect computations and vice versa.
A  peculiar class of attacks in such systems is the so-called \emph{physics-based attacks}: attacks targeting the physical devices (sensors and actuators)
of \CPSs~\cite{ACM-survey2018,lanotte2020formal}. 
For instance,  \emph{sensor attacks}, such  as DoS or integrity attacks on sensors,  may lead to crashing the system under attack~\cite{son2015rocking}, or allow an adversary to control the system~\cite{cao2019adversarial, davidson2016controlling}.
\looseness=-1

The importance of ensuring the \emph{safety} of CPSs motivates a growing body of work on formal verification for embedded and hybrid systems~\cite{alur2015principles, larsen2009verification, lee2016introduction, tabuada2009verification, tiwari2011logic, Nigam-Esorics2016, Nigam2019,Hankin2020,Villa2020,Sirjani2021}, 
some of which focus on the analysis of sensor-based attacks \cite{xiang2021relational, lanotte2020formal}.
Existing work often treats satisfaction of safety as a boolean predicate: either a system satisfies a desired safety property or it does not.
However, a simple yes/no answer doesn't fit the setting of CPSs, which interact with continuous and quantitative entities, such as measurements of the controlled physical process. 
For example, under the same road conditions, a vehicle with a shorter braking distance toward an obstacle is considered safer than a vehicle with a longer braking distance, even if both of them can brake in time. 
Thus, when working with \CPSs, a  \emph{quantitative} notion of safety can be much more informative than standard safety.
\looseness=-1

 However, knowing the degree of safety of a CPS is not enough to analyze the effect of  attacks targeting its sensors. For example, a vehicle with a very short braking distance may not be able to tolerate certain attacks on the
\emph{obstacle detection system}, resulting in unsafe runtime behaviors.
%
Here, it is important to understand the \emph{robustness} of a system's safety under sensor attacks, that is, how the safety may change because of  sensor attacks. 
 For example, consider a vehicle equipped with a self-braking system whose \emph{safety requirement} is to  brake from the speed of 100 km/h 
 when  an obstacle is detected  40 meters away.
 And suppose the vehicle, at that speed, starts braking when the obstacle is detected 60 meters away.
	Assume that an adversary is able to perturb the readings of the distance to an obstacle by 10 meters without being detected. Then, the vehicle is still safe as it starts braking,  at the speed of 100 km/h, when the obstacle is 50 meters away; 10 meters more than the safety requirements. 
        The degree of safety loss is a clear indicator of the vehicle's robustness against such an attack. 

In this work, we define two notions of \emph{quantitative safety} for \CPSs and use them to analyze a system's robustness under sensor attacks. Our threat model assumes \emph{bounded sensor attacks},  that is, attacks that may compromise a subset of sensors and offset their readings by some degree. 
We do not model or discover the mechanisms by which  attackers manipulate sensor values; we simply assume they are able to do so. We also assume every system has a known \emph{precondition} and \emph{postcondition}. The precondition specifies the initial conditions and environment when the system starts operating, and the postcondition specifies the desired condition that the system should always satisfy for it to be safe.

The first notion is \emph{forward quantitative safety}, which estimates the room for maneuver to ensure that the system remains safe after any execution starting from a state satisfying the precondition. It basically estimates how strong the strongest postcondition is  with respect to the  desired safety postcondition. Said in other words, given a precondition,  forward safety provides  a quantification of the margins on possible strengthening of the safety postcondition with respect to the strongest  postcondition. Technically, it is defined as the \emph{shortest} distance between the set of states satisfying the strongest postcondition and the set of unsafe states. 
The larger this distance is, the further away the system's reachable states are from unsafe states, and thus the safer the system is.
Built upon forward safety, we introduce 
\emph{forward robustness}, which characterizes the impact of a given sensor attack as a ratio: the degree of forward safety of the compromised system over the degree of forward safety of the original system. Intuitively, the closer the ratio gets to 1, the more robust the original system is against the attack. A ratio of 1 means the attack doesn't weaken the safety guarantee at all.    
\looseness=-1

The second notion is \emph{backward quantitative safety}, which  provides a degree of safety by estimating the room for maneuver to ensure that the system remains safe with respect to a given postcondition by weakening the precondition. It basically estimates how strong the specified precondition is with respect to the weakest precondition needed to ensure the safety of the system after its execution. Said in other words, given a safety  postcondition,  backward safety provides a quantification 
of the precondition with respect to the weakest  precondition.
Technically, it is defined as the \emph{shortest} distance between the set of states satisfying the {precondition} 
and the set of ``bad'' initial states that may lead the system to  unsafe states. 
%
The larger this distance is, the further away the system's states that satisfy the precondition are from ``bad'' initial states, and thus the safer the system is.
Built upon backward safety, we introduce \emph{backward robustness} that characterizes the impact of a sensor attack as a ratio: the degree of backward safety of the compromised system over the original system. Similar to forward robustness, the closer the ratio gets to 1, the more robust the original system is against the attack. 

The two robustness notions together give system designers a good way to understand and compare different design candidates by focusing either on preconditions or on postconditions. 
For example, if a system is likely to suffer from sensor attacks, a designer may simply choose a candidate design with better degrees of robustness.
If one degree of robustness (e.g., forward robustness) is identical or similar among different designs, the designers may use the other (e.g., backward robustness) to compare the designs.

To reason about forward (and backward) robustness, we introduce a forward (and backward) simulation distance to, respectively, provide an upper bound of the loss of forward (and backward) safety caused by sensor attacks. The simulation distances are defined based on the behavioral distances \cite{GJS90} between the original system and the system with compromised sensors. In particular, the forward simulation distance characterizes the \emph{forward distance} between the two systems by quantifying the distance between their reachable states, given the same set of initial states. Thus, the forward distance between the original and the compromised system returns an upper bound on the admissible perturbations introduced by a sensor attack on the safety of the behaviors originating from a desired precondition.
Analogously, the backward simulation distance characterizes the \emph{backward distance} between the two systems by quantifying the distance between their sets of safe initial states, i.e., those states that never lead the system to an unsafe state, given the same set of safe final states. Thus, the backward distance between the original and the compromised system returns an upper bound on the admissible perturbations introduced by a sensor attack on the initial states  leading to possible violations of safety, given a desired postcondition. 
We prove that the forward (and backward) simulation distance represents a sound proof technique for calculating upper bounds of  forward (and backward) robustness as it returns upper bounds of the loss of forward (and backward) safety caused by sensor attacks.

In the paper, we work within the formalism of hybrid programs and
\emph{differential dynamic logic} (\dL)  \cite{platzer2008differential, Platzer18book, 
platzer2017complete}.
Hybrid programs are a formalism for modeling 
systems that have both continuous and discrete dynamic behaviors. Hybrid programs can express continuous evolution (as differential equations) as well as discrete transitions.
Differential dynamic logic is the logic of hybrid programs, which is used to specify and verify safety properties.

To easily reason with the forward and backward distances,  we express them as \dL formulas and use existing \dL axioms and proof rules to verify the formulas. Moreover, we introduce an ad-hoc \emph{modality} and an associated \emph{proof system} to efficiently reason with the modality. The proposed modality can be used to encode both forward and backward simulation distances in an intuitive and concise way. We formally prove the soundness of the encodings and the associated proof system. We demonstrate the applicability of the proof system on a case study on a water tank system. 



\medskip
The main contributions of this paper are the following:
\smallskip

\begin{itemize}
\item The notions of forward and backward quantitative safety in the context of differential dynamic logic, which models the safety properties of CPSs (Section~\ref{sec:q-safety}).

\item The notions of forward and backward quantitative robustness for systems under bounded sensor attacks, defined using the two notions of quantitative safety (Section~\ref{sec:robust-safety}).
  
\item Two simulation distances, forward and backward simulation distances over hybrid programs, to reason about robustness 
  (Section~\ref{sec:h-eq}). 

\item  
Reasoning techniques for the two simulation distances based on an ad-hoc modality and its proof system (Section~\ref{sec:proof-calculus}). 

\item Two case studies. The first one showcases the usefulness of the proposed notions on a case study on autonomous vehicles (Section~\ref{sec:vehicle}). And the second one demonstrates the proof system for simulation distances in a case study of a water tank (Section~\ref{sec:casestudy}). 
\end{itemize}
\smallskip

We introduce preliminaries in Section~\ref{sec:background}. Section~\ref{sec:relatedwork} discusses related work, and Section~\ref{sec:conclusion} concludes.     

\medskip
{This is the journal version that extends and generalizes a conference paper~\cite{hscc2023}. The new contribution of this article are Section~\ref{sec:proof-calculus} and 
 Section~\ref{sec:casestudy}. 
Here, we also add  detailed proofs of all theorems presented in the conference version and of the new results  introduced in this journal version. }
  
\section{Preliminaries} \label{sec:background}

In this section, we recall the formalisms of hybrid programs~\cite{Platzer18book} and Differential Dynamic Logic~\cite{platzer2008differential, Platzer18book, 
platzer2017complete}, the model of sensor attacks from~\cite{xiang2021relational} and a notion of distance between states from~\cite{fainekos2009robustness, boyd2004convex}.

\subsection{Differential Dynamic Logic}
\label{sec:DDL}

\emph{Hybrid programs} \cite{Platzer18book} are a formalism for modeling systems that have both continuous and discrete dynamic behaviors. 
Figure~\ref{fig:dl:syntax} gives the syntax for hybrid programs. Variables are real-valued and can be deterministically assigned (\DLassign{\DLvarsym}{\DLtermsym}, where \DLtermsym is a real-valued term) or nondeterministically assigned (\DLassignN{\DLvarsym}).
A hybrid program \dLprogODE{\dLODE{\DLvarsym}{\DLtermsym}}{\evolconstraint} expresses the continuous evolution of variables: given the current value of variable \DLvarsym, the system follows the differential equation \dLODE{\DLvarsym}{\DLtermsym} for some (nondeterministically chosen) amount of time so long as the formula \evolconstraint,  the \emph{evolution domain constraint}, holds for all of that time. Note that \DLvarsym can be a vector of variables and then \DLtermsym is a vector of terms of the same dimension.  

Hybrid programs also include the operations of Kleene algebra with tests \cite{kozen1997kleene}: testing whether a formula holds, sequential composition, nondeterministic choice, and nondeterministic repetition.

\emph{Differential dynamic logic} (\dL) \cite{platzer2008differential, Platzer18book, 
platzer2017complete} is the dynamic logic~\cite{dynamiclogic} of hybrid programs. 
Figure~\ref{fig:dl:syntax} also gives the syntax for 
\dL formulas. 
In addition to the standard logical connectives of first-order logic, \dL includes primitive propositions that allow for comparisons of real-valued terms (which may include derivatives) and the \emph{modality of necessity} $\DLfmlModalA{\DLprogsym}{\DLfmlsym}$, which holds in a state if and only if after any possible execution of hybrid program \DLprogsym, formula \DLfmlsym holds. The modality of necessity can be used to encode the \emph{modality of existence}, i.e.,  $\DLfmlModalE{\DLprogsym}{\DLfmlsym} = \DLfmlneg{ \DLfmlModalA{\DLprogsym}{\DLfmlneg{\DLfmlsym}}}$.
Common abbreviations for other logical connectives apply, e.g., $\DLfmldisj{\DLfmlsym}{\DLfmlsymp} = \DLfmlneg{(\DLfmlconj{\DLfmlneg{\DLfmlsym}}{\DLfmlneg{\DLfmlsymp}})}$
and
$\DLfmlsym  \hostE{\, \rightarrow \,} \DLfmlsymp = 
\DLfmldisj{\DLfmlneg{\DLfmlsym}}{\DLfmlsymp}$.

\begin{figure}
  \centering
  \begin{tabular}{r@{\;\;}l}
   {\DLtermsym}, {\DLtermsymp} & \, ::= \, \DLvarsym $\mid$ \DLconst $\mid$ \DLtermcomp{\DLtermsym}{\DLtermsymp} \\[0.5 ex]
  \DLprogsym, \DLprogsymp & \, ::= \, \DLassign{\DLvarsym}{\DLtermsym} $\mid$ \DLassignN{\DLvarsym} $\mid$ \dLprogODE{\dLODE{\DLvarsym}{\DLtermsym}}{\evolconstraint} $\mid$ \DLtest{\DLfmlsym} $\mid$ \DLseq{\DLprogsym}{\DLprogsymp} $\mid$ \DLchoice{\DLprogsym}{\DLprogsymp} $\mid$ \DLloop{\DLprogsym} 
  \\[0.5 ex]
    \DLfmlsym, \DLfmlsymp & \, ::= \, \F $\mid$ \DLfmlcomp{\DLtermsym}{\DLtermsymp} $\mid$ \DLfmlneg{\DLfmlsym} $\mid$ \DLfmlconj{\DLfmlsym}{\DLfmlsymp} $\mid$ \DLfmlUQ{\DLvarsym}{\DLfmlsym} $\mid$ $\DLfmlModalA{\DLprogsym}{\DLfmlsym}$
    \\[0.9 ex]
  \end{tabular}
  \caption{Syntax of hybrid programs and \dL}
  \label{fig:dl:syntax}
\end{figure}

The semantics of $\dL$ \cite{platzer2008differential, platzer2017complete} is a Kripke semantics in which the Kripke model's worlds are the states of the system. Let $\realSet$ denote the set of real numbers and $\allvariableSet$ denote the set of variables. A \emph{state} is a map $\statesym$ : $\allvariableSet$ $\mapsto$ $\realSet$ assigning a real value $\statesym(x)$ to each variable $x \in \allvariableSet$. The set of all states is denoted by $\allstates$.
The semantics of hybrid programs and $\dL$ are shown in Figure~\ref{fig:dlsem}.
We write $\DLsem{\DLfmlsym}$ to denote the set of states that satisfy formula $\DLfmlsym$. The value of term $\DLtermsym$ at state $\statesym$ is denoted $\DLtermsem{\statesym}{\DLtermsym}$.
The semantics of a program $\DLprogsym$ is expressed as a transition relation $\DLsem{\DLprogsym}$ between states. If $\DLsemPair{\statesym}{\statesymp}$ $\in$ $\DLsem{\DLprogsym}$ then there is an execution of  $\DLprogsym$ that starts in state $\statesym$ and ends in state $\statesymp$.


\emph{Safety properties} of a system are often defined as follows: 
\begin{definition}[Safety] \label{def:safety} 
A hybrid program $\DLprogsym$ is \emph{safe for $\phi_{post}$ assuming $\phi_{pre}$}, denoted $\safety{\DLprogsym}{\dLfmlpre}{\dLfmlpost}$,
if $\dLfmlpre \hostE{\, \rightarrow \,} \WP{\DLprogsym}{\dLfmlpost}$ holds.
\end{definition}

$\safety{\DLprogsym}{\dLfmlpre}{\dLfmlpost}$ means if $\dLfmlpre$ is true then $\dLfmlpost$ holds after any possible execution of $\DLprogsym$.  The program $\DLprogsym$ often has the form \dLHPgeneralform, where \hostE{ctrl} models atomic actions of the controller and does not contain continuous parts; and
\hostE{plant} models evolution of the physical environment and has the form of $\dLprogODE{\dLODE{\DLvarsym}{\DLtermsym}}{\evolconstraint}$.  That is, the system is modeled as unbounded repetitions of a controller action followed by an update to the physical environment.

For example, consider a simple cooling system that operates in an environment where temperature grows at the rate of 1 degree per minute, shown in Figure~\ref{fig:eg-cooling}. 
Let $\hostE{temp}$ be the current temperature of the environment in degrees. The \emph{safety condition} that we would like to enforce (\dLfmlpost) is that $\hostE{temp}$ is no greater than 105 degrees. Let $\hostE{delta}$ be the rate of change of the temperature (degrees per minute). Let $\hostE{t}$ be the time elapsed since the controller was last invoked.

The program $\dLplant$ describes how the physical environment evolves over time interval ($\hostE{1}$ minute): temperature changes according to $\hostE{delta}$ (i.e., $\dLODE{temp}{delta}$) and time passes constantly (i.e., $\dLODE{t}{1}$). The differential equations evolve only within the time interval $\hostE{t \le 1}$ and if $\hostE{temp}$ is non-negative 
(i.e., $\hostE{temp \geq 0}$).

The hybrid program \dLctrl models the system's controller. If the temperature is above 100 degrees, the system activates cooling and the temperature drops at a rate of 0.5 degrees per time unit (i.e., \DLassign{delta}{-0.5}).
The controller doesn't activate cooling under other temperatures. Then the temperature would grow at the rate of 1 degree per minute (i.e., \DLassign{delta}{1}). 

\renewcommand{\arraystretch}{1.2}

\begin{figure}[t]
  \centering
  \begin{tabular}{r@{\;\;}c@{\;\;}l}
\multicolumn{3}{l}{\textbf{Term semantics}} \\
    $ \DLtermsem{\statesym}{\DLvarsym} $ & =& $\statesym(\varsym)$  \\
    $ \DLtermsem{\statesym}{\DLconst} $ & = & $c$  \\
    $ \DLtermsem{\statesym}{\DLtermcomp{\DLtermsym}{\DLtermsymp}} $ & =& $\statesym\DLsem{\DLtermsym} ~\oplus~ \statesym\DLsem{\DLtermsymp}$ where $\oplus$ denotes corresponding arithmetic operations for  $\DLtermcompop \in \{ \hostE{+}, \hostE{\times} \}$  \\
\multicolumn{3}{l}{\textbf{Program semantics}} \\
    $\DLsem{\DLassign{\DLvarsym}{\DLtermsym}}$ & = &$\{\DLsemPair{\statesym}{\statesymp} ~|~ \statesymp(\varsym) = \statesym\DLsem{\DLtermsym}$ and for all other variables $\varsymp \not=\varsym$, $\statesymp(\varsymp) = \statesym(\varsymp)\}$  \\
    
    $\DLsem{\DLassignN{\DLvarsym}}$ & = &$\{ \DLsemPair{\statesym}{\statesymp} ~|~ \statesymp(\varsymp) = \statesym(\varsymp)$ for all variables $\varsymp \not= \varsym \}$  \\

    $\DLsem{\dLprogODE{\dLODE{\DLvarsym}{\DLtermsym}}{\evolconstraint}}$ & = &$\{\DLsemPair{\statesym}{\statesymp} ~|$ exists solution $\varphi:[0,r] \mapsto \allstates$ of $x'=\theta$ with $\varphi(0)=\statesym$ and $\varphi(r)=\statesymp$, \\& &\qquad and $\varphi(t) \in \dLfmlsemstate{\evolconstraint}$ for all $t \in [0,r] \}$  \\
    
    $\DLsem{\DLtest{\DLfmlsym}}$ & =& $\{\DLsemPair{\statesym}{\statesym} ~|~ \statesym \in \dLfmlsemstate{\DLfmlsym} \}$ \\
    
    $\DLsem{\DLseq{\DLprogsym}{\DLprogsymp}}$ & = & $\{\DLsemPair{\statesym}{\statesymp} ~|~ \exists \mu, \DLsemPair{\statesym}{\mu} \in \DLsem{\DLprogsym} \text{~and~}
     \DLsemPair{\mu}{\statesymp} \in \DLsem{\DLprogsymp} \}$\\

   $\DLsem{\DLchoice{\DLprogsym}{\DLprogsymp}}$ & = & $\DLsem{\DLprogsym} \cup \DLsem{\DLprogsymp}$  \\
                                     
    $\DLsem{\DLloop{\DLprogsym}}$ & =& $\DLsem{\DLprogsym}^*$,  the transitive, reflexive closure of $\DLsem{\DLprogsym}$ \\
    
\multicolumn{3}{l}{\textbf{Formula semantics}}  \\
    $\dLfmlsemstate{\F}$  & =& $\emptyset$       \\

    $\dLfmlsemstate{\DLfmlcomp{\DLtermsym}{\DLtermsymp}}$ & = & $\{ \statesym ~|~ 
    \DLtermsem{\statesym}{\DLtermsym}$ $\fmlcompop$ $\DLtermsem{\statesym}{\DLtermsymp} \}$, where $\fmlcompop$ denotes comparison for $\DLfmlcompop$ $\in$ $\{ \hostE{=}, \hostE{\leq}, \hostE{<}, \hostE{\geq}, \hostE{>} \}$ \\

    $\dLfmlsemstate{\DLfmlneg{\DLfmlsym}}$ & = & $\allstates \setminus \dLfmlsemstate{\DLfmlsym}$ \\

    $\dLfmlsemstate{\DLfmlconj{\DLfmlsym}{\DLfmlsymp}}$ & = & $\dLfmlsemstate{\DLfmlsym} \cap \dLfmlsemstate{\DLfmlsymp}$ \\


    $\dLfmlsemstate{\DLfmlUQ{\DLvarsym}{\DLfmlsym}}$ & = & $\dLfmlsemstate{\DLfmlModalA{\DLassignN{\DLvarsym}}{\DLfmlsym}}$ \\


    $\dLfmlsemstate{\DLfmlModalA{\DLprogsym}{\DLfmlsym}}$ & = & $\{~ \statesym ~|~ \forall \statesymp \text{~if~}  \DLsemPair{\statesym}{\statesymp} \in \DLsem{\DLprogsym} \text{~then~} \statesymp \in \dLfmlsemstate{\DLfmlsym} \}$ \\
  \end{tabular}
    \caption{Semantics of hybrid programs and \dL formulas}
    \label{fig:dlsem}
\end{figure}

The formula to be verified, $\dLsafety$, is shown at the last line of Figure~\ref{fig:eg-cooling}. Given an appropriate precondition \dLfmlpre, the axioms and proof rules of $\dL$ can be used to prove that the safety condition \dLfmlpost holds. For this model, assuming the precondition of initial temperature of 100 degrees, i.e., $\dLfmlpre$, we want to ensure the temperature stays no greater than 105 degrees, i.e., $\dLfmlpost$.
The tactic-based theorem prover KeYmaera X \cite{fulton2015keymaera} provides tool support.

To present some of our definitions, we need to refer to the variables that occur in a hybrid program~\cite{Platzer18book,platzer2017complete}. 
We write $\VAR{\DLprogsym}$ and $\VAR{\DLfmlsym}$ to denote, respectively, the set of all variables of program $\DLprogsym$ and formula $\DLfmlsym$. Their definitions can be found in~\ref{appendix:definitions}.   


\begin{figure}
  \begin{align*}
    \dLfmlpre  & \equiv \hostE{temp = 100} \\
    \dLfmlpost & \equiv \hostE{temp \leq 105} \\
    \dLctrl & \equiv {\DLassign{t}{0}} \, \hostE{ ;} \\ 
& \phantom{\qquad \hostE{\cup}} \,  \hostE{(} \DLseq{\DLtest{temp>100}}{\DLassign{delta}{-0.5}} \, \hostE{)}  \\
& \qquad \hostE{\cup} \, \hostE{(} \DLseq{\DLtest{temp \leq 100}}{\DLassign{delta}{1}} \hostE{)}  \\
    \dLplant & \equiv \dLprogODE{\dLODE{temp}{delta}, \dLODE{t}{1}}{(temp \geq 0 \land t \leq 1)} \\                                             
    \dLsafety & \equiv \hostE{\dLfmlpre \rightarrow [\DLloop{(\DLseq{\dLctrl}{\dLplant})}]\dLfmlpost}
  \end{align*}
  \caption{\dL model of a cooling system}
  \label{fig:eg-cooling}
\end{figure}

\subsection{Modeling Sensor Attacks}

Recent work introduces a framework for modeling and analyzing sensor attacks in the setting of hybrid programs and \dL~\cite{xiang2021relational}.  
It models sensing by separately representing physical values and their sensor reads, and then requires that variables holding sensor   reads  are equal to the underlying sensor's value.  See, for instance,  Figure~\ref{fig:eg-cooling} and Figure~\ref{fig:eg-cooling-sensing}. 
Here, \hostE{temp_p} represents the actual physical  temperature and it changes according to \hostE{delta}, while \hostE{temp_s} represents the variable  into which the sensor's value is read. The controller program \hostE{ctrl} sets the sensed values equal to the physical values, i.e., \DLassign{temp_s}{temp_p}, to indicate the sensor is working correctly.

Models of a system under sensor attack can be then derived by manipulating the variables representing the sensor reads.  
For example, with the model shown in Figure~\ref{fig:eg-cooling-sensing}, an attack on the temperature sensor can be modeled by replacing 
\hostE{\DLassign{temp_s}{temp_p}}
with \hostE{\DLassign{temp_s}{*}}, allowing \hostE{temp_s} to take arbitrary values.

We later extend this approach to model \emph{bounded sensor attacks}.

\subsection{Distance Metrics}

To conduct quantitative analysis, we define a notion of \emph{distance between states}, using the \emph{Euclidean distance} $\distsym : \allstates \times \allstates \rightarrow \realSet$:
\begin{equation}
\label{eq:distance}
\diststate{\statesym}{\statesymp} = \sqrt{\sum_{x \in \allvariableSet}(\statesym(x) - \statesymp(x))^2}
\end{equation}

Notice that $\distsym$ is a \emph{metric}, namely, it satisfies the following properties:
$\diststate{\statesym}{\statesymp} = 0$ if and only if $\statesym = \statesymp$;
$\diststate{\statesym}{\statesymp}  = \diststate{\statesymp}{\statesym}$; and
$\diststate{\statesym}{\statesymp} \leq \diststate{\statesym}{\statesympp} + \diststate{\statesympp}{\statesymp}$ for $\statesym, \statesymp, \statesympp \in \allstates$.

For a state $\statesym$ and a real $\epsilon > 0$, the ball of ray $\epsilon$ centered in $\statesym$ is the set of states 
$\balldist{\statesym}{\epsilon} = \{ \statesymp ~|~ \diststate{\statesym}{\statesymp} \leq \epsilon \}$. 

We adopt existing notions~\cite{fainekos2009robustness, boyd2004convex} to specify the distance between a state and a set of states:
\begin{itemize}
\item The distance between a state $\statesym$ and a set of states $\statesetsym \subseteq \allstates$ is the \emph{shortest} distance between $\statesym$ and all states in $\statesetsym$, that is, \[\distStateToSet{\statesym}{\statesetsym} = \infsym\{ \diststate{\statesym}{\statesymp} ~|~ \statesymp \in \statesetsym \}\]

\item The \emph{depth} of $\statesym$ in $\statesetsym \subseteq \allstates$ is the \emph{shortest} distance between $\statesym$ and the \emph{boundary} of $\statesetsym$, that is, \[\depthStateToSet{\statesym}{\statesetsym} = \infsym\{ \diststate{\statesym}{\statesymp} ~|~ \statesymp \in (\allstates \setminus \statesetsym) \}\]

\item The \emph{signed distance} between $\statesym$ and a set of states $\statesetsym \subseteq \allstates$ is defined as follows:
  \begin{align*}
    \signdistStateToSet{\statesym}{\statesetsym} =
    \begin{cases}
      \depthStateToSet{\statesym}{\statesetsym}, & \text{if } \statesym \in \statesetsym \\
      -\distStateToSet{\statesym}{\statesetsym}, & \text{if } \statesym \not\in \statesetsym  \\
    \end{cases}
  \end{align*}
  Note that in the first case the signed distance is a positive real number, while in the second case the signed distance is negative. Thus, $\signdistStateToSet{\statesym}{\statesetsym} > 0$ implies that $\balldist{\statesym}{\epsilon} \subseteq \statesetsym$ for all $\epsilon < \signdistStateToSet{\statesym}{\statesetsym}$, whereas $\signdistStateToSet{\statesym}{\statesetsym} < 0$ implies that $\balldist{\statesym}{\epsilon} \subseteq (\allstates \setminus \statesetsym)$ for all $\epsilon < -\signdistStateToSet{\statesym}{\statesetsym}$.
$\signdistStateToSet{\statesym}{\statesetsym} = 0$ is not very informative. 
\end{itemize}

Here, we assume $\infsym~\emptyset = \infty$ and $\infsym~\realSet = -\infty$, and we let operator $\infsym$ in the set of ${\realSet} \cup \{\infty , -\infty \}$, thus every set has an infimum. 

\section{Quantitative Safety}  \label{sec:q-safety}

The Boolean notion of safety in \dL, e.g., $\safety{\Pwoattack}{\dLfmlpre}{\dLfmlpost}$, does not provide any quantitative information on how ``good'' (i.e., safe) the system is. In this section, we introduce two quantitative notions of safety. The two notions are the foundation of defining forward and backward robustness. They are, respectively, built on the \emph{strongest postcondition} and \emph{weakest precondition} in the setting of \dL. In defining quantitative safety, we use hybrid program $\Pwoattack$ to model a system of interest, formula $\dLfmlpre$ as the precondition of the system, and $\dLfmlpost$ as the postcondition.

\subsection{Extended \dL}

To help define quantitative safety, we extend \dL with another syntactic structure: $\SP{\Pwoattack}{\DLfmlsym}$, which intuitively represents the \emph{strongest postcondition} after the execution of the program $\Pwoattack$ in a state satisfying the precondition $\DLfmlsym$. Its formal definition is the following:
\[
\dLfmlsemstate{\SP{\Pwoattack}{\DLfmlsym}} = \{~ \statesymp ~|~ \exists \statesym \text{~such that~} \statesym \in \dLfmlsemstate{\DLfmlsym} \text{~and~} \DLsemPair{\statesym}{\statesymp}  \in  \DLsem{\Pwoattack}~ \}
\]

Its dual is the modality of necessity $\WP{\Pwoattack}{\DLfmlsym}$, which represents the \emph{weakest precondition} to ensure that $\DLfmlsym$ is satisfied after any execution of program $\Pwoattack$. Its formal definition was shown above in Figure~\ref{fig:dlsem}.

\subsection{Forward Quantitative Safety}

\begin{figure}[t]
  \begin{align*}
    & ... \\
    \dLctrl & \equiv \DLseq{\DLassign{temp_s}{temp_p}}{\DLassign{t}{0}} \, \hostE{ ;} \\ 
& \phantom{\qquad \hostE{\cup}} \,  \hostE{(} \DLseq{\DLtest{temp_s>100}}{\DLassign{delta}{-0.5}} \,  \hostE{)}  \\
& \qquad \hostE{\cup} \, \hostE{(} \DLseq{\DLtest{temp_s\leq 100}}{\DLassign{delta}{1}} \hostE{)}   \\
    \dLplant & \equiv \dLprogODE{\dLODE{temp_p}{delta}, \dLODE{t}{1}}{(temp_p \geq 0 \land t \leq 1)} 
  \end{align*}
  \caption{\dL model of a cooling system with sensing}
  \label{fig:eg-cooling-sensing}
\end{figure}

  A quantitative variation to the Boolean notion of safety,  e.g.,  $\safety{\Pwoattack}{\dLfmlpre}{\dLfmlpost}$, is \emph{forward quantitative safety}, which provides a degree of safety by estimating the room of maneuver to ensure that the system remains in the safety region after any admissible execution. It basically estimates how strong the strongest postcondition $\SP{\Pwoattack}{\dLfmlpre}$ (obtained by the execution of program $\Pwoattack$ in the precondition $\dLfmlpre$) is
  with respect to the postcondition $\dLfmlpost$. 
  In other words, this degree of safety gives an indication of the margins on  possible strengthening of the postcondition $\dLfmlpost$.

  %

\begin{definition}[Forward quantitative safety] 
\label{def:fwd-safety}
Given a real $\Qdegsymu \in \realSet$ and formulas $\dLfmlpre$ and $\dLfmlpost$, a hybrid program $\Pwoattack$ is forward $\Qdegsymu$-safe for $\dLfmlpre$ and $\dLfmlpost$, denoted $\Fsafety{\Pwoattack}{\dLfmlpre}{\dLfmlpost}{\Qdegsymu}$, if 
\[
\Qdegsymu = \infsym \{ \signdistStateToSet{\statesymp}{\dLfmlsemstate{\dLfmlpost}} ~|~ 
\statesymp \in \dLfmlsemstate{\SP{\Pwoattack}{\dLfmlpre}} \} 
\]
\end{definition}

Given a system $\Pwoattack$ and a precondition $\dLfmlpre$, the real number $\Qdegsymu$
measures the \emph{shortest} distance between the set of states satisfying the strongest postcondition $\SP{\Pwoattack}{\dLfmlpre}$ and the set of unsafe states. 
If $\Qdegsymu$ is positive, then all reachable states by the system $\Pwoattack$ from initial states satisfying the precondition $\dLfmlpre$ stay safe. The bigger $\Qdegsymu$ is, the safer the system is. Conversely, if $\Qdegsymu$ is negative, then some reachable states violate the safety condition $\dLfmlpost$. If $\Qdegsymu$ is 0, then the system cannot be considered safe as its safety may depend on very small perturbations of the system's variables~\cite{fainekos2009robustness}.

\begin{example}
\label{ex:fw-safety}
    Consider the cooling system shown in Figure~\ref{fig:eg-cooling} and Figure~\ref{fig:eg-cooling-sensing}, where $\hostE{\dLfmlpre \equiv temp_p = 100}$ and $\hostE{\dLfmlpost \equiv temp_p \leq 105}$. 
During the execution of the system the temperature lies in the real interval $(99.5, 101]$, namely the strongest postcondition  is
$\SP{\Pwoattack}{\dLfmlpre} \equiv \hostE{ temp_p > 99.5 \land temp_p \le 101 }$.
Then, we have $\Fsafety{\Pwoattack}{\dLfmlpre}{\dLfmlpost}{\Qdegsymu}$, where $\Pwoattack = \DLloop{(\DLseq{\dLctrl}{\dLplant})}$, for $\Qdegsymu=4$.
So, $\Qdegsymu$ is our ``degree of safety'' with respect to $\dLfmlpost$: the system will always satisfy the postcondition $\hostE{temp_p \leq 105}$ with a margin of at least 4 degrees.
  
  Now, consider a different postcondition $\hostE{\posts \equiv temp_p <= 101}$, then we have $\Fsafety{\Pwoattack}{\dLfmlpre}{\posts}{\Qdegsymu}$, for $\Qdegsymu = 0$, and the system is actually safe, as  $\safety{\Pwoattack}{\dLfmlpre}{\posts}$ holds.
  However, for a slightly different postcondition $\hostE{\postss \equiv temp_p < 101}$,  we still have $\Fsafety{\Pwoattack}{\dLfmlpre}{\postss}{\Qdegsymu}$, for $\Qdegsymu = 0$,  but the system is actually unsafe, as  $\safety{\Pwoattack}{\dLfmlpre}{\postss}$ is false. This shows that when the degree of safety is 0 we cannot  assess the safety of the system.
  \end{example}

\subsection{Backward Quantitative Safety}

{
  Another quantitative safety notion 
  is \emph{backward quantitative safety}, which estimates how strong the precondition is with respect to the required initial condition for the system to be safe. 
  It provides quantitative
  information on how ``good'' (i.e., strong) the precondition  $\dLfmlpre$ is with respect to the  weakest precondition $\WP{\Pwoattack}{\dLfmlpost}$,  while ensuring safety (i.e., $\dLfmlpost$) after executions of the system \nolinebreak $\Pwoattack$.
  In other words, this degree of safety gives an indication \nolinebreak  of the margins on a possible weakening of the precondition $\dLfmlpre$.
  It is defined as the \emph{shortest} distance from states that satisfy the precondition to any ``bad'' initial states that can lead the system to unsafe states. 
}

\begin{definition}[Backward quantitative safety] 
\label{def:bkw-safety}
Given a real $\Qdegsymr \in \realSet$ and formulas $\dLfmlpre$ and $\dLfmlpost$, a hybrid program $\Pwoattack$ is backward $\Qdegsymr$-safe for $\dLfmlpre$ and $\dLfmlpost$, denoted $\Bsafety{\Pwoattack}{\dLfmlpre}{\dLfmlpost}{\Qdegsymr}$, if 
\[ 
\Qdegsymr = \infsym \{ \signdistStateToSet{\statesym}{\dLfmlsemstate{\WP{\Pwoattack}{\dLfmlpost}}} ~|~ 
\statesym \in \dLfmlsemstate{\dLfmlpre} \}
\]
\end{definition}

Here, if $\Qdegsymr$ is positive then any execution of the system that starts from initial states in $\dLfmlpre$ shall always stay safe. The bigger $\Qdegsymr$ is, the safer the system is. On the contrary, if $\Qdegsymr$ is negative, then some initial states in $\dLfmlpre$ can lead the system's execution to an unsafe state. Similar to the forward quantitative safety, if $\Qdegsymr$ is 0 the system cannot be considered safe. 

\begin{example}
\label{ex:back-safety}
Assuming the precondition ($\hostE{temp_p = 100}$) and the postcondition ($\hostE{temp_p <= 105}$), we have $\Bsafety{\Pwoattack}{\dLfmlpre}{\dLfmlpost}{\Qdegsymr}$, for $\Qdegsymr=5$,
since the weakest precondition is $\WP{\Pwoattack}{\dLfmlpost} \equiv \hostE{temp_p <= 105}$.
Then $\Qdegsymr = 5$ is our ``degree of safety'' with respect to $\dLfmlpre$:  we have a room of maneuver of $5$ on the precondition to ensure
the postcondition  after the execution of $\Pwoattack$.
\end{example}


{
The Boolean version of safety, \safety{\Pwoattack}{\dLfmlpre}{\dLfmlpost} of Definition~\ref{def:safety}, can be expressed in terms of backward quantitative safety. 
\begin{proposition}
   \label{prop:backward}
Given a program $\Pwoattack$ and formula $\dLfmlpre$ and $\dLfmlpost$. 
\begin{itemize}
\item
  If there is $\Qdegsymr > 0$ such that  $\Bsafety{\Pwoattack}{\dLfmlpre}{\dLfmlpost}{\Qdegsymr}$, then $\safety{\Pwoattack}{\dLfmlpre}{\dLfmlpost}$. 
\item  If  $\safety{\Pwoattack}{\dLfmlpre}{\dLfmlpost}$ then there is  $\Qdegsymr  \geq 0$ such  that 
$\Bsafety{\Pwoattack}{\dLfmlpre}{\dLfmlpost}{\Qdegsymr}$. 
\end{itemize}
\end{proposition}
}
The results follow directly from the definitions.

  Note that the two quantitative notions of safety never contradict each other, i.e., if one degree of safety is positive, the other is non-negative. And if one degree is negative, the other is non-positive.
%
\begin{proposition} \label{prop:coherency}
Given a program $\Pwoattack$ and formula $\dLfmlpre$ and $\dLfmlpost$. 
\begin{itemize}
  \item If $\Fsafety{\Pwoattack}{\dLfmlpre}{\dLfmlpost}{\Qdegsymu}$ for some $\Qdegsymu > 0$, then $\Bsafety{\Pwoattack}{\dLfmlpre}{\dLfmlpost}{\Qdegsymr}$ for some $\Qdegsymr \geq 0$;
  \item If $\Bsafety{\Pwoattack}{\dLfmlpre}{\dLfmlpost}{\Qdegsymr}$ for some $\Qdegsymr > 0$, then $\Fsafety{\Pwoattack}{\dLfmlpre}{\dLfmlpost}{\Qdegsymu}$ for some $\Qdegsymu \geq 0$.
  \end{itemize}
\end{proposition}
Intuitively, the two properties hold since positive values of both forward safety and backward safety ensure that by running $\Pwoattack$ from states satisfying the preconditions, only  states satisfying the postcondition can be reached. 
A formal proof is given in~\ref{app:sec:q-safety}.

However, the degree of safety of the two notions are not quantitatively related, i.e., given formula $\dLfmlpre$, $\dLfmlpost$, and a hybrid program $\Pwoattack$, if $\Bsafety{\Pwoattack}{\dLfmlpre}{\dLfmlpost}{\Qdegsymu}$ and $\Fsafety{\Pwoattack}{\dLfmlpre}{\dLfmlpost}{\Qdegsymr}$ for some $\Qdegsymu > 0$ and $\Qdegsymr > 0$, the relationship between 
$\Qdegsymu$ and $\Qdegsymr$ can be arbitrary.

{
  Notice that given a system $\Pwoattack$, a precondition  $\dLfmlpre$ and a postondition $\dLfmlpost$, forward quantitative safety, i.e., $\Fsafety{\Pwoattack}{\dLfmlpre}{\dLfmlpost}{\Qdegsymu}$ \emph{always} holds for some $\Qdegsymu$, since the infimum always exists (even for unsafe systems whose $\Qdegsymu$ is non-positive). The same for backward  safety.  
}



\section{Quantitative Robustness} \label{sec:robust-safety}

This section first introduces the \emph{threat model} on sensor attacks and then provides two notions of robustness to measure the resilience of a CPS, in terms of forward and backward safety, respectively, with respect to a specific sensor attack. 

\subsection{Bounded Sensor Attacks}
\label{sec:bounded-sensor-attacks}

Existing work~\cite{xiang2021relational} considers a threat model of sensor attacks that the attackers can arbitrarily manipulate the sensor readings, e.g., 
compromised temperature sensor is modeled by $\DLassignN{temp_s}$. 
The threat model is too coarse and strong, in particular, when the system under attacks is equipped with some sort of IDS (for instance, anomaly detection IDSs~\cite{ACM-survey2018}) that the attacker would like to evade.

In this work, we consider more refined sensor attacks in which the measurement deviation is bounded. Such finer attacks can be modeled by assignments of the form $\hostE{q_{s} = q_{p} + \offsetsym}$, where $\hostE{q_s}$ and $\hostE{q_p}$ respectively represent sensor and physical values of a real-world quantity, and $\hostE{\offsetsym}$ represents a suitable offset. The idea is that for low values of $| \hostE{\offsetsym} |$ the attack may remain \emph{stealthy}, i.e., undetected by  IDSs. The attack can be formalized as follows: 

\begin{definition}[Bounded \sensorset-sensor attack]
Given a hybrid program $\Pwoattack$, 
a set of sensors $\sensorset \subseteq \VAR{\Pwoattack}$ and an offset function $\offsetsym : \sensorset \rightarrow  \realSetP$, we write 
$\Qattacked{\Pwoattack}{\sensorset}{\offsetsym}$ to denote the program obtained by replacing in $\Pwoattack$  all assignments to variables $\hostE{q_{s}}$ in $\sensorset$, with programs of the form 
\[
\DLassignN{q_{s}} \hostE{;} ~\DLtest{(q_{s} \ge q_{p} -o(q_{s})  \wedge q_{s} \le q_{p} +o(q_{s})} )
\]
\end{definition}

\newcommand{\dLctrlp}{\hostE{ctrl^\prime}}
For example, for the cooling system shown in Figure~\ref{fig:eg-cooling-sensing}, consider a sensor attack introducing an offset $\hostE{0.3}$ to the temperature sensor. Figure~\ref{fig:eg-cooling-sensing-attacked} shows a model of the system with compromised sensors.
\begin{figure}[t]
  \begin{align*}
    \dLctrlp & \equiv \DLassignN{temp_{s}} \hostE{;} \\
             & \qquad \DLtest{(temp_{s} \ge temp_{p} -0.3  \wedge temp_{s} \le
               temp_{p} + 0.3} ) \hostE{;} \\
             & \qquad \quad \cdots
  \end{align*}
\vspace{-5mm}
\caption{\dL model of a cooling system under sensor attack (the omitted part is the same as the model in Figure~\ref{fig:eg-cooling-sensing})} \label{fig:eg-cooling-sensing-attacked}
\end{figure}

The following theorem states that forward safety is affected
by bounded sensor attacks in a proportional manner: the stronger the attack is, the lower the degree of safety of the attacked system.

\newcommand{\Qdegratio}{\delta}
\newcommand{\Qdegsymup}{\Qdegsymu_1}
\newcommand{\Qdegsymrp}{\Qdegsymr_1}

  \begin{theorem} 
 \label{theorem:forward-rob-mono} 
	Assume a hybrid program $\Pwoattack$, a set of sensors $\sensorset \subseteq \VAR{\Pwoattack}$ and two offset functions $\offsetsym_1 : \sensorset \rightarrow  \realSetP$ and
	$\offsetsym_2 : \sensorset \rightarrow  \realSetP$, with $\offsetsym_1(s) \leq \offsetsym_2(s)$ for any  $s \in \sensorset$,
	real numbers $\Qdegsymu, \Qdegsymu_1, \Qdegsymu_2 \in \realSet$, and properties $\dLfmlpre$ and $\dLfmlpost$. Then,  if
  \begin{itemize}
	\item $\Fsafety{\Pwoattack}{\dLfmlpre}{\dLfmlpost}{\Qdegsymu}$
  \item $\Fsafety{\Qattacked{\Pwoattack}{\sensorset}{\offsetsym_1}}{\dLfmlpre}{\dLfmlpost}{\Qdegsymu_1}$
  \item $\Fsafety{\Qattacked{\Pwoattack}{\sensorset}{\offsetsym_2}}{\dLfmlpre}{\dLfmlpost}{\Qdegsymu_2}$
  \end{itemize}
  then $\Qdegsymu_2 \leq \Qdegsymu_1 \leq  \Qdegsymu$. 
\end{theorem}
The detailed proof can be found in~\ref{app:sec:robust-safety}.
A similar theorem for backward safety can be proven:

\begin{theorem} \label{theorem:backward-rob-mono}
	Assume a hybrid program $\Pwoattack$, a set of sensors $\sensorset \subseteq \VAR{\Pwoattack}$ and two offset functions $\offsetsym_1 : \sensorset \rightarrow  \realSetP$ and
$\offsetsym_2 : \sensorset \rightarrow  \realSetP$, with $\offsetsym_1(s) \leq \offsetsym_2(s)$ for any  $s \in \sensorset$,
real numbers $\Qdegsymr, \Qdegsymr_1, \Qdegsymr_2 \in \realSet$, and properties $\dLfmlpre$ and $\dLfmlpost$. Then, if
\begin{itemize} 
\item $\Bsafety{\Pwoattack}{\dLfmlpre}{\dLfmlpost}{\Qdegsymr}$ 
\item $\Bsafety{\Qattacked{\Pwoattack}{\sensorset}{\offsetsym_1}}{\dLfmlpre}{\dLfmlpost}{\Qdegsymr_1}$
 \item $\Bsafety{\Qattacked{\Pwoattack}{\sensorset}{\offsetsym_2}}{\dLfmlpre}{\dLfmlpost}{\Qdegsymr_2}$
\end{itemize} 
then $\Qdegsymr_2 \leq \Qdegsymr_1 \leq  \Qdegsymr$. 
\end{theorem}
Also the proof of this result can be found in~\ref{app:sec:robust-safety}.


\subsection{Quantitative Robustness}

{
  With the definitions of quantitative safety, we can characterize the robustness of a system against sensor attacks as the loss of safety. In particular, the robustness notions are defined by comparing the degree of safety of the original system and the system whose sensors have been compromised.
  We introduce two notions of quantitative robustness, forward and backward robustness, which are built on the notions of forward and backward safety,  respectively. 
}

\paragraph{Forward Robustness}

The first robustness notion, \emph{\Frobustn},  measures, intuitively, how much an attack affects the system's reachable states if the system starts with the expected precondition. 
Forward robustness characterizes the impact of a sensor attack as a ratio: the degree of safety of the compromised system over the degree of safety of the original system. 

\begin{definition}[Quantitative forward robustness] \label{def:forward-rob}
  Given a hybrid program $\Pwoattack$, a set of sensors $\sensorset \subseteq \VAR{\Pwoattack}$,  an offset function  $\offsetsym: \sensorset \rightarrow  \realSetP$, real numbers $\Qdegsymu, \Qdegsymup, \Qdegratio\in \realSet$, and properties $\dLfmlpre$ and $\dLfmlpost$, 
we say that \emph{\Pwoattack is forward {$\Qdegratio$}-robust under \offsetsym-bounded \sensorset-attacks},
written 
$\Frobust{\Pwoattack}{\dLfmlpre}{\dLfmlpost}{\sensorset}{\offsetsym}{\Qdegratio}{}$,  if
\begin{itemize}
\item $\Fsafety{\Pwoattack}{\dLfmlpre}{\dLfmlpost}{\Qdegsymu}$, with $\Qdegsymu > 0$
\item $\Fsafety{\Qattacked{\Pwoattack}{\sensorset}{\offsetsym}}{\dLfmlpre}{\dLfmlpost}{\Qdegsymup}$
\item $\Qdegratio = \Qdegsymup/\Qdegsymu$.
\end{itemize}
\end{definition}

 As expected, forward robustness  applies only to systems that are safe when not exposed to sensor attacks, i.e., $\Qdegsymu > 0$.   
The value of ratio $\Qdegratio$ indicates the system's robustness under the sensor attack. Note that by Theorem~\ref{theorem:forward-rob-mono}, we know that $\Qdegsymup \le \Qdegsymu$. We can analyze $\Qdegratio$ via the following cases: 

 \begin{itemize}
 \item $\Qdegratio = 1$: the attack doesn't affect the system's forward safety.

 \item $0 < \Qdegratio < 1$: 
then $0 < \Qdegsymup < \Qdegsymu$. Given initial states where the precondition holds, reachable states of both the original system and the compromised system stay safe.  
  The value of ($1-\Qdegratio$) quantifies the \emph{percentage of forward safety that is lost} due to the attack. The closer $\Qdegratio$ is to $1$, the more robust the system is.

\item $\Qdegratio \le 0$: then $\Qdegsymu >0$ and $\Qdegsymup \le 0$.
  Executions of the original system stay safe, but the attack may be able to ``break'' the system: some of its executions under attack may run into unsafe states. The lower the value of $\Qdegratio $, the more effective the attack can be.
 If $\Qdegratio = 0$ the attacked system can no longer be considered safe. 
\looseness=-1




 \end{itemize}


\begin{example}
\label{ex:fw-robustness}    
Consider again the cooling system described before. We already know {from Example~\ref{ex:fw-safety}} that $\Fsafety{\Pwoattack}{\dLfmlpre}{\dLfmlpost}{4}$ for $\dLfmlpre \equiv \hostE{temp_p = 100}$ and $\dLfmlpost \equiv \hostE{temp_p \le 105}$, where $\Pwoattack$ models the original system shown in Figure~\ref{fig:eg-cooling} and Figure~\ref{fig:eg-cooling-sensing}.  For the compromised system shown in Figure~\ref{fig:eg-cooling-sensing-attacked}, starting again from $\dLfmlpre$, during executions of $\Qattacked{\Pwoattack}{\sensorset}{\offsetsym}$, the temperature lies in the interval $(99.2,101.3]$. Thus {we have
$\SP{\Qattacked{\Pwoattack}{\sensorset}{\offsetsym}}{\dLfmlpre} \equiv \hostE{99.2 \le temp_p \le 101.3}$ and}
$\Fsafety{\Qattacked{\Pwoattack}{\sensorset}{\offsetsym}}{\dLfmlpre}{\dLfmlpost}{3.7}$. The degree of forward robustness of the original system with  respect to the attack is: $\Qdegratio = 3.7/4 = 0.925$. 
\end{example}

The value of $\Qdegratio$ can help engineers evaluate or compare different defense mechanisms against potential attacks. For a specific set of attacks, a mechanism with less safety loss, i.e., bigger $\Qdegratio$, may be considered better than another one with more safety loss.

Note that using a ratio for $\Qdegratio$ is a better indicator of robustness than using an absolute value, e.g., $\Qdegsymu - \Qdegsymup$: it is consistent regardless of the units of measurement used for safety. For example, the ratio of robustness for a braking system with respect to a sensor attack would be the same whether the safety is measured in feet or in meters.


\paragraph{Backward Robustness}

{
  The second robustness notion, \emph{\Brobustn},
  measures, intuitively, how resilient the initial states that satisfy the precondition are  to sensor attacks whose goal is to drag the system into unsafe states.
%
It characterizes the impact of a sensor attack as a ratio: the degree of safety of the compromised system over the degree of safety of the original system. 
}

\begin{definition}[Quantitative backward robustness] \label{def:backward-rob}
  Given a hybrid program $\Pwoattack$, a set of sensors $\sensorset \subseteq \VAR{\Pwoattack}$,  an offset function  $\offsetsym: \sensorset \rightarrow  \realSetP$,
   real numbers $\Qdegsymr, \Qdegsymrp, \Qdegratio \in \realSet$, and  properties $\dLfmlpre$ and $\dLfmlpost$, 
we say that \emph{\Pwoattack is backward $\Qdegratio$-robust under \offsetsym-bounded \sensorset-attacks},
written
$\Brobust{\Pwoattack}{\dLfmlpre}{\dLfmlpost}{\sensorset}{\offsetsym}{\Qdegratio}{}$,  if
\begin{itemize}
\item $\Bsafety{\Pwoattack}{\dLfmlpre}{\dLfmlpost}{\Qdegsymr}$,  with $\Qdegsymr>0$
\item $\Bsafety{\Qattacked{\Pwoattack}{\sensorset}{\offsetsym}}{\dLfmlpre}{\dLfmlpost}{\Qdegsymrp}$
\item $\Qdegratio = \Qdegsymrp/\Qdegsymr$.
\end{itemize}
\end{definition}

 Again, backward robustness applies only to systems that are safe when not exposed to sensor attacks (i.e., $\Qdegsymr>0$). The meanings of different values of $\Qdegratio$ are analogous to values of $\Qdegratio$ in Definition~\ref{def:forward-rob}.  

\begin{example}
\label{ex:bw-robustness}
Consider again the cooling system. Given $\dLfmlpre \equiv \hostE{temp_p = 100}$ and $\dLfmlpost \equiv \hostE{temp_p \le 105}$, we already know {from Example~\ref{ex:back-safety}} that $\Bsafety{\Pwoattack}{\dLfmlpre}{\dLfmlpost}{\Qdegsymr}$, where $\Pwoattack$ models the original system, for $\Qdegsymr=5.0$. Consider a sensor attack that offsets 0.3 degrees of sensor readings, formula $\WP{\Qattacked{\Pwoattack}{\sensorset}{\offsetsym}}{\dLfmlpost}$ is $\hostE{temp_p  \le 105.0}$. So we know $\Bsafety{\Qattacked{\Pwoattack}{\sensorset}{\offsetsym}}{\dLfmlpre}{\dLfmlpost}{\Qdegsymrp}$
for $\Qdegsymrp = 5.0$. Therefore, the degree of backward robustness of the original system with respect to the attack is: $\Qdegratio = 5.0/5.0 = 1$. Meaning the attack doesn't affect the backward safety of the system. 
\end{example}

\section{Reasoning about Robustness with Simulation Distances} \label{sec:h-eq}

Using Definition~\ref{def:forward-rob} (and~\ref{def:backward-rob}), we can compute  forward (and backward) robustness of a system in terms of the forward (and backward) safety of the system, before and after a bounded sensor attack. 
However, the computation of forward and backward safety may be difficult, as they consider all admissible values to compute the infimum. This is particularly difficult for a system with compromised sensors, due to the complications caused by the offset function. 
%
%
%

In  this  section, we introduce two \emph{simulation distances} between hybrid programs, called \emph{forward simulation distance} (or \emph{forward distance}) and \emph{backward simulation  distance} (or \emph{backward distance}). They quantify the behavioral distance between the original system and the compromised
one, according to a forward and backward flavor, respectively. These distances allow us to compute an upper bound on the loss of forward (and backward) safety. 
{The computed upper bounds are not necessarily tight bounds, but they are easier to reason with and can be verified with existing tools or via the  proof systems proposed in Section \ref{sec:proof-calculus}.}
\looseness=-1

\newcommand{\Bsimulation}[4]{\ensuremath{#1 \sqsubseteq^{\scriptsize\rm{B}}_{{#3,#4}} #2}\xspace}
\newcommand{\Fsimulation}[4]{\ensuremath{#1 \sqsubseteq^{\scriptsize\rm{F}}_{{#3,#4}} #2}\xspace}

\newcommand{\HsigndistStateToSet}[3]{\mathemphfont{Dist}_{\scalebox{0.7}{#3}}(#1,#2)}
\newcommand{\HdepthStateToSet}[3]{\mathemphfont{depth}_{\scalebox{0.7}{#3}}(#1,#2)}

To define forward (and backward) simulation distance between two programs, we extend the notion of distance between states, i.e., $\diststate{\statesym}{\statesymp}$ given in Equation~\ref{eq:distance}, to support computing distance on a set $\Hsymbol$ of variables~\cite{xiang2021relational}.
Intuitively, variables in $\Hsymbol$ are the ones that are relevant to the specified precondition and postcondition. And thus computing distance over these variables gives us the quantitative distance of interest. Consider the cooling system example, we are interested in the behavioral distance between the original program and the compromised one with respect to the variable $\hostE{temp_p}$, rather than $\hostE{temp_s}$.

We introduce a new notion of distance between states with respect to a set $\Hsymbol$ of variables, as follows: 

\begin{definition}
For a set of variables $\Hsymbol \subseteq \allvariableSet$, two states $\statesym$ and $\statesymp$ are at \Hdistance \Hdistdegsym, written
$\Hdiststate{\statesym}{\statesymp}{\Hsymbol} = \Hdistdegsym$, if 
$\sqrt{ \sum_{x \in \Hsymbol} \left(\statesym(x) - \statesymp(x)\right)^2}= d$.
We write $\HsigndistStateToSet{\statesym}{\statesetsym}{\Hsymbol}$ to denote 
$\signdistStateToSet{\statesym}{\statesetsym}$ where  $\Hdiststate{\statesym}{\statesymp}{\Hsymbol}$  is used instead of $\diststate{\statesym}{\statesymp} $.
Then, $\HdepthStateToSet{\statesym}{\statesetsym}{\Hsymbol}$ is defined in the same manner. 
\end{definition}

The following proposition shows that computing the forward and backward safety (using $\diststate{\statesym}{\statesymp}$) can be reduced to a computation using $\Hdiststate{\statesym}{\statesymp}{\Hsymbol}$ with the appropriate variable sets $\Hsymbol$, i.e., $\VAR{\dLfmlpre}$ or $\VAR{\dLfmlpost}$.   
\newcommand{\Qdegsymupp}{\Qdegsymu_2}

{
\begin{proposition} \label{lemma:distance-h}
  For $\Qdegsymu, \Qdegsymup, \Qdegsymupp \in \realSet$, and formula $\DLfmlsym$, $\DLfmlsymp$, if
  \begin{itemize} 
  \item $\Qdegsymu$ = $\infsym \{ \signdistStateToSet{\statesym}{\dLfmlsemstate{\DLfmlsym}} ~|~ 
\statesym \in \DLsem{\DLfmlsymp} \}$
\item $\Qdegsymup$ = $\infsym \{ \HsigndistStateToSet{\statesym}{\DLsem{\DLfmlsym}}{\VAR{\DLfmlsym}} ~|~ 
  \statesym \in \DLsem{\DLfmlsymp} \}$
\item $\Qdegsymupp$ = $\infsym \{ \HsigndistStateToSet{\statesym}{\DLsem{\DLfmlsym}}{\VAR{\DLfmlsymp}} ~|~ 
  \statesym \in \DLsem{\DLfmlsymp} \}$
  \end{itemize}
then $\Qdegsymu = \Qdegsymup = \Qdegsymupp$.
\end{proposition}
}
Intuitively, the proposition holds because the infimum value of $\Qdegsymu$ is essentially decided by the distance calculated with respect to the relevant variables in $\DLfmlsym$ or $\DLfmlsymp$. The detailed proof can be found in~\ref{app:sec:h-eq}. 
\looseness=-1

\subsection{Forward Simulation Distance}

We introduce the notion of \emph{forward simulation distance}.
Intuitively, programs $\Pwoattacka$ and $\Pwoattackb$ are in forward simulation at distance $ \Hdistdegsym$ if given the same initial condition, $\Pwoattackb$ can mimic the behaviors of $\Pwoattacka$, i.e., $\Pwoattackb$ is able to reach states whose distance from those reached by $\Pwoattacka$ is at most  $\Hdistdegsym$.

{
\begin{definition}[Forward simulation distance] \label{def:forward-sim}
  For hybrid programs $\Pwoattacka$, $\Pwoattackb$, formula
${\dLfmlpre}$ and a set of variables $\Hsymbol$, 
$\Pwoattacka$ and $\Pwoattackb$ are at forward simulation distance $\Hdistdegsym$ with respect to ${\dLfmlpre}$ and $\Hsymbol$, written
\[\FHsimulation{\Pwoattacka}{\Pwoattackb}{\dLfmlpre}{\Hsymbol}{\Hdistdegsym}
\]
if for each state $\statesymp_1 \in \dLfmlsemstate{\SP{\Pwoattacka}{\dLfmlpre}}$
 there exists a state $\statesymp_2 \in \dLfmlsemstate{\SP{\Pwoattackb}{\dLfmlpre}}$ 
 such that $\Hdiststate{\statesymp_1}{\statesymp_2}{\Hsymbol} \le \Hdistdegsym$.
\end{definition}
}

{
Here, for programs $\Pwoattack$ and  $\Qattacked{\Pwoattack}{\sensorset}{\offsetsym}$,  the forward simulation 
$\FHsimulation{\Qattacked{\Pwoattack}{\sensorset}{\offsetsym}}{\Pwoattack}{\dLfmlpre}{\Hsymbol}{\Hdistdegsym}\,$  expresses that for each state $\statesymp_1$ reachable by $\Qattacked{\Pwoattack}{\sensorset}{\offsetsym}$, from some initial states in 
$\dLfmlsemstate{\dLfmlpre}$, there is a state $\statesymp_2$ reachable by $\Pwoattack$, from some initial state in 
$\dLfmlsemstate{\dLfmlpre}$, such that $\statesymp_1$  and $\statesymp_2$ are at distance at most $\Hdistdegsym$, for a fixed variable set $\Hsymbol$. The distance $\Hdistdegsym$ gives an upper bound on the perturbation introduced by the attack on the safety  of the behaviors originating from $\dLfmlsemstate{\dLfmlpre}$. The set $\Hsymbol$ here often refers to variables that are relevant to the system's postcondition, i.e., $\VAR{\dLfmlpost}$.

}

\begin{example}

Let $\Pwoattack$ be the program modeling the cooling system shown in Figure~\ref{fig:eg-cooling} and Figure~\ref{fig:eg-cooling-sensing}
and $\Qattacked{\Pwoattack}{\sensorset}{\offsetsym}$ the attacked version shown in Figure~\ref{fig:eg-cooling-sensing-attacked}.
Let $\Hsymbol$ be $\VAR{\dLfmlpost} = \{ \hostE{temp_p} \}$. The forward distance between $\Qattacked{\Pwoattack}{\sensorset}{\offsetsym}$ and $\Pwoattack$ with respect to $\dLfmlpre$ and $\Hsymbol$ is $0.3$, as proven  in the next section. 
Actually, by Example~\ref{ex:fw-robustness} 
we already know that $0.3$ is indeed an upper bound of the loss of forward safety. 
{In Example~\ref{ex:fw-robustness}   we showed that $0.3$ is the ''exact'' loss of forward safety. Therefore, in this case, the upper bound is tight.}
\end{example}
\looseness=-1


The following theorem states that the forward simulation distance $\Hdistdegsym$ between $\Qattacked{\Pwoattack}{\sensorset}{\offsetsym}$ and $\Pwoattack$ with respect to $\VAR{\dLfmlpost}$, is indeed an \emph{upper bound} to the loss of forward safety due to the attack.

{	
\begin{theorem}
\label{thm:forward_rob}
For a hybrid program $\Pwoattack$,  a set of variables $\sensorset \subseteq \VAR{\Pwoattack}$, formulas $\dLfmlpre$ and $\dLfmlpost$,  an offset function $\hostE{\offsetsym}$,  and $\Hdistdegsym, \Qdegsymu \in  \realSet$,
if
\begin{itemize}
\item
$\Fsafety{\Pwoattack}{\dLfmlpre}{\dLfmlpost}{\Qdegsymu}$,  {with $\Qdegsymu>0$}
\item 
$\FHsimulation{\Qattacked{\Pwoattack}{\sensorset}{\offsetsym}}{\Pwoattack}{\dLfmlpre}{\VAR{\dLfmlpost}}{\Hdistdegsym}$ 
\end{itemize}
then $\Frobust{\Pwoattack}{\dLfmlpre}{\dLfmlpost}{\sensorset}{\offsetsym}{\Qdegratio}{}$, for some $\Qdegratio$ such that
{  $\Qdegratio \ge (\Qdegsymu  -  \Hdistdegsym )  / \Qdegsymu $.}
\end{theorem}

The theorem says that $\Hdistdegsym$ is an upper bound of the loss of forward safety, meaning that for some $\Qdegsymup$ such that $\Hdistdegsym \ge \Qdegsymu - \Qdegsymup$ we have $\Fsafety{\Qattacked{\Pwoattack}{\sensorset}{\offsetsym}}{\dLfmlpre}{\dLfmlpost}{\Qdegsymup}$.
The detailed proof can be found in~\ref{app:sec:h-eq}.

\subsection{Backward Simulation Distance}

Symmetrically, we introduce \emph{backward simulation distance} to reason with upper bounds of loss \nolinebreak  of backward safety caused by 
sensor attacks. Intuitively, programs  \nolinebreak  $\Pwoattacka$ and $\Pwoattackb$ are in backward simulation distance $\Hdistdegsym$ if for the same postcondition, $\Pwoattackb$ can mimic the behaviors of $\Pwoattacka$ that may violate the postcondition. This 
means that initial states that can lead to violation of safety condition of the two systems  are distant at most \nolinebreak $\Hdistdegsym$.  
\looseness=-1

\begin{definition}[Backward simulation distance] \label{def:backward-sim}
For hybrid programs $\Pwoattacka$ and $\Pwoattackb$, formula
${\dLfmlpost}$ and a set of variables $\Hsymbol$,
$\Pwoattacka$ and $\Pwoattackb$ are at backward simulation distance 
$\Hdistdegsym$ with respect to
${\dLfmlpost}$ and $\Hsymbol$,
formally written as 
\[
\BHsimulation{\Pwoattacka}{\Pwoattackb}{\dLfmlpost}{\Hsymbol}{\Hdistdegsym}
\]
if for each state 
$\statesym_1 \in \dLfmlsemstate{\DLfmlModalE{\Pwoattacka}{\DLfmlneg{\dLfmlpost}}}$ there exists a state
$\statesym_2 \in \dLfmlsemstate{\DLfmlModalE{\Pwoattackb}{\DLfmlneg{\dLfmlpost}}}$ such that $\Hdiststate{\statesym_1}{\statesym_2}{\Hsymbol} \le \Hdistdegsym$.
\end{definition}

Here, property $\BHsimulation{\Qattacked{\Pwoattack}{\sensorset}{\offsetsym}}{\Pwoattack}{\dLfmlpost}{\Hsymbol}{\Hdistdegsym}$ means that for each initial state  $\statesym_1$, from which $\Qattacked{\Pwoattack}{\sensorset}{\offsetsym}$ can reach an unsafe state in $\dLfmlsemstate{\neg\dLfmlpost}$,  there is an initial state $\statesym_2$, from which $\Pwoattack$ can reach a state in $\dLfmlsemstate{\neg\dLfmlpost}$, such that $\statesym_1$ and $\statesym_2$ are  at distance at most $\Hdistdegsym$, with respect to a set of variables $\Hsymbol$.
Thus, the backward distance between the original and the compromised system returns an upper bound on the admissible perturbations introduced by a sensor attack on the initial states leading to possible violations of safety, fixed a desired postcondition $\dLfmlpost$.  The set $\Hsymbol$ often is the set of variables that are relevant to the system's precondition, i.e., $\VAR{\dLfmlpre}$.

\begin{example}
    Consider again
be the program  $\Pwoattack$, modeling the cooling system shown in Figure~\ref{fig:eg-cooling} and Figure~\ref{fig:eg-cooling-sensing}, 
and $\Qattacked{\Pwoattack}{\sensorset}{\offsetsym}$ the attacked version shown in Figure~\ref{fig:eg-cooling-sensing-attacked}.
Let $\Hsymbol$ be $\VAR{\dLfmlpre} = \{ \hostE{temp_p} \}$. The backward distance between $\Qattacked{\Pwoattack}{\sensorset}{\offsetsym}$ and $\Pwoattack$ with respect to $\dLfmlpost$ and $\Hsymbol$ is $0$ (as proven in the next section). From Example~\ref{ex:bw-robustness} we know that $0$ is indeed an upper bound of the loss of backward safety. 
{Also in this case the upper bound is tight, since in Example~\ref{ex:bw-robustness}  we showed that 0 is the "exact" loss of backward safety.}
\end{example}

The following theorem states that the backward simulation distance $\Hdistdegsym$ between $\Qattacked{\Pwoattack}{\sensorset}{\offsetsym}$ and $\Pwoattack$ with respect to variable set $\VAR{\dLfmlpre}$, is indeed an \emph{upper bound} to the loss of backward safety due to the attack. 
\looseness=-1

{

\begin{theorem}
\label{thm:backward_rob}
For a hybrid program $\Pwoattack$,  a set of variables $\sensorset \subseteq \VAR{\Pwoattack}$, formulas $\dLfmlpre$ and $\dLfmlpost$, an offset function $\hostE{\offsetsym}$ and $\Hdistdegsym, \Qdegsymr \in  \realSet$, if
\begin{itemize}
\item
$\Bsafety{\Pwoattack}{\dLfmlpre}{\dLfmlpost}{\Qdegsymr}${, with $\Qdegsymr>0$}
\item 
$\BHsimulation{\Qattacked{\Pwoattack}{\sensorset}{\offsetsym}}{\Pwoattack}{\dLfmlpost}{\VAR{\dLfmlpre}}{\Hdistdegsym}$ 
\end{itemize}
then $\Brobust{\Pwoattack}{\dLfmlpre}{\dLfmlpost}{\sensorset}{\offsetsym}{\Qdegratio}{}$ for some $\Qdegratio$ such that {
 $\Qdegratio \ge (\Qdegsymr-\Hdistdegsym )/ \Qdegsymr$. }
\end{theorem}

The theorem says that $\Hdistdegsym$ is an upper bound to the loss of backward safety, meaning that we have $\Bsafety{\Qattacked{\Pwoattack}{\sensorset}{\offsetsym}}{\dLfmlpre}{\dLfmlpost}{\Qdegsymrp}$ for some $\Qdegsymrp$ such that $\Hdistdegsym \ge \Qdegsymr - \Qdegsymrp$. 
}
%
The detailed proof can be found in~\ref{app:sec:h-eq}.

Notice that, while both simulation distances are helpful to provide an estimation of the robustness of a \CPS under attack, expressing and verifying simulation distances for a system under sensor attacks may be non-trivial. In the next section, we introduce approaches to encode and verify both simulation distances for systems that may suffer from sensor attacks. 

\newcommand{\SAModality}{{D}-modality\xspace}
\section{Expressing and Verifying Simulation Distances}  \label{sec:proof-calculus}

The content of this section is organized as follows. With the help of a use case on autonomous vehicles, 
we first demonstrate how to encode both forward and backward simulation distances as \dL formulas, in particular via modalities. We then propose an ad-hoc modality, called \SAModality, and formally prove that it can be used to encode both forward and backward simulation distances (Theorem~\ref{thm:proof-calculus-forward-soundness} and Theorem~\ref{thm:proof-calculus-backward-soundness}). Furthermore, we define a proof system to reasoning with D-modality and prove its soundness (Proposition~\ref{lem:proof-calculus-forward-soundness}). 




\subsection{Encoding Simulation Distances as \dL Formulas}

The forward and backward simulation distance are defined upon distance between states that respectively satisfy two formulas. For example, the forward distance (Definition~\ref{def:forward-sim}) is computed on states satisfying, respectively, $\SP{\Pwoattacka}{\dLfmlpre}$ and $\SP{\Pwoattackb}{\dLfmlpre}$. Moreover, both distances are formalized in a ``forall exists'' manner. 
Therefore, a direct way to verify them, is to compute the relevant two formulas, and then verify the distance between states that satisfy the two formulas. 

Based on this insight, the following formula can be instantiated with different formulas to verify both simulation 
distances: 
\[
\hostE{(\DLfmlsym \land (\overline{y} = \overline{x})) \rightarrow \exists \overline{x}.~(\DLfmlsymp \land {(\Hdiststate{\overline{y}}{\overline{x}}{\Hsymbol} \leq \Hdistdegsym))}}
\]
	where $\DLfmlsym$ and $\DLfmlsymp$ are formulas specifying, respectively, conditions of the compromised system and the genuine system. They share the same set of variables.    
  Here $\hostE{\overline{x}}$ are variables used by $\DLfmlsym$ and $\DLfmlsymp$, and $\hostE{\overline{y}}$ are a list of \emph{fresh} variables whose dimension is the same as $\hostE{\overline{x}}$. Variables in $\hostE{\overline{y}}$ are (implicitly) universally quantified. The fresh variables are used to store values of $\hostE{\overline{x}}$ that satisfy the first formula. 
  The notation $\hostE{\Hdiststate{\overline{y}}{\overline{x}}{\Hsymbol}}$ computes the distance between two vectors of variables with respect to the set \Hsymbol:
	$\sqrt{ \sum_{\overline{x}(i) \in \Hsymbol} \left(\overline{x}(i) - \overline{y}(i)\right)^2}$, where $\overline{x}(i)$ and $\overline{y}(i)$ represent, respectively, the $i$th element in vector $\overline{x}$ and $\overline{y}$.

The encoding can be used to verify forward distance by letting $\DLfmlsym$ and $\DLfmlsymp$, respectively, be the formula $\SP{\Qattacked{\Pwoattack}{\sensorset}{\offsetsym}}{\dLfmlpre}$ and  $\SP{\Pwoattack}{\dLfmlpre}$.

\begin{example}
 \label{ex:sec6.1}

Consider the usual cooling system. From Example~\ref{ex:fw-safety}, 
we know the $\SP{\Pwoattack}{\dLfmlpre}$ is given by the following:  $\hostE{99.5 < temp_p \leq 101}$.
For the compromised system shown in Figure~\ref{fig:eg-cooling-sensing-attacked}, 
from Example~\ref{ex:fw-robustness} we know that 
  $\SP{\Qattacked{\Pwoattack}{\sensorset}{\offsetsym}} {\dLfmlpre}$ is  $\hostE{99.2 < temp_p \leq 101.3}$. 

Thus, we can express that $\Qattacked{\Pwoattack}{\sensorset}{\offsetsym}$ and $\Pwoattack$ are at forward distance $0.3$ with respect to $\dLfmlpre$ and $\Hsymbol = \VAR{\dLfmlpre} = \{ \hostE{temp_p} \}$ with the following formula:

\[
    \hostE{(99.2 < temp_p \leq 101.3 \, \land \, \freshV{v_p} = temp_p )} 
    \quad \hostE{\rightarrow} \quad
  \hostE{( \exists temp_p. ~ 99.5 < temp_p \leq 101 \, \land \, (\sqrt{(temp_p-\freshV{v_p})^2} \leq 0.3))}                  
\]
\end{example}


The encoding can also be instantiated for verifying backward simulation distance by letting $\DLfmlsym$ and $\DLfmlsymp$ be
$\DLfmlModalE{\Qattacked{\Pwoattack}{\sensorset}{\offsetsym}}{\DLfmlneg{\dLfmlpost}}$ and $\DLfmlModalE{\Pwoattack}{\DLfmlneg{\dLfmlpost}}$, respectively.

\begin{example}
    \label{ex:sec6.2}

For the cooling system example, {from Example~\ref{ex:back-safety} and Example~\ref{ex:bw-robustness}} we know that $\DLfmlModalE{\Pwoattack}{\DLfmlneg{\dLfmlpost}}$ and $\DLfmlModalE{\Qattacked{\Pwoattack}{\sensorset}{\offsetsym}}{\DLfmlneg{\dLfmlpost}}$ are both $\hostE{temp_p > 105}$. 
We can express $\Qattacked{\Pwoattack}{\sensorset}{\offsetsym}$ and $\Pwoattack$ are at backward simulation distance $0$ with respect to $\dLfmlpost$ and $\Hsymbol = \{ \hostE{temp_p} \}$, with the following formula: 
\[ 
\hostE{(temp_p > 105.0 \land \freshV{v_p} = temp_p)} \quad\hostE{\rightarrow} 
       \quad \hostE{( \exists temp_p. ~temp_p > 105.0 \land \sqrt{(temp_p-\freshV{v_p})^2} \leq 0  } )
\]
  
\end{example}

Both formulas in Example     \ref{ex:sec6.1} and Example     \ref{ex:sec6.2} can be easily  verified with KeYmaera X. 

%

\subsection{Simulation Distances at Work on an Autonomous Vehicle}
\label{sec:vehicle}
We now apply the encoding above to a non-trivial case study. In this case study, we pre-compute the weakest precondition and strongest postcondition involved, in particular, we compute $\SP{\Qattacked{\Pwoattack}{\sensorset}{\offsetsym}}{\dLfmlpre}$ and $\SP{\Pwoattack}{\dLfmlpre}$ for forward distance, and  $\DLfmlModalE{\Qattacked{\Pwoattack}{\sensorset}{\offsetsym}}{\DLfmlneg{\dLfmlpost}}$ and $\DLfmlModalE{\Pwoattack}{\DLfmlneg{\dLfmlpost}}$ for backward distance.

Consider an autonomous vehicle that needs to stop
before hitting an obstacle 
  (Platzer introduces this autonomous vehicle example~\cite{Platzer18book}.) 
For simplicity, we model the vehicle in just one dimension. 
Figure~\ref{fig:eg-vehicle-sensing} shows a \dL model of such an autonomous vehicle with sensing. 
Let \hostE{d_p} and \hostE{d_s}, respectively, be the vehicle's physical and sensed distances from the obstacle.  
The safety condition that we would like to enforce (\dLfmlpost) is that \hostE{d_p} is positive.
Let \hostE{v_p} be the vehicle's velocity towards the obstacle in meters per second (m/s) and \hostE{v_s} be its sensed value. Let \hostE{a} be the vehicle's acceleration (m/s${}^2$). 
Let \hostE{t} be the time elapsed since the controller was last invoked. 

The hybrid program \dLplant describes how
the physical environment evolves over time interval $\hostE{\epsilon}$: 
distance changes according to \hostE{-v_p} (i.e., \dLODE{d_p}{-v_p}), 
velocity changes according to the acceleration (i.e., \dLODE{v_p}{a}), 
and time passes at a constant rate (i.e., \dLODE{t}{1}). 
The differential equations evolve only within the time interval $\hostE{t \le \epsilon}$ 
and if \hostE{v_p} is non-negative 
(i.e., \hostE{v_p \geq 0}).

Program \dLctrl models the vehicle's controller. The
vehicle can either accelerate at \hostE{A} m/$s^2$ or brake at \hostE{-B} m/$s^2$. For the purposes of the model, the controller chooses nondeterministically between these options. Hybrid programs \hostE{accel} and \hostE{brake} express
the controller accelerating or braking (i.e., setting \hostE{a} to \hostE{A} or \hostE{-B} respectively). The controller can accelerate only
if condition \hostE{\DLfmlsymp} is true, which captures that the
vehicle can accelerate for the next $\hostE{\epsilon}$ seconds only if doing
so would still allow it to brake in time to avoid the obstacle.
{In other words, \hostE{\DLfmlsymp} ensures that postcondition $\dLfmlpost \equiv \hostE{d_p} >0$ is satisfied by all runs of the system.}

For the quantitative analysis of this model, we treat symbolic variables $\hostE{A, B, \epsilon}$ as the parameters of the system and set them as constants: \hostE{A =1}, \hostE{B = 1}, and $\hostE{\epsilon = 1}$. In addition, in this case study, we verify the forward and backward distance using the \dL encoding with formulas, after computing the relevant weakest preconditions and strongest postconditions using these constants.
\begin{figure}[H]
	\begin{align*}
		& (\textit{System Constants~:~} \hostE{A = 1}, \hostE{B = 1}, \hostE{\epsilon=1}) \\
		\dLfmlpre  & \equiv \hostE{(2Bd_p > v_p^2) \land v_p \ge 0} \\
		\dLfmlpost & \equiv \hostE{d_p > 0} \\
		\DLfmlsymp & \equiv \hostE{2Bd_s > v_s^2 + (A+B)(A\epsilon^2 + 2v_s\epsilon)} \\
		\hostE{accel} & \equiv \DLseq{\DLtest{\DLfmlsymp}}{\DLassign{a}{A}}  \\
		\hostE{brake} & \equiv \DLassign{a}{-B}  \\
		\dLctrl & \equiv \DLseq{\DLassign{d_s}{d_p}}{\DLassign{v_s}{v_p}} \, ; \hostE{(}\DLchoice{accel}{brake}\hostE{)} 
		\\
		\dLplant & \equiv \dLprogODE{\dLODE{d_p}{-v_p}, \dLODE{v_p}{a}, \dLODE{t}{1}}{(v_p \geq 0 \land t \leq \epsilon)} \\
		\dLsafety & \equiv \hostE{\dLfmlpre \rightarrow [\DLloop{(\DLseq{\dLctrl}{\dLplant})}]\dLfmlpost}
	\end{align*}
	\caption{\dL model of an autonomous vehicle with sensing} \label{fig:eg-vehicle-sensing}
\end{figure}

\paragraph{Bounded Sensor Attack}
The formula $\dLsafety$ specifies the desired (Boolean) safety property: given an appropriate precondition \dLfmlpre,  the safety condition \dLfmlpost holds after any execution of the system. 
The safety property for system $\Pwoattack = \DLloop{(\DLseq{\dLctrl}{\dLplant})}$ indeed holds. 
{However, there is no margin for strengthening the postcondition or for weakening the precondition.
More precisely, the testing \DLtest{\DLfmlsymp} guarding the acceleration ensures that the strongest postcondition is $\SP{\Pwoattack}{\dLfmlpre} \equiv \hostE{d_p > 0}$, which coincides with the postcondition $\dLfmlpost$.  Thus,  by
Definition~\ref{def:fwd-safety}
the system $\Pwoattack$ satisfies $\Fsafety{\Pwoattack}{\hostE{\dLfmlpre}}{\dLfmlpost}{0}$.
Analogously, the weakest precondition is $\WP{\Pwoattack}{\dLfmlpost} \equiv \hostE{(2Bd_p > v_p^2) \land v_p \ge 0}$, which coincides with the precondition $\dLfmlpre$. Thus, 
by Definition~\ref{def:bkw-safety}, system $\Pwoattack$ satisfies $\Bsafety{\Pwoattack}{\hostE{\dLfmlpre}}{\dLfmlpost}{0}$.}

Then,  the system's safety has no room for sensing errors. Any sensor attacks that offset the readings can compromise the safety.

Consider a bounded sensor attack on the velocity sensor that deviates the readings of $\hostE{v_s}$ from $\hostE{v_p}$ up to 1 m/s. We can model it by replacing $\DLassign{v_s}{v_p}$ with 
\[
\DLassignN{v_s}; \DLtest{(v_s \leq v_p+1 \land v_s \geq v_p-1)}
\]
in Figure~\ref{fig:eg-vehicle-sensing}.
The system is not robust against this attack, i.e., the safety property no longer holds when the sensor is compromised.

\paragraph{A Safer Controller}
Now, consider a different controller $\dLctrls$ whose condition for acceleration is designed to tolerate the inaccuracy of sensed velocity at a maneuver of 2 m/s, then the system can then be modeled as follows: \\
\begin{align*}
	\psis & \equiv \hostE{2Bd_s > (v_s+2)^2 + (A+B)(A\epsilon^2 + 2(v_s+2)\epsilon)} \\
	\dLctrls & \equiv \DLseq{\DLassign{d_s}{d_p}}{\DLassign{v_s}{v_p}} \, ; \hostE{(}\DLchoice{(\DLseq{\DLtest{\psis}}{\DLassign{a}{A}})}{\DLassign{a}{-B}} \hostE{)} \\
	& ...
\end{align*}
Let $\Ps$ denote the new system, i.e., $\Ps = \DLloop{(\DLseq{\dLctrls}{\dLplant})}$. It still holds that $\Fsafety{\Ps}{\hostE{\dLfmlpre}}{\dLfmlpost}{0}$ and   $\Bsafety{\Ps}{\hostE{\dLfmlpre}}{\dLfmlpost}{0}$. 

Consider a different precondition:
\begin{align*}
	\pres  & \equiv \hostE{(2Bd_p > (v_p+2)^2) \land v_p \ge 0}        
\end{align*}
Executing $\Ps$ given precondition $\pres$, we get a strongest postcondition $
\SP{\Ps}{\pres} \equiv
\hostE{(2d_p > (v_p+2)^2) \land v_p \ge 0}$. So $\Ps$ is forward safe for a degree of $2$ with respect to $\dLfmlpost$, i.e., $\Fsafety{\Ps}{\hostE{\phi_{pre}^\prime}}{\dLfmlpost}{2}$.

\paragraph{Forward Distance}
We can prove $\Qattacked{\Ps}{\sensorset}{\offsetsym}$ and $\Ps$ are at forward distance $1.5$ with respect to $\pres$ and $\Hsymbol = \VAR{\dLfmlpost} = \{ \hostE{d_p} \}$.  
Formula $\SP{\Qattacked{\Ps}{\sensorset}{\offsetsym}}{\pres}$ is $\hostE{(2d_p > (v_p+1)^2) \land v_p \ge 0}$, so the forward distance can be expressed as: 
\[ \hostE{2d_p > (v_p+1)^2 \land v_p \ge 0 \land d_p = \freshV{d_p}} \quad \hostE{\rightarrow } \quad
	 \hostE{\exists d_p. ((2d_p > (v_p+2)^2) \land v_p \ge 0 \, \land \, \sqrt{(d_p-\freshV{d_p})^2} \leq 1.5))}
\]
Here, $\freshV{d_p}$ is a fresh variable. KeYmaera X easily verifies this formula.
So,  $\FHsimulation{\Qattacked{\Ps}{\sensorset}{\offsetsym}}{\Ps}{\pres}{\{\hostE{d_p}\}}{1.5}$, being $1.5$  the  upper bound of the loss of forward safety.
  Since $\Fsafety{\Ps}{\hostE{\phi_{pre}^\prime}}{\dLfmlpost}{2}$, by Theorem~\ref{thm:forward_rob} it follows:
  \looseness=-1
\begin{center}
$ \Frobust{\Ps}{\pres}{\dLfmlpost}{\sensorset}{\offsetsym}{\Qdegratio}{} $
\end{center}
for some $\Qdegratio \geq \frac{0.5}{2} = 0.25$. So the system is still safe under the attack, and the percentage of forward safety loss is at most $75\%$.

\paragraph{Backward Distance}
%
%
%
%
%
We already know it holds that
$\Bsafety{\Ps}{\dLfmlpre}{\dLfmlpost}{0}$, so there is not much we can learn from backward simulation distance here. 

Now consider the backward safety of $\Ps$ with respect to $\pres$ and a different postcondition $\hostE{\posts \equiv d_p > 0.5}$. We can compute that formula $\DLfmlModalE{\Ps}{\neg\posts}$ is $\hostE{d_p <= 0.5 \lor (2(d_p-0.5) <= v_p^2 \land v_p >=0)}$, and further compute $\Bsafety{\Ps}{\pres}{\posts}{\sqrt{2}}$. Then,  $\DLfmlModalE{\Qattacked{\Ps}{\sensorset}{\offsetsym}}{\neg\posts}$ is
$\hostE{d_p <= 0.5 \lor (2d_p <= (v_p+1)^2 \land v_p >=0)}$. 
 We can express that program $\Qattacked{\Ps}{\sensorset}{\offsetsym}$ and $\Ps$ are at backward distance $1$ with respect to $\posts$ and $\VAR{\pres}$:
\begin{gather*}
   \hostE{((d_p <= 0.5 \lor (2d_p <= (v_p+1)^2 \land v_p >=0))  \land~ \freshV{d_p} = d_p \land \freshV{v_p} = v_p)}
   \\
   \hostE{\rightarrow } \\
   \hostE{\exists d_p~v_p.} \hostE{(d_p <= 0.5 \lor (2(d_p-0.5) <= v_p^2 \land v_p >=0)~\land~ \sqrt{(d_p-\freshV{d_p})^2 + (v_p-\freshV{v_p})^2} \leq 1)}
\end{gather*}
Again, the formula can be verified by KeYmaera X. Then by Theorem~\ref{thm:backward_rob} and $\Bsafety{\Ps}{\pres}{\posts}{\sqrt{2}}$ it follows:
\begin{center}
$ \Brobust{\Ps}{\pres}{\posts}{\sensorset}{\offsetsym}{\Qdegratio}{} $
\end{center}
for some $\Qdegratio \geq \frac{\sqrt{2}-1}{\sqrt{2}}$. So the system is still backward safe under the attack, and the percentage of loss of backward safety due to the attack is at most $1/\sqrt{2} \approx 71\%$.


\subsection{A Modality-based Approach}
\label{sec:modality-approach}

In the case study above, we have pre-computed the strongest postcondition and weakest precondition for the compromised system and the genuine system. However, such a computation may be difficult for certain systems. To alleviate this problem, we introduce another approach for encoding and reasoning with simulation distances using modalities. 


\paragraph{Encoding Simulation Distances with Modalities}

An alternative way to encode the two simulation distances is through modalities (c.f.\ Section~\ref{sec:DDL}) which directly express program executions.  
In the case of  forward distance, i.e., $\FHsimulation{\Qattacked{\Pwoattack}{\sensorset}{\offsetsym}}{\Pwoattack}{\dLfmlpre}{\Hsymbol}{\Hdistdegsym}$, we can express it as the following \dL formula: 

\[  
\hostE{(\dLfmlpre \land \DLfmlModalE{\Qattacked{\Pwoattack}{\sensorset}{\offsetsym}}{(\overline{y} = \overline{x})})} \quad \hostE{\rightarrow} \quad
		\hostE{( \exists \overline{x}. ~\dLfmlpre \land \DLfmlModalE{\Pwoattack}{(\Hdiststate{\overline{y}}{\overline{x}}{\Hsymbol} \leq \Hdistdegsym)} })
\]

  The left-hand-side formula encodes ``for each state that can be reached from precondition $\dLfmlpre$ after an execution of the compromised program''. The fresh variables of $\hostE{\overline{y}}$ are used to record the reachable states.
The right-hand-side formula encodes ``there is an execution of the genuine program under precondition $\dLfmlpre$ such that the distance between the corresponding final
	states is bound by \hostE{\Hdistdegsym}.''

	We can similarly express the backward distance 
 $\BHsimulation{\Qattacked{\Pwoattack}{\sensorset}{\offsetsym}}{\Pwoattack}{\dLfmlpost}{\Hsymbol}{\Hdistdegsym}$:

 \[
\hostE{( (\overline{y} = \overline{x})  \, \land \, \DLfmlModalE{\Qattacked{\Pwoattack}{\sensorset}{\offsetsym}}{\DLfmlneg{\dLfmlpost}})}
\quad \hostE{\rightarrow}
\quad
 \hostE{(\exists \overline{x}.~  (\Hdiststate{\overline{x}}{\overline{y}}{\Hsymbol} \leq \Hdistdegsym) \, \land \, \DLfmlModalE{\Pwoattack}{\DLfmlneg{\dLfmlpost}} ) }
 \]


The left-hand-side formula encodes ``for each initial state that can lead the compromised system to unsafe states''. The fresh variables of $\hostE{\overline{y}}$ are used to record the initial states.
The right-hand-side formula encodes ``there is an initial state that can lead the genuine program to unsafe states such that the distance between the two initial	states is bound by \hostE{\Hdistdegsym}".
\looseness=-1
	 
  


\paragraph{\SAModality}
\label{sec_super_modality}
The modality-based encodings above closely follow the definitions of forward and backward distances. However, they may be cumbersome and difficult to reason with due to the quantifiers. 
To address this problem, we introduce a more efficient modality, denoted \SAModality, to readily encode both forward and backward distances. We develop a proof system to reason with this new modality in an effective manner. 
The proof system contains a set of proof rules  which can be derived from existing \dL axioms and proof rules~\cite{Platzer18book,platzer2017complete}. 
\looseness=-1
%
%

The \SAModality 
is designed to capture the forall exists relationship between the compromised program and the genuine program. To capture such a relationship, the modality needs to refer to variables in both programs. Since the compromised and genuine programs often refer to the same set of variables, the \SAModality renames variables used by the genuine program into a fresh set of variables, and then captures the relationship between the two programs using both the original and renamed variables.


To rename variables we use renaming functions. 
\begin{definition}[Renaming function for \DLprogsym] \label{def:var-rename}  For a \dL program \DLprogsym and a set of variables $V$ such that $V \cap \VAR{\DLprogsym} = \emptyset$, a function $\renamingsym : \VAR{\DLprogsym} \rightarrow V$ is a \emph{renaming function} for $\DLprogsym$ if it is a bijection.
\end{definition}

We write $\renaming{\DLprogsym}{\renamingsym}$ for the program equivalent to $\DLprogsym$ but whose variables have been renamed according to $\renamingsym$. Renaming functions similarly apply to \dL formulas and states. 
Moreover, for a state $\statesym \colon \VAR{\DLprogsym} \mapsto \realSet$, 
we let $\renaming{\statesym}{\renamingsym}$ denote the state 
$\statesymp \colon \renaming{\VAR{\DLprogsym}}{\renamingsym} \mapsto \realSet$ such that 
$\statesymp(\renaming{x}{\renamingsym}) = \statesym(x)$ for all $x \in \VAR{\DLprogsym}$. 

\medskip

We now introduce \SAModality, a special forall exists modality, written  $\ModalATT{\Pwoattack}{\sensorset}{\offsetsym}{\renamingsym}{\DLfmlsym}$, and defined as 
\[
\ModalATT{\Pwoattack}{\sensorset}{\offsetsym}{\renamingsym}{\DLfmlsym} \equiv \DLfmlModalA{ \Qattacked{\Pwoattack}{\sensorset}{\offsetsym} }{ \DLfmlModalE{\renaming{\Pwoattack}{\renamingsym}}{\DLfmlsym}}
\] 
for a program $\Pwoattack$, a set of sensors $\sensorset \subseteq \VAR{\Pwoattack}$, an offset function $\offsetsym$, and a renaming function $\renamingsym$.  

Intuitively, $\ModalATT{\Pwoattack}{\sensorset}{\offsetsym}{\renamingsym}{\DLfmlsym}$ expresses the ability by the genuine program to simulate the compromised program with respect to a formula $\DLfmlsym$, which models a relational property on the states reachable by the two programs.
More precisely, $\ModalATT{\Pwoattack}{\sensorset}{\offsetsym}{\renamingsym}{\DLfmlsym}$ formalises that for each state
$\statesymp$ reachable by the compromised program $\Qattacked{\Pwoattack}{\sensorset}{\offsetsym}$ there is a state $\statesymp'$ reachable by the genuine program $\Pwoattack$ such that $\statesymp \uplus \renaming{\statesymp'}{\renamingsym}$ satisfy 
$\DLfmlsym$\footnote{Here, as expected, the symbol $\uplus$ denotes disjoint union.}.
\looseness=-1

\begin{proposition}
\label{lem:newMod}
 For a hybrid program $\Pwoattack$,  a set of variables $\sensorset \subseteq \VAR{\Pwoattack}$, an offset function $\hostE{\offsetsym}$, a formula $\DLfmlsym$ and a renaming function $\renamingsym$, the set of states  
$\dLfmlsemstate{         \ModalATT{\Pwoattack}{\sensorset}{\offsetsym}{\renamingsym}{\DLfmlsym}} $ is equal to 
\[
\left\{\statesym \uplus
\renaming{\statesym'}{\renamingsym}
 ~|~ \forall \statesymp \text{~if~}  \DLsemPair{\statesym}{\statesymp} \in \DLsem{\Qattacked{\Pwoattack}{\sensorset}{\offsetsym} } \text{~then~}
     \exists \statesymp' \text{~s.t.~}  (\renaming{\statesym'}{\renamingsym} ,
 \renaming{\statesymp'}{\renamingsym})
 \in \DLsem{\renaming{\Pwoattack}{\renamingsym}} \text{~and~} \statesymp \uplus \renaming{\statesymp'}{\renamingsym}
 \in \dLfmlsemstate{\DLfmlsym}
\right\} .
\]
\end{proposition}

Note that $\ModalATT{\Pwoattack}{\sensorset}{\offsetsym}{\renamingsym}{\DLfmlsym}$ is a \dL formula, thus the sound and complete proof calculus of \dL~\cite{platzer2017complete} can be used to reason with this modality.

\paragraph{Encoding Forward Distances using the  \SAModality}

In the following, we show how we can use the \SAModality $\ModalATT{\Pwoattack}{\sensorset}{\offsetsym}{\renamingsym}{\DLfmlsym}$ to encode the forward distance. 
The following theorem introduces the \SAModality based encoding of the forward distance and states its soundness. 
%
Intuitively, if both genuine and compromised programs start under a precondition $\DLfmlsymp$ then 
for any reachable state of the compromised program, there exists a reachable state of the genuine program satisfying $\DLfmlsym$, which ensures an upper bound of distance between the two reachable states. 
The formulas $\DLfmlsym$ and $\DLfmlsymp$ satisfy certain requirements to match the definition of the forward distance. Formally speaking, 

 \begin{theorem}[\SAModality based Encoding of Forward Distance]
    \label{thm:proof-calculus-forward-soundness}
    For a hybrid program $\Pwoattack$,  a set of variables $\sensorset \subseteq \VAR{\Pwoattack}$, an offset function $\hostE{\offsetsym}$,  formulas $\DLfmlsymp$, $\DLfmlsym$ and $\dLfmlpre$, a renaming function $\renamingsym$, a set of variables $\Hsymbol$ and $\Hdistdegsym \in  \realSet$,
      if we have
  \begin{itemize}[topsep=3pt]
    \setlength\itemsep{3pt}
    \item 
    $
        \DLfmlsymp
        \hostimply
        \ModalATT{\Pwoattack}{\sensorset}{\offsetsym}{\renamingsym}{\DLfmlsym} 
    $
    \item 
     $ \dLfmlpre \hostimply
         \exists 
         \renaming{\VAR{\dLfmlpre}}{\renamingsym}. \, (\DLfmlsymp
         \land
    \renaming{\dLfmlpre}{\renamingsym})$ 
   
    \item $\DLfmlsym \hostimply \simuHdist{\renamingsym}{\Hsymbol}{\Hdistdegsym}
    $, for $\HdistRenaming{\Hsymbol}{\renamingsym} = \sqrt{ \sum_{x \in \Hsymbol} \left(x - \renaming{x}{\renamingsym}\right)^2}$
  \end{itemize}
   then  
   $\FHsimulation{\Qattacked{\Pwoattack}{\sensorset}{\offsetsym}}{\Pwoattack}{\dLfmlpre}{\Hsymbol}{\Hdistdegsym}$. 
 \end{theorem}

Let us go in some detail of the three requirements of the theorem. 
Here, the formula $\DLfmlsymp$ captures the relationship between the initial states of the two programs $\Qattacked{\Pwoattack}{\sensorset}{\offsetsym}$ and $\renaming{\Pwoattack}{\renamingsym}$. 
The first requirement expresses the following:  if an execution of $\Qattacked{\Pwoattack}{\sensorset}{\offsetsym}$ starting from an initial state satisfying $\DLfmlsymp$ 
reaches a state where $\DLfmlsym$ holds, then there exists an execution of $\renaming{\Pwoattack}{\renamingsym}$ starting from the same condition that reaches a state where $\DLfmlsym$ holds as well. 
%
The second requirement says that each state in $\DLsem{\dLfmlpre}$ 
is in relation  with at least one state in $\DLsem{\renaming{\dLfmlpre}{\renamingsym}}$ under the condition $\DLfmlsymp$. 
The first two requirements together capture the description ``for each state $\statesymp_1 \in \dLfmlsemstate{\SP{\Pwoattacka}{\dLfmlpre}}$, there exists a state $\statesymp_2 \in \dLfmlsemstate{\SP{\Pwoattackb}{\dLfmlpre}}$'' in the definition of the forward distance.
The third requirement specifies that the postcondition $\DLfmlsym$ ensures an upper bound of distance between reachable states of the two programs. It matches the description ``$\Hdiststate{\statesymp_1}{\statesymp_2}{\Hsymbol} \le \Hdistdegsym$'' in the definition of the forward distance. 

The detailed proof of this theorem can be found in~\ref{app:sec:proof-calculus}.

\paragraph{Encoding Backward Distances using the  \SAModality}

We can also use the \SAModality 
$\ModalATT{\Pwoattack}{\sensorset}{\offsetsym}{\renamingsym}{\DLfmlsym}$ for encoding the backward distance. The following theorem introduces the \SAModality based encoding of backward distances and states its soundness.  
Intuitively, if both genuine and compromised programs start from a precondition $\DLfmlsym$ ensuring an upper bound $d$ on its states, then for any reachable state of the compromised program that violates the postcondition $\dLfmlpost$, there exists a reachable state of the genuine program that violates $\dLfmlpost$. The precondition $\DLfmlsym$ ensures that all possible initial states of the compromised program are considered. Formally,


\begin{theorem}[\SAModality based Encoding of Backward Distance]
  \label{thm:proof-calculus-backward-soundness}
  For a hybrid program $\Pwoattack$, a set of variables $\sensorset \subseteq \VAR{\Pwoattack}$,  an offset function $\hostE{\offsetsym}$, formulas $\DLfmlsym$ and $\dLfmlpost$, a renaming function $\renamingsym$, a set of variables $\Hsymbol$, and $\Hdistdegsym \in \realSet$, if we have
  \begin{itemize}[topsep=3pt]
    \setlength\itemsep{3pt}
    \item 
    ${ 
       \DLfmlsym \hostimply  \ModalATT{\Pwoattack}{\sensorset}{\offsetsym}{\renamingsym}{(\DLfmlneg{\dLfmlpost} \hostimply \renaming{\DLfmlneg{\dLfmlpost}}{\renamingsym})} 
     }$
     \item $ \forall~ \VAR{\Pwoattack}~\exists \renaming{\VAR{\Pwoattack}}{\renamingsym}.\, \DLfmlsym $ 
     
   
    \item $ \DLfmlsym \hostimply \simuHdist{\renamingsym}{\Hsymbol}{\Hdistdegsym} $, for $\HdistRenaming{\Hsymbol}{\renamingsym} = \sqrt{ \sum_{x \in \Hsymbol} \left(x - \renaming{x}{\renamingsym}\right)^2}$
  \end{itemize}
   then 
   $\BHsimulation{\Qattacked{\Pwoattack}{\sensorset}{\offsetsym}}{\Pwoattack}{\dLfmlpost}{\Hsymbol}{\Hdistdegsym}$.
\end{theorem}

Let us explain the three requirements of the theorem. 
The formula $\DLfmlsym$ captures the relationship between the initial states of the two programs $\Qattacked{\Pwoattack}{\sensorset}{\offsetsym}$ and $\renaming{\Pwoattack}{\renamingsym}$. 
The first requirement  expresses the following: if an execution of $\Qattacked{\Pwoattack}{\sensorset}{\offsetsym}$ starts  
from a state satisfying $\DLfmlsym$ and it  reaches a state where $\DLfmlneg{\dLfmlpost}$ holds, 
 then there exists an execution of  $\renaming{\Pwoattack}{\renamingsym}$ starting from the same condition 
 that reaches a state where 
 $\renaming{\DLfmlneg{\dLfmlpost}}{\renamingsym}$ holds. 
%
The second requirement specifies that for any initial state of the compromised program, there always exists an initial state of the genuine program which can make the precondition $\DLfmlsym$ true. That is, the formula $\DLfmlsym$ does not restrict the possible values of variables in \VAR{\Pwoattack}, thus the initial states of the compromised program $\Qattacked{\Pwoattack}{\sensorset}{\offsetsym}$ can be arbitrary.   
Finally, similarly to the previous theorem, the third requirement  specifies that the precondition $\DLfmlsym$ ensures that $d$ is an upper bound of distance between the initial states of the two programs. 

The detailed proof of this theorem can be found in~\ref{app:sec:proof-calculus}.
\looseness=-1

\medskip

We have presented the \SAModality based encodings. To reason with the encodings, especially with item 1 in both Theorem~\ref{thm:proof-calculus-forward-soundness} and \ref{thm:proof-calculus-backward-soundness}, we develop a proof system, i.e., a set of proof rules  for the \SAModality. 

\begin{table}[t]
  \begin{mathpar}

    \label{rule:Fdef} \inferrule*[Lab=\rulenameFdef]
    {}
    {  
      \DLfmlModalA{ \Qattacked{\Pwoattack}{\sensorset}{\offsetsym} }{ \DLfmlModalE{\renaming{\Pwoattack}{\renamingsym}}{\DLfmlsym} }  \hostlrarrow \ModalATT{\DLprogsym}{\sensorset}{\offsetsym}{\renamingsym}{\DLfmlsym}
    }

    \\\\

    \label{rule:Fseq} \inferrule*[Lab=\rulenameFseq ]
    {
    \sequent{\Gamma}
       { 
         \ModalATT{\DLprogsym}{\sensorset}{\offsetsym}{\renamingsym}{ 
           \ModalATT{\DLprogsymp}{\sensorset}{\offsetsym}{\renamingsym}{\DLfmlsym} 
         }
       }
    }
    { 
      \sequent
      {\Gamma}
      {
        \ModalATT{\DLseq{\DLprogsym}{\DLprogsymp}}{\sensorset}{\offsetsym}{\renamingsym}{\DLfmlsym}
      }
    }

    \label{rule:Fmr} \inferrule*[Lab=\rulenameFmr ]
    {
      \sequent{\Gamma}{ \ModalATT{\Pwoattack}{\sensorset}{\offsetsym}{\renamingsym}{\DLfmlsym} } 
      \\
      \sequent{\DLfmlsym}{\DLfmlsymp}
    }
    { 
      \sequent{\Gamma}{ \ModalATT{\Pwoattack}{\sensorset}{\offsetsym}{\renamingsym}{\DLfmlsymp} }
    }
    
    \label{rule:Fand} \inferrule*[Lab=\rulenameFand]
    {
    \sequent{\Gamma}{\ModalATT{\DLprogsym}{\sensorset}{\offsetsym}{\renamingsym}{(\DLfmlsym \hostand \DLfmlsymp)}}
    }
    {  
       \sequent{\Gamma}{\ModalATT{\DLprogsym}{\sensorset}{\offsetsym}{\renamingsym}{\DLfmlsym} \hostand 
       \ModalATT{\DLprogsym}{\sensorset}{\offsetsym}{\renamingsym}{\DLfmlsymp} }       
    }

    \label{rule:For} \inferrule*[Lab=\rulenameFor]
    {
      \sequent
      {\Gamma}
      {\ModalATT{\DLprogsym}{\sensorset}{\offsetsym}{\renamingsym}{\DLfmlsym} \hostor \ModalATT{\DLprogsym}  
            {\sensorset}{\offsetsym}{\renamingsym}{\DLfmlsymp}}
    }
    {  
      \sequent
      {\Gamma}
      { \ModalATT{\DLprogsym}{\sensorset}{\offsetsym}{\renamingsym}{(\DLfmlsym \hostor \DLfmlsymp)} }
    }
    
    \label{rule:Fchoice} \inferrule*[Lab=\rulenameFchoice]
    {
     \sequent{\Gamma}{\ModalATT{\DLprogsym}{\sensorset}{\offsetsym}{\renamingsym}{\DLfmlsym}}
     \\ 
     \sequent{\Gamma}{\ModalATT{\DLprogsymp}{\sensorset}{\offsetsym}{\renamingsym}{\DLfmlsym}}
    }
    {  
      \sequent{\Gamma}{ \ModalATT{\DLchoice{\DLprogsym}{\DLprogsymp}}{\sensorset}{\offsetsym}{\renamingsym}{\DLfmlsym} }
    }

    \label{rule:Ftest} \inferrule*[Lab=\rulenameFtest]
    {}
    {
      (\DLfmlsymtest \rightarrow (\renaming{\DLfmlsymtest}{\renamingsym} \land \DLfmlsym))
      \leftrightarrow   
      \ModalATT{\DLtest{\DLfmlsymtest}}{\sensorset}{\offsetsym}{\renamingsym}{\DLfmlsym} 
    }


    \label{rule:Finv} \inferrule*[Lab=\rulenameFinv]
    {
      \sequent{\Gamma}{ \DLfmlsymInv } \\ 
      \sequent{\DLfmlsymInv}{ \ModalATT{{\DLprogsym}}{\sensorset}{\offsetsym}{\renamingsym}{\DLfmlsymInv} } \\
      \sequent{\DLfmlsymInv}{\DLfmlsymp}
    }
    { 
      \sequent{\Gamma}{\ModalATT{\DLloop{\DLprogsym}}{\sensorset}{\offsetsym}{\renamingsym}{\DLfmlsymp}}
    }

    \\\\

    \label{rule:Fv} \inferrule*[Lab=\rulenameFv]
    {
      \sequent{\DLfmlsym}{ \DLfmlModalA{ \Qattacked{\Pwoattack}{\sensorset}{\offsetsym} }{\hostE{\DLfmlsym_1}}} \\
      \sequent{\DLfmlsymp}{ \DLfmlModalE{ \Qattacked{\Pwoattack}{\sensorset}{\offsetsym} }{\hostE{\DLfmlsymp_1}} }
    }
    {  
      \sequent{(\DLfmlsym \hostand \DLfmlsymp)}{\ModalATT{\DLprogsym}{\sensorset}{\offsetsym}{\renamingsym}{(\DLfmlsym_1 \hostand \DLfmlsymp_1)}} 
    }    
    \\
    \text{~where~} (\VAR{\DLfmlsym} \cup \VAR{\DLfmlsym_1}) \cap \VAR{\renaming{\Pwoattack}{\renamingsym}}  = \emptyset \text{~and~} (\VAR{\DLfmlsymp} \cup \VAR{\DLfmlsymp_1}) \cap \VAR{\Pwoattack}  = \emptyset

    \\\\
    
    \label{rule:FodeForall}  \inferrule*[Lab=\rulenameFodeForall]
    {
      \sequent
      {
        \Gamma
      }
      {
        \DLfmlModalA{ \dLODE{\DLvarsym}{\DLtermsym}}{\DLfmlModalA{\renaming{\dLODE{\DLvarsym}{\DLtermsym}}{\renamingsym} }{( \DLfmlsymp(t, \renaming{t}{\renamingsym}) \hostimply \DLfmlsym )}} 
      }
    }
    { \sequent
      {
        \Gamma
      }
      {
        \ModalATT{\dLODE{\DLvarsym}{\DLtermsym}, \dLODE{t}{1}}{\sensorset}{\offsetsym}{\renamingsym}{\DLfmlsym}
      }
    }

    \label{rule:Fodemerge}  \inferrule*[Lab=\rulenameFodemerge]
    {
      \sequent
      {
        \Gamma
      }
      {
        \DLfmlModalA{ \hostE{\dLODE{\DLvarsym}{\DLtermsym}}, ~ \renaming{ \dLODE{\DLvarsym}{\DLtermsym} }{\renamingsym} }{\DLfmlsym} 
      }
    }
    { \sequent
      {
        \Gamma
      }
      {
        \ModalATT{ \dLprogODE{\dLODE{\DLvarsym}{\DLtermsym}, \dLODE{t}{1}}{ t \leq \epsilon } }{\sensorset}{\offsetsym}{\renamingsym}{\DLfmlsym}
      }
    }	


  \end{mathpar}
  \caption{\label{rules-fwd} Proof system for the derived modality}
  \label{fig:calculus-rules}
  \vspace{3mm}
\end{table}   

\subsection{A Proof System for \SAModality}
\label{sec:calculus-proof-rules}

Our proof system allows us to derive  \emph{sequents} \cite{Platzer18book} of the form $\sequent{\Gamma}{\Delta}$, where \emph{antecedent} $\Gamma$ and \emph{succedent} $\Delta$ are finite sets of $\dL$ formulas. 
The semantics of $\sequent{\Gamma}{\Delta}$ is that of the \dL formula
  $\bigwedge_{\DLfmlsym \in \Gamma} \DLfmlsym \rightarrow \bigvee_{\DLfmlsymp \in \Delta} \DLfmlsymp$.

The set of proof rules are given in
Table \ref{fig:calculus-rules}. 
For completeness, in Figure~\ref{psc} and Figure~\ref{other-rules}
of the appendix we report the original axioms and proof rules of Platzer's 
$\dL$ formulas~\cite{platzer2017complete,Platzer18book}.
Essentially, we extend the existing axioms and proof rules from~\cite{Platzer18book,platzer2017complete} with a set of proof rules
to \emph{effectively} reason with the modality $\ModalATT{\Pwoattack}{\sensorset}{\offsetsym}{\renamingsym}{\DLfmlsym}$.




%
Rule \rulenameFdef is defined based on the definition of the \SAModality. This rule can be applied on every program construct. 
Rule \rulenameFseq decomposes the reasoning for a sequential composition of two programs into reasoning for both of them.
Rule \textsc{\rulenameFmr} specifies that the \SAModality holds for a postcondition $\DLfmlsymp$ if the modality holds for a stronger postcondition $\DLfmlsym$.   
Rule \rulenameFand and \rulenameFor reason with, respectively, the \SAModality on a conjunction and disjunction of two formulas. 
Rule \rulenameFchoice deals with the nondeterministic choice of two subprograms. 
Rule \rulenameFtest follows the semantics of the \SAModality and $\DLtest{\DLfmlsymtest}$. 
Rule \rulenameFinv is the loop invariant rule. We can reason with a program with loop $\DLloop{\DLprogsym}$ by identifying an invariant $\DLfmlsymInv$. 
Rule \rulenameFv reduces the reasoning of a \SAModality into two separate proof obligations of the compromised program and the genuine program. 
%
Rule \rulenameFodeForall 
reduces the reasoning with a \SAModality to reasoning with the modality of necessity. Intuitively, the formula $\DLfmlsym$ holds for all executions of \hostE{\dLODE{\DLvarsym}{\DLtermsym}} and \hostE{\renaming{\dLODE{\DLvarsym}{\DLtermsym}}{\renamingsym}}  if a timing condition $\DLfmlsymp$  holds. Note that formula $\DLfmlsymp$ only refers to variables \hostE{t} and \hostE{\renaming{t}{\renamingsym}}; and there always exists a value of \hostE{\renaming{t}{\renamingsym}} that can make $\DLfmlsymp(\hostE{t}, \hostE{\renaming{t}{\renamingsym}})$ true.  
Rule \rulenameFodemerge is a simplified version of \rulenameFodeForall, obtained by merging the two modalities of necessity in \rulenameFodeForall into a single modality of necessity on the dynamics \hostE{ \dLODE{\DLvarsym}{\DLtermsym}} and \hostE{\renaming{ \dLODE{\DLvarsym}{\DLtermsym} }{\renamingsym} }.
With this merging, all solutions of \hostE{\dLODE{\DLvarsym}{\DLtermsym}} and \hostE{\renaming{\dLODE{\DLvarsym}{\DLtermsym}}{\renamingsym}}  evolve for the same amount of time. 
%
Note that a proof obligation like $\DLfmlModalA{ \hostE{\dLODE{\DLvarsym}{\DLtermsym}}, ~ \renaming{ \dLODE{\DLvarsym}{\DLtermsym} }{\renamingsym} }{\DLfmlsym}$ is often easier to verify.

\medskip

 The soundness proof of these rules can be found in~\ref{app:sec:proof-calculus}. Now, we can state the soundness of the proof system. 
\begin{proposition}[Soundness of the rules in Table\ref{fig:calculus-rules}]
\label{lem:proof-calculus-forward-soundness}
If a sequent 
$\sequent{\Gamma}{\Delta}$ can be derived through the proof system 
then the formula 
  $\bigwedge_{\DLfmlsym \in \Gamma} \DLfmlsym \rightarrow \bigvee_{\DLfmlsymp \in \Delta} \DLfmlsymp$ holds.
\end{proposition}

 \medskip

\section{Proof System at Work on a Water Tank System} 
\label{sec:casestudy}

%


\newcommand{\WTmaxL}{x_L}
\newcommand{\WTfstMode}{\textit{on}}
\newcommand{\WTsndMode}{\textit{off}}
\newcommand{\WTborder}{25}

In this case study, we demonstrate how to use the proof rules introduced in Table~\ref{fig:calculus-rules} to reason with the forward and backward distances.

Consider an example of a water tank shown in Figure~\ref{fig:eg-watertank-sensing}, which is inspired by literature~\cite{boyce2012elementary}. It mixes the salt and water inside the tank. Initially, it contains 50 lb of salt (\hostE{Q_p = 50}) dissolved in 100 gal of water (\hostE{\WTmaxL = 100}). An inflow of water containing 1/4 lb of salt/gal is entering the tank at a rate of $\hostE{\WTR}$ gal/min. The well-stirred mixture is draining from the tank at the same rate $\hostE{\WTR}$. 
The tank has two modes of control: a switch-on mode with $\hostE{r}$ set to \hostE{10} if the measured salt level is high ($\DLtest{\WTsalts \ge \WTborder}$) and a switch-off mode with with $\hostE{\WTR}$ set to \hostE{0} if the measured salt level is low ($\DLtest{\WTsalts < \WTborder}$).   
The rate of change of salt in the tank $\hostE{\WTsaltp^\prime}$ is equal to the rate at which salt is flowing in minus the rate at which is flowing out: $\dLODE{\WTsaltp}{(\WTR/4 - \WTR*\WTsaltp/\WTmaxL)}$, where $\hostE{\WTsaltp/\WTmaxL}$ computes the concentration of the salt. 
The postcondition we are interested is that the salt level stays above a certain level ($\hostE{\WTsaltp \ge 20}$). Note that $\hostE{\WTsaltp} = 25$ is an equilibrium of the dynamics if $\WTR >0 $. 
\begin{figure}[H]
	\begin{align*}
		& (\textit{System Constants~:~} \hostE{\epsilon=1 \land \WTmaxL =100}) \\
		\dLfmlpre  & \equiv \hostE{\WTsaltp = 50} \\
		\dLfmlpost & \equiv \hostE{\WTsaltp \ge 20} \\
		\hostE{\WTfstMode} & \equiv \DLseq{\DLtest{\WTsalts \ge \WTborder}}{\DLassign{\WTR}{10}}  \\
		\hostE{\WTsndMode} & \equiv \DLseq{\DLtest{\WTsalts < \WTborder}}{\DLassign{\WTR}{0}}  \\
		\dLctrl & \equiv {\DLassign{\WTsalts}{\WTsaltp}} \, ; \hostE{(}\DLchoice{\WTfstMode}{\WTsndMode}\hostE{\,)} 
		\\
		\dLplant & \equiv \dLprogODE{( \dLODE{\WTsaltp}{\WTR/4 - \WTR*\WTsaltp/\WTmaxL}, \dLODE{t}{1})}{(t \leq \epsilon)} \\
		\dLsafety & \equiv \hostE{\dLfmlpre \rightarrow [\DLloop{(\DLseq{\dLctrl}{\dLplant})}]\dLfmlpost}
	\end{align*}
	\caption{\dL model of a water tank with sensing} \label{fig:eg-watertank-sensing} 
 \vspace{2mm}
\end{figure}

\paragraph{Bounded Sensor Attack} 

Consider a bounded sensor attack on the sensor of salt level that deviates the readings of $\hostE{\WTsalts}$ from $\hostE{\WTsaltp}$ up to 3 lb. We can model it by replacing $\DLassign{\WTsalts}{\WTsaltp}$ in Figure~\ref{fig:eg-watertank-sensing} with: 
\[
\DLassignN{\WTsalts}; \DLtest{(\WTsalts \leq \WTsaltp+3 \land \WTsalts \geq \WTsaltp-3)}
\]

\paragraph{Forward Distance}
We can use the proof rules in Table~\ref{fig:calculus-rules} and other \dL axioms and rules to prove that $\Qattacked{\Ps}{\sensorset}{\offsetsym}$ and $\Ps$ are at forward distance $3$ with respect to $\dLfmlpre$ and $\Hsymbol = \VAR{\dLfmlpost} = \{ \hostE{\WTsaltp} \}$.

We show such a proof in Figure~\ref{fig:water-tank-proof-forward}. Figure~\ref{fig:water-tank-proof-formulas-forward} presents the definitions of formulas and programs used in Figure~\ref{fig:water-tank-proof-forward}. The proof rules used are shown at the left of each derivation step. A key step is applying invariant rule \rulenameFinv, which allows the proof to focus on an invariant $\DLfmlsymInv$. The invariant specifies that, intuitively, given the initial condition ($\dLfmlpre$), for any salt level the compromised system can reach, there exists an execution of the genuine system whose reachable salt level would be within the same regions. The proof considers two regions of the salt level: the first one is that the initial (physical) salt level stays above 28, captured by the formula \DLfmlsymInvb. If both the genuine system and the compromised system begin executions from states specified by \DLfmlsymInvb, then for any salt level reachable by the compromised system, the genuine system can reach the same level by taking the same value of $\WTR$ and then evolving for the same period. Figure~\ref{fig:water-tank-proof-heavy-forward} and Figure~\ref{fig:water-tank-proof-plant-forward} show, respectively, the proofs for the controller and plant in this scenario. 
The second region is where the initial (physical) salt level stays between 25 (non-inclusive) and 28, captured by the formula \DLfmlsymInva. If both the genuine system and the compromised system begin executions from this region, then the reachable states of both systems stay within this region, no matter whether $\WTR$ is set to $10$ or $0$. The derivation for the obligation $\sequent{ \DLfmlsymInva }{ \ModalATT{{\DLprogsym}}{\sensorset}{\offsetsym}{\renamingsym}{\DLfmlsymInva} }$ (left branch of Figure~\ref{fig:water-tank-proof-forward}) shows the proof for this scenario. 
In addition to the two scenarios, Figure~\ref{fig:water-tank-proof-light-forward} shows the proof that the two systems keep $\WTR$ at $10$ if they both start with a salt level above $28$. 
A notable step in Figure~\ref{fig:water-tank-proof-plant-forward} is applying the proof rule \rulenameFodemerge. It reduces the proof obligation into another one with a modality of necessity on the combined dynamics. The new obligation can be proven using the regular \dL rules. 


\begin{figure}[H]
  \vspace{-3mm}
  \begin{align*}
    \dLfmlpre & \equiv~  \hostE{\WTsaltp=50 \land \WTsaltprn =50}\\
    \dLfmlpost & \equiv~ \hostE{\sqrt{(\WTsaltp - \WTsaltprn)^2} \le 3} \text{~(same as~} \hostE{\simuHdist{\renamingsym}{\Hsymbol}{3}}) \\
    \renamingsym & \equiv~ \{ \WTsaltp \mapsto \WTsaltprn, \WTsalts \mapsto \WTsaltsrn, \WTR \mapsto \WTRrn \} \\
    \DLfmlsym & \equiv~  \DLfmlsymInvb \hostand (\WTsaltp - 3 \le \WTsalts \le \WTsaltp + 3) \\
    \DLfmlsymp & \equiv~  \DLfmlsymInvb \hostand \hostE{ (\WTR = \WTRrn = 10) } \\
    \DLfmlsympp  & \equiv~ \hostE{(\WTsaltp = \WTsaltprn \land \WTsaltp > 25)}\\
    \DLfmlsymInv & \equiv~  \DLfmlsymInva \hostE{\lor\,} \DLfmlsymInvb \\
    \DLfmlsymInval & \equiv~ \hostE{25 < \WTsaltp < 28} \\
    \DLfmlsymInvar & \equiv~ \hostE{25 < \WTsaltprn < 28} \\
    \DLfmlsymInva & \equiv~ \DLfmlsymInval \hostand \DLfmlsymInvar \\
    \DLfmlsymInvb & \equiv~ \hostE{(\WTsaltp = \WTsaltprn \land \WTsaltp \ge 28)}\\
    \DLprogsym & \equiv~ (\DLseq{\dLctrl}{\dLplant}) \text{~in Figure~\ref{fig:eg-watertank-sensing}} \\
    \Patt & \equiv~ \Qattacked{\Pwoattack}{\sensorset}{\offsetsym}
  \end{align*}
  \caption{Formulas used in the example proof of a forward distance for the water tank case study}
  \label{fig:water-tank-proof-formulas-forward}
\end{figure}

\def\labelSpacing{0pt}
\def\defaultHypSeparation{\quad}

\begin{figure}[H]
\prooftreesize
  \begin{prooftree}

   \AxiomC{ $ \sequent{\dLfmlpre}{ \DLfmlsymInv} $}
    
   \AxiomC{By \dL rules}
   \UnaryInfC{$\sequent{ \DLfmlsymInval }{ \DLfmlModalA{\Patt}{\DLfmlsymInval} }$}
    
   \AxiomC{By \dL rules}
   \UnaryInfC{$\sequent{ \DLfmlsymInvar }{ \DLfmlModalE{\renaming{\DLprogsym}{\renamingsym}}{\DLfmlsymInvar} }$}

   \LeftLabel{\prooflabel{\hyperref[rule:Fv]{\rulenameFv}}}
   \BinaryInfC{$\sequent{ \DLfmlsymInva }{ \ModalATT{{\DLprogsym}}{\sensorset}{\offsetsym}{\renamingsym}{\DLfmlsymInva} }$}
   
   \LeftLabel{\prooflabel{\hyperref[rule:wr]{\rulenameWR}}}   
   \UnaryInfC{$\sequent{ \DLfmlsymInva }{ \ModalATT{{\DLprogsym}}{\sensorset}{\offsetsym}{\renamingsym}{\DLfmlsymInva} 
      \hostor 
      \ModalATT{{\DLprogsym}}{\sensorset}{\offsetsym}{\renamingsym}{\DLfmlsymInvb}
   }$}

   \LeftLabel{\prooflabel{\hyperref[rule:For]{\rulenameFor}}}
   \UnaryInfC{$\sequent{ \DLfmlsymInva }{ \ModalATT{{\DLprogsym}}{\sensorset}{\offsetsym}{\renamingsym}{\DLfmlsymInv} }$}
   
   \AxiomC{Figure~\ref{fig:water-tank-proof-continued-forward}}
   \UnaryInfC{$\sequent{ \DLfmlsymInvb }{ \ModalATT{{\DLprogsym}}{\sensorset}{\offsetsym}{\renamingsym}{\DLfmlsymInv} }$}
   
   \LeftLabel{\prooflabel{\hyperref[rule:orL]{\rulenameOrL}}}     
   \BinaryInfC{$\sequent{\DLfmlsymInv}{\ModalATT{{\DLprogsym}}{\sensorset}{\offsetsym}{\renamingsym}{\DLfmlsymInv}}$}
   
   \AxiomC{$ \sequent{\DLfmlsymInv}{\dLfmlpost} $ }
   \LeftLabel{\prooflabel{\hyperref[rule:Finv]{\rulenameFinv}}} 
   \TrinaryInfC{$\sequent{}{\dLfmlpre \hostimply \ModalATT{\DLloop{\DLprogsym}}{\sensorset}{\offsetsym}{\renamingsym}{\dLfmlpost}}$}
 \end{prooftree}
 \vspace{-1mm}
 \caption{Proof of a forward distance for the water tank case study}
   \label{fig:water-tank-proof-forward}
 \end{figure}

\begin{figure}[H]
\prooftreesize
 \begin{prooftree}

  \AxiomC{By \rulenameFdef and then \dL rules}
  \UnaryInfC{$\sequent{\DLfmlsymInvb}{  \ModalATT{\DLassign{\WTsalts}{\WTsaltp}}{\sensorset}{\offsetsym}{\renamingsym}{ 
    \DLfmlsym
   } 
   }$
  }

  \AxiomC{Figure~\ref{fig:water-tank-proof-heavy-forward}}
  \AxiomC{Figure~\ref{fig:water-tank-proof-light-forward}}
  \LeftLabel{\prooflabel{\hyperref[rule:Fchoice]{\rulenameFchoice}}} 
  \BinaryInfC{$\sequent{\DLfmlsym}{  
    \ModalATT{\DLchoice{\WTfstMode}{\WTsndMode}}{\sensorset}{\offsetsym}{\renamingsym}{\DLfmlsymp} 
   }$
  }

  \AxiomC{Figure~\ref{fig:water-tank-proof-plant-forward}}
   \UnaryInfC{$\sequent{\DLfmlsymp}{  
    \ModalATT{\dLplant}{\sensorset}{\offsetsym}{\renamingsym}{\DLfmlsympp} 
   }$ 
   }

   \AxiomC{*}
   \UnaryInfC{$\sequent{\DLfmlsympp}{\DLfmlsymInv}$}
   
   \LeftLabel{\prooflabel{\hyperref[rule:Fmr]{\rulenameFmr}}} 
   \BinaryInfC{$\sequent{\DLfmlsymp}{  
    \ModalATT{\dLplant}{\sensorset}{\offsetsym}{\renamingsym}{\DLfmlsymInv} 
   }$ 
   }
   
  \LeftLabel{\prooflabel{\hyperref[rule:Fmr]{\rulenameFmr}}} 
  \BinaryInfC{$\sequent{\DLfmlsym}{  
    \ModalATT{(\DLchoice{\WTfstMode}{\WTsndMode})}{\sensorset}{\offsetsym}{\renamingsym}{ 
      \ModalATT{\dLplant}{\sensorset}{\offsetsym}{\renamingsym}{\DLfmlsymInv} 
     }
   }$
  }
  \LeftLabel{\prooflabel{\hyperref[rule:Fseq]{\rulenameFseq}}} 
  \UnaryInfC{$\sequent{\DLfmlsym}{  
    \ModalATT{\DLseq{ (\DLchoice{\WTfstMode}{\WTsndMode}) } {\dLplant} }{\sensorset}{\offsetsym}{\renamingsym}{\DLfmlsymInv} 
   }$
  }
  \LeftLabel{\prooflabel{\hyperref[rule:Fmr]{\rulenameFmr}}} 
  \BinaryInfC{$\sequent{\DLfmlsymInvb}{  \ModalATT{\DLassign{\WTsalts}{\WTsaltp}}{\sensorset}{\offsetsym}{\renamingsym}{ 
    \ModalATT{\DLseq{ (\DLchoice{\WTfstMode}{\WTsndMode}) } {\dLplant}}{\sensorset}{\offsetsym}{\renamingsym}{\DLfmlsymInv} 
   } 
   }$
  }
  \LeftLabel{\prooflabel{\hyperref[rule:Fseq]{\rulenameFseq}}}
  \UnaryInfC{$\sequent{\DLfmlsymInvb}{\ModalATT{{\DLprogsym}}{\sensorset}{\offsetsym}{\renamingsym}{\DLfmlsymInv}}$}

\end{prooftree}
\vspace{-1mm}
\caption{Proof of a forward distance for the water tank case study (continued)}
  \label{fig:water-tank-proof-continued-forward}
\end{figure}

\begin{figure}[H]
\prooftreesize
  \begin{prooftree}

    \AxiomC{By \rulenameFdef and then \dL rules}
    \UnaryInfC{$\sequent{\DLfmlsym}{  \ModalATT{\DLtest{( \WTsalts \ge \WTborder)}}{\sensorset}{\offsetsym}{\renamingsym}{ 
        \DLfmlsymInvb
    } 
    }$
    }

    \AxiomC{By \rulenameFdef and then \dL rules}
    \UnaryInfC{$\sequent{\DLfmlsymInvb}{  
        \ModalATT{\DLassign{\WTR}{10}}{\sensorset}{\offsetsym}{\renamingsym}{\DLfmlsymp} 
       }$ 
    }
    \LeftLabel{\prooflabel{\hyperref[rule:Fmr]{\rulenameFmr}}} 
    \BinaryInfC{$\sequent{\DLfmlsym}{  \ModalATT{\DLtest{( \WTsalts \ge \WTborder )}}{\sensorset}{\offsetsym}{\renamingsym}{ 
        \ModalATT{\DLassign{\WTR}{10}}{\sensorset}{\offsetsym}{\renamingsym}{\DLfmlsymp} 
       } 
       }$
    }
    \LeftLabel{\prooflabel{\hyperref[rule:Fseq]{\rulenameFseq}}}
    \UnaryInfC{$\sequent{\DLfmlsym}{  
     \ModalATT{\WTheavy}{\sensorset}{\offsetsym}{\renamingsym}{\DLfmlsymp} 
     }$
    }
 \end{prooftree}
 \vspace{-1mm}
 \caption{Proof of a forward distance for the water tank case study (switch-on mode)}
   \label{fig:water-tank-proof-heavy-forward}
 \end{figure}


 \begin{figure}[H]
 \prooftreesize
  \begin{prooftree}
   \AxiomC{*}
   \LeftLabel{\prooflabel{\hyperref[rule:closefalse]{\rulenameFalse}}} 
   \UnaryInfC{$\sequent{\DLfmlsym \hostand \hostE{( \WTsalts < \WTborder)} }{  
     (\renaming{\DLfmlsymtest}{\renamingsym} \land (\ModalATT{ \DLassign{\WTR}{5} }{\sensorset}{\offsetsym}{\renamingsym}{\DLfmlsymp}))  
    }$
   }
   \LeftLabel{\prooflabel{\hyperref[rule:implyR]{
    \rulenameImplyR}}} 
   \UnaryInfC{$\sequent{\DLfmlsym}{  
    { \hostE{( \WTsalts < \WTborder)}  } \rightarrow (\renaming{\DLfmlsymtest}{\renamingsym} \land (\ModalATT{ \DLassign{\WTR}{5} }{\sensorset}{\offsetsym}{\renamingsym}{\DLfmlsymp}))  
    }$
   }
   \LeftLabel{\prooflabel{\hyperref[rule:Ftest]{\rulenameFtest}}} 
   \UnaryInfC{$\sequent{\DLfmlsym}{  \ModalATT{\DLtest{( \WTsalts < \WTborder)}}{\sensorset}{\offsetsym}{\renamingsym}{ 
    \ModalATT{\DLassign{\WTR}{5}}{\sensorset}{\offsetsym}{\renamingsym}{\DLfmlsymp} 
   } 
   }$
   }
   \LeftLabel{\prooflabel{\hyperref[rule:Fseq]{\rulenameFseq}}} 
   \UnaryInfC{$\sequent{\DLfmlsym}{  
    \ModalATT{\WTlight}{\sensorset}{\offsetsym}{\renamingsym}{\DLfmlsymp} 
    }$
   }
 \end{prooftree}
 \vspace{-1mm}
 \caption{Proof of a forward distance for the water tank case study (off mode)}
   \label{fig:water-tank-proof-light-forward}
 \end{figure}

 \begin{figure}[H]
 \prooftreesize
 \begin{prooftree}
 
  \AxiomC{By \dL rules}
  \UnaryInfC{$\sequent{\DLfmlsymp}{  
   \DLfmlModalA{\dLODE{\WTsaltp}{\WTR/4 - \WTR*\WTsaltp/100},~ \dLODE{\WTsaltprn}{\WTRrn/4 - \WTRrn*\WTsaltprn/100}}{\hostE{\DLfmlsympp}} 
   }
  $ 
  }
  \LeftLabel{\prooflabel{\hyperref[rule:Fodemerge]{\rulenameFodemerge}}} 
  \UnaryInfC{$\sequent{\DLfmlsymp}{  
   \ModalATT{\dLplant}{\sensorset}{\offsetsym}{\renamingsym}{\DLfmlsympp} 
   }$ 
  }

\end{prooftree}
 \vspace{-1mm}
 \caption{Proof of a forward distance for the water tank case study (plant)}
   \label{fig:water-tank-proof-plant-forward}
 \end{figure}

The proof above establishes that $\sequent{}{\dLfmlpre \hostimply \ModalATT{\DLloop{\DLprogsym}}{\sensorset}{\offsetsym}{\renamingsym}{\dLfmlpost}}$ for $\dLfmlpre \equiv \hostE{\WTsaltp=50 \land \WTsaltprn =50}$ and 
$\dLfmlpost \equiv \hostE{\simuHdist{\renamingsym}{\Hsymbol}{3}})$. Then by Proposition~\ref{lem:proof-calculus-forward-soundness} and  Theorem~\ref{thm:proof-calculus-forward-soundness} (the other conditions in Theorem~\ref{thm:proof-calculus-forward-soundness} are trivially satisfied), we know that 
$\FHsimulation{\Qattacked{\Pwoattack}{\sensorset}{\offsetsym}}{\Pwoattack}{\dLfmlpre}{\Hsymbol}{\Hdistdegsym}$ for $\Hdistdegsym = 3$ and $\Hsymbol = \{\WTsaltp\}$.



\paragraph{Backward Distance}

We can also use the proof rules in Table~\ref{fig:calculus-rules} and other \dL axioms and rules to prove that $\Qattacked{\Ps}{\sensorset}{\offsetsym}$ and $\Ps$ are at backward distance $0$ with respect to $\dLfmlpost$ and $\Hsymbol = \VAR{\dLfmlpost} = \{ \hostE{\WTsaltp} \}$.

We present a proof in Figure~\ref{fig:water-tank-proof-backward}. Figure~\ref{fig:water-tank-proof-formulas-backward} presents the definitions of formulas and programs used in Figure~\ref{fig:water-tank-proof-backward}. 
The proof is similar to the example proof for forward distance. The key step is applying invariant rule \rulenameFinv, with an invariant $\DLfmlsymInv$. The invariant specifies that, intuitively, given the initial condition $\DLfmlsym$, for any salt level the compromised system can reach, there exists an execution of the genuine system whose reachable salt level would be within the same regions. The proof considers two regions of the salt level: the first one is that the initial (physical) salt level stays above 20, captured by the formula \DLfmlsymInvb. If both the genuine system and the compromised system begin executions from states specified by \DLfmlsymInvb, then for any salt level reachable by the compromised system, the genuine system can reach the same level by taking the same value of $\WTR$ and then evolving for the same period. Figure~\ref{fig:water-tank-proof-light-backward} and Figure~\ref{fig:water-tank-proof-plant-backward} show, respectively, the proofs for the controller and plant in this scenario. 
The second region is where the initial (physical) salt level stays above 20 (inclusive), captured by the formula \DLfmlsymInva. If both the genuine system and the compromised system begin executions from this region, then the reachable states of both systems stay within this region, no matter whether $\WTR$ is set to $10$ or $0$. The left branch of Figure~\ref{fig:water-tank-proof-backward} shows the proof for this scenario. 
In addition to the two scenarios, Figure~\ref{fig:water-tank-proof-heavy-backward} shows the proof that the two systems would keep $\WTR=0$ if they both start with a salt level below $20$.


\begin{figure}[H]
  \vspace{-2mm}
  \begin{align*}
    \DLfmlsym & \equiv~  \hostE{\WTsaltp = \WTsaltprn}\\
    \dLfmlpost & \equiv~ \hostE{\WTsaltp \ge 20} \hostor \hostE{(\WTsaltp < 20 \hostand \WTsaltprn < 20)} \text{~(same as~} \hostE{\WTsaltp < 20 \hostimply \WTsaltprn < 20})\\ 
    \renamingsym & \equiv~ \{ \WTsaltp \mapsto \WTsaltprn, \WTsalts \mapsto \WTsaltsrn, \WTR \mapsto \WTRrn \} \\
    \DLfmlsympp & \equiv~  \DLfmlsymInvb \hostand (\WTsaltp - 3 \le \WTsalts \le \WTsaltp + 3) \\
    \DLfmlsymp & \equiv~  \DLfmlsymInvb \hostand \hostE{ (\WTR = \WTRrn = 5) } \\
    \DLfmlsymInv & \equiv~  \DLfmlsymInva \hostE{\lor\,} \DLfmlsymInvb \\
    \DLfmlsymInval & \equiv~ \hostE{\WTsaltp \ge 20} \\
    \DLfmlsymInvar & \equiv~ \hostE{\WTsaltprn \ge 20} \\
    \DLfmlsymInva & \equiv~ \DLfmlsymInval \hostand \DLfmlsymInvar \\
    \DLfmlsymInvb & \equiv~ \hostE{(\WTsaltp = \WTsaltprn \land \WTsaltp < 20)}\\
    \DLprogsym & \equiv~ (\DLseq{\dLctrl}{\dLplant}) \text{~in Figure~\ref{fig:eg-watertank-sensing}} \\
    \Patt & \equiv~ \Qattacked{\Pwoattack}{\sensorset}{\offsetsym} \\  
  \end{align*}
  \caption{Formulas used in the example proof of a backward distance for the water tank case study}
  \label{fig:water-tank-proof-formulas-backward}
\end{figure}

\def\labelSpacing{0pt}
\def\defaultHypSeparation{\quad}

\begin{figure}[H]
\prooftreesize
  \begin{prooftree}

   \AxiomC{ $ \sequent{\DLfmlsym}{ \DLfmlsymInv} $}
    
   \AxiomC{By \dL rules}
   \UnaryInfC{$\sequent{ \DLfmlsymInval }{ \DLfmlModalA{\Patt}{\DLfmlsymInval} }$}
    
   \AxiomC{By \dL rules}
   \UnaryInfC{$\sequent{ \DLfmlsymInvar }{ \DLfmlModalE{\renaming{\DLprogsym}{\renamingsym}}{\DLfmlsymInvar} }$}

   \LeftLabel{\prooflabel{\hyperref[rule:Fv]{\rulenameFv}}}
   \BinaryInfC{$\sequent{ \DLfmlsymInva }{ \ModalATT{{\DLprogsym}}{\sensorset}{\offsetsym}{\renamingsym}{\DLfmlsymInva} }$}
   
   \LeftLabel{\prooflabel{\hyperref[rule:wr]{\rulenameWR}}}   
   \UnaryInfC{$\sequent{ \DLfmlsymInva }{ \ModalATT{{\DLprogsym}}{\sensorset}{\offsetsym}{\renamingsym}{\DLfmlsymInva} 
      \hostor 
      \ModalATT{{\DLprogsym}}{\sensorset}{\offsetsym}{\renamingsym}{\DLfmlsymInvb}
   }$}

   \LeftLabel{\prooflabel{\hyperref[rule:For]{\rulenameFor}}}
   \UnaryInfC{$\sequent{ \DLfmlsymInva }{ \ModalATT{{\DLprogsym}}{\sensorset}{\offsetsym}{\renamingsym}{\DLfmlsymInv} }$}
   
   \AxiomC{Figure~\ref{fig:water-tank-proof-continued-backward}}
   \UnaryInfC{$\sequent{ \DLfmlsymInvb }{ \ModalATT{{\DLprogsym}}{\sensorset}{\offsetsym}{\renamingsym}{\DLfmlsymInv} }$}
   
   \LeftLabel{\prooflabel{\hyperref[rule:orL]{\rulenameOrL}}}     
   \BinaryInfC{$\sequent{\DLfmlsymInv}{\ModalATT{{\DLprogsym}}{\sensorset}{\offsetsym}{\renamingsym}{\DLfmlsymInv}}$}
   
   \AxiomC{$ \sequent{\DLfmlsymInv}{\dLfmlpost} $ }

   \LeftLabel{\prooflabel{\hyperref[rule:Finv]{\rulenameFinv}}} 
   \TrinaryInfC{$\sequent{}{\DLfmlsym \hostimply \ModalATT{\DLloop{\DLprogsym}}{\sensorset}{\offsetsym}{\renamingsym}{\dLfmlpost}}$}
 \end{prooftree}
 \vspace{-1mm}
 \caption{Proof of a backward distance for the water tank case study}
   \label{fig:water-tank-proof-backward}
 \end{figure}

\begin{figure}[H]
\prooftreesize
 \begin{prooftree}

  \AxiomC{By \rulenameFdef and then \dL rules}
  \UnaryInfC{$\sequent{\DLfmlsymInvb}{  \ModalATT{\DLassign{\WTsalts}{\WTsaltp}}{\sensorset}{\offsetsym}{\renamingsym}{ 
    \DLfmlsympp
   } 
   }$
  }

  \AxiomC{Figure~\ref{fig:water-tank-proof-heavy-backward}}
  \AxiomC{Figure~\ref{fig:water-tank-proof-light-backward}}
  \LeftLabel{\prooflabel{\hyperref[rule:Fchoice]{\rulenameFchoice}}} 
  \BinaryInfC{$\sequent{\DLfmlsympp}{  
    \ModalATT{\DLchoice{\WTfstMode}{\WTsndMode}}{\sensorset}{\offsetsym}{\renamingsym}{\DLfmlsymp} 
   }$
  }

  \AxiomC{Figure~\ref{fig:water-tank-proof-plant-backward}}
   \UnaryInfC{$\sequent{\DLfmlsymp}{  
    \ModalATT{\dLplant}{\sensorset}{\offsetsym}{\renamingsym}{\DLfmlsymInv} 
   }$ 
   }
   
  \LeftLabel{\prooflabel{\hyperref[rule:Fmr]{\rulenameFmr}}} 
  \BinaryInfC{$\sequent{\DLfmlsympp}{  
    \ModalATT{(\DLchoice{\WTfstMode}{\WTsndMode})}{\sensorset}{\offsetsym}{\renamingsym}{ 
      \ModalATT{\dLplant}{\sensorset}{\offsetsym}{\renamingsym}{\DLfmlsymInv} 
     }
   }$
  }
  \LeftLabel{\prooflabel{\hyperref[rule:Fseq]{\rulenameFseq}}} 
  \UnaryInfC{$\sequent{\DLfmlsympp}{  
    \ModalATT{\DLseq{ (\DLchoice{\WTfstMode}{\WTsndMode}) } {\dLplant} }{\sensorset}{\offsetsym}{\renamingsym}{\DLfmlsymInv} 
   }$
  }
  \LeftLabel{\prooflabel{\hyperref[rule:Fmr]{\rulenameFmr}}} 
  \BinaryInfC{$\sequent{\DLfmlsymInvb}{  \ModalATT{\DLassign{\WTsalts}{\WTsaltp}}{\sensorset}{\offsetsym}{\renamingsym}{ 
    \ModalATT{\DLseq{ (\DLchoice{\WTfstMode}{\WTsndMode}) } {\dLplant}}{\sensorset}{\offsetsym}{\renamingsym}{\DLfmlsymInv} 
   } 
   }$
  }
  \LeftLabel{\prooflabel{\hyperref[rule:Fseq]{\rulenameFseq}}}
  \UnaryInfC{$\sequent{\DLfmlsymInvb}{\ModalATT{{\DLprogsym}}{\sensorset}{\offsetsym}{\renamingsym}{\DLfmlsymInv}}$}

\end{prooftree}
\vspace{-1mm}
\caption{Proof of a backward distance for the water tank case study (continued)}
  \label{fig:water-tank-proof-continued-backward}
\end{figure}

\begin{figure}[H]
  \prooftreesize
  \begin{prooftree}
   \AxiomC{*}
   \LeftLabel{\prooflabel{\hyperref[rule:closefalse]{\rulenameFalse}}} 
   \UnaryInfC{$\sequent{\DLfmlsympp \hostand \hostE{( \WTsalts \ge 25)}}{  
     (\renaming{\DLtest{\WTsalts \ge \WTborder}}{\renamingsym} \land (\ModalATT{ \DLassign{\WTR}{10} }{\sensorset}{\offsetsym}{\renamingsym}{\DLfmlsymp}))  
    }$
   }
   \LeftLabel{\prooflabel{\hyperref[rule:implyR]{
    \rulenameImplyR}}}
   \UnaryInfC{$\sequent{\DLfmlsympp}{  
    \hostE{( \WTsalts \ge 25)} \hostimply (\renaming{\DLtest{\WTsalts \ge \WTborder}}{\renamingsym} \land (\ModalATT{ \DLassign{\WTR}{10} }{\sensorset}{\offsetsym}{\renamingsym}{\DLfmlsymp}))  
    }$
   }
   \LeftLabel{\prooflabel{\hyperref[rule:Ftest]{\rulenameFtest}}} 
   \UnaryInfC{$\sequent{\DLfmlsympp}{  \ModalATT{\DLtest{( \WTsalts \ge \WTborder)}}{\sensorset}{\offsetsym}{\renamingsym}{ 
    \ModalATT{\DLassign{\WTR}{10}}{\sensorset}{\offsetsym}{\renamingsym}{\DLfmlsymp} 
   } 
   }$
   }
   \LeftLabel{\prooflabel{\hyperref[rule:Fseq]{\rulenameFseq}}} 
   \UnaryInfC{$\sequent{\DLfmlsympp}{  
    \ModalATT{\WTheavy}{\sensorset}{\offsetsym}{\renamingsym}{\DLfmlsymp} 
    }$
   }
 \end{prooftree}
 \vspace{-1mm}
 \caption{Proof of the backward distance for the water tank case study (Switch-on mode)}
   \label{fig:water-tank-proof-heavy-backward}
 \end{figure}

 \begin{figure}[H]
  \prooftreesize
  \begin{prooftree}
    \AxiomC{By \rulenameFdef and \dL rules}
    \UnaryInfC{$\sequent{\DLfmlsympp}{  \ModalATT{\DLtest{( \WTsalts < \WTborder)}}{\sensorset}{\offsetsym}{\renamingsym}{ 
        \DLfmlsymInvb
    } 
    }$
    }

    \AxiomC{By \rulenameFdef and \dL rules}
    \UnaryInfC{$\sequent{\DLfmlsymInvb}{  
        \ModalATT{\DLassign{\WTR}{5}}{\sensorset}{\offsetsym}{\renamingsym}{\DLfmlsymp} 
       }$ 
    }
    \LeftLabel{\prooflabel{\hyperref[rule:Fmr]{\rulenameFmr}}}
    \BinaryInfC{$\sequent{\DLfmlsympp}{  \ModalATT{\DLtest{( \WTsalts < \WTborder)}}{\sensorset}{\offsetsym}{\renamingsym}{ 
        \ModalATT{\DLassign{\WTR}{5}}{\sensorset}{\offsetsym}{\renamingsym}{\DLfmlsymp} 
       } 
       }$
    }
    \LeftLabel{\prooflabel{\hyperref[rule:Fseq]{\rulenameFseq}}} 
    \UnaryInfC{$\sequent{\DLfmlsympp}{  
     \ModalATT{\WTlight}{\sensorset}{\offsetsym}{\renamingsym}{\DLfmlsymp} 
     }$
    }
 \end{prooftree}
 \vspace{-1mm}
 \caption{Proof of the backward distance for the water tank case study (off mode)}
   \label{fig:water-tank-proof-light-backward}
 \end{figure}

 \begin{figure}[H]
  \small
\begin{prooftree}

 \AxiomC{By \dL rules}
 \UnaryInfC{$\sequent{\DLfmlsymp}{  
  \DLfmlModalA{\DLseq{\dLODE{\WTsaltp}{\WTR/4 - \WTR*\WTsaltp/100}} {\dLODE{\WTsaltprn}{\WTRrn/4 - \WTRrn*\WTsaltprn/100}}}{\DLfmlsym} 
 }$ 
 }
 \LeftLabel{\prooflabel{\hyperref[rule:Fodemerge]{\rulenameFodemerge}}} 
 \UnaryInfC{$\sequent{\DLfmlsymp}{  
  \ModalATT{\dLplant}{\sensorset}{\offsetsym}{\renamingsym}{\DLfmlsym} 
 }$ 
 }
 
 \AxiomC{}
 \UnaryInfC{$\sequent{\DLfmlsym}{\DLfmlsymInv}$}
 \LeftLabel{\prooflabel{\hyperref[rule:Fmr]{\rulenameFmr}}} 
 \BinaryInfC{$\sequent{\DLfmlsymp}{  
  \ModalATT{\dLplant}{\sensorset}{\offsetsym}{\renamingsym}{\DLfmlsymInv} 
 }$ 
 }

\end{prooftree}
\vspace{-1mm}
\caption{Proof of the backward distance for the water tank case study (plant)}
 \label{fig:water-tank-proof-plant-backward}
\end{figure}

The proof establishes that $\sequent{}{\DLfmlsym \hostimply \ModalATT{\DLloop{\DLprogsym}}{\sensorset}{\offsetsym}{\renamingsym}{\dLfmlpost}}$ for 
$\dLfmlpost \equiv \hostE{\WTsaltp < 20 \hostimply \WTsaltprn < 20}$ and $\DLfmlsym \equiv \hostE{\WTsaltp = \WTsaltprn}$. Since $\DLfmlsym \hostimply \simuHdist{\renamingsym}{\Hsymbol}{0}$ and $\forall \WTsaltp~\exists \WTsaltprn. \DLfmlsym$, then by Proposition~\ref{lem:proof-calculus-forward-soundness} and  Theorem~\ref{thm:proof-calculus-backward-soundness}, 
$\BHsimulation{\Qattacked{\Pwoattack}{\sensorset}{\offsetsym}}{\Pwoattack}{\dLfmlpost}{\Hsymbol}{\Hdistdegsym}$ holds for $\Hdistdegsym = 0$ and $\Hsymbol = \{\WTsaltp\}$.

\section{Related Work}
\label{sec:relatedwork}

\paragraph*{Robustness of CPSs}
{
  Our work is a quantitative generalization of Xiang et at.~\cite{xiang2021relational}, in the setting of hybrid programs and \dL.
  In that paper, the authors propose two notions of robustness for CPSs: \emph{robustness of safety}, when (unbound) sensor attacks are unable to affect the system under attack, and \emph{robustness of high-integrity state}, when high-integrity parts of the system cannot be compromised. In the current paper, we generalize the first of the two relations.
  }
  
Fr{\"{a}}nzle et al. \cite{FKP16} classify the notions of robustness for CPSs as follows: 
(i) input/output robustness;
(ii) robustness with respect to system parameters;
(iii) robustness in real-time system implementation;
(iv) robustness due to unpredictable environment;
(v) robustness to faults.
The notion of robustness considered in this paper falls in category (iv), where the attacks are the source of environment's unpredictability.
Other works study robustness properties for CPSs~\cite{HuLVHMPZLLX16, HZTYQ18, TCRM14, RT16}. Some of them focus on robustness against attacks~\cite{HuLVHMPZLLX16, HZTYQ18}, even adopting quantitative reasonings~\cite{TCRM14, RT16}.
%


Our notion of forward robustness shares similarities with some existing notions of robustness, such as invariance~\cite{ames2016control} and input-to-state stability~\cite{agrachev2008input}. These notions concern if a system stays in a safe region when small changes happen to initial conditions,  while forward robustness concerns if a system stays in a safe region when under attack. Although it might be possible to reformulate  existing notions of robustness to characterize our forward robustness, our formulation focuses on modeling attacks which makes it easier to analyze their impact.

The current work extends and generalizes the paper by Chong et al.~\cite{hscc2023}. 
In particular, in Section \ref{sec:proof-calculus}, we first demonstrate how it is possible to encode both forward and backward simulation distances via  modalities. 
Then, we propose a new modality, called \SAModality, to encode both forward and backward distances (Theorem~\ref{thm:proof-calculus-forward-soundness} and Theorem~\ref{thm:proof-calculus-backward-soundness}). Finally, we 
define a proof system to help reasoning   on D-modality and prove its soundness (Proposition~\ref{lem:proof-calculus-forward-soundness}).
 The proof system has been applied in a new case study on the field of water tank (Section~\ref{sec:casestudy}). Last but not least, unlike the conference version~\cite{hscc2023}, 
the current paper provides in the appendix the full formal proofs of all results. 

\looseness=-1

  \emph{Signal Temporal Logic} (STL) \cite{Maler04} is a specification formalism for expressing real-time temporal \emph{safety} and performance properties, such as \emph{robustness},  of CPSs.
  Ferr\`ere et al.\ \cite{ferrere2019interface} study a quantitative extension of STL that classifies signals as inputs and outputs  to specify the system-under-test as an input/output relation instead  \nolinebreak of a set of correct execution traces. The idea behind their approach is quite similar  to that followed in our forward robustness, as they express families of admissible patterns of both the model inputs and  the model preconditions that guarantee the desired behavior of the model output. Mohammadinejad et al.~\cite{mohammadinejad2020mining} adopt a dual approach, similar to that followed in our backward robustness. Given an output  requirement they propose an algorithm to mine an environment assumption,  consisting  of a large subset  of  input signals for which the corresponding output signals \nolinebreak  satisfy \nolinebreak  the \nolinebreak  output \nolinebreak   requirement.

  }

\paragraph*{Formal Analysis of Sensor Attacks}
Lanotte et al. \cite{ifm2018,lanotte2020formal,LanotteMMT21} propose process-calculus approaches to model and analyze the   impact of
physics-based attacks, as sensor attacks in CPSs. \nolinebreak  Their threat models consider attacks that may manipulate both sensor readings and  control commands. 
Their model of physics is discrete and they focus on crucial timing aspects of attacks, such as beginning  and duration. 
Bernardeschi et al. \cite{bernardeschi2020formalization} introduce a framework to analyze the effects of attacks on sensors and actuators. Controllers of systems are specified using the formalism PVS~\cite{owre1992pvs}. The physics is described by other modeling tools.
Their threat  model  is similar to ours: the effect of an attack is a set of assignments to  the variables defined in the controller. Simulation is used to analyze the effects of attacks. 
	Huang et al.~\cite{HZTYQ18}  proposed a risk assessment method that uses a Bayesian network to model the attack propagation process and infers the probabilities of sensors and actuators to be compromised. These probabilities are fed into a stochastic hybrid system  model \nolinebreak to predict the evolution of the physical process. Then, the security risk is quantified by evaluating the system availability with the model.


\section{Conclusion} \label{sec:conclusion} 

A formal framework for \emph{quantitative} analysis of bounded sensor attacks on CPSs is introduced.
Given a precondition $\dLfmlpre$ and postcondition $\dLfmlpost$ of a system $\Pwoattack$, we formalize two safety notions, \emph{quantitative forward safety}, $\Fsafety{\Pwoattack}{\dLfmlpre}{\dLfmlpost}{\Qdegsymu}$, and \emph{quantitative backward safety}, $\Bsafety{\Pwoattack}{\dLfmlpre}{\dLfmlpost}{\Qdegsymu}$, where $\Qdegsymu \in \mathbb{R}$ respectively express:  (1) how strong the strongest postcondition $\SP{\Pwoattack}{\dLfmlpre}$ 
is  with respect to the postcondition $\dLfmlpost$, 
  and (2) how  strong the precondition $\dLfmlpre$ is with respect to the  weakest precondition $\WP{\Pwoattack}{\dLfmlpost}$. 
  The bigger $\Qdegsymu$ is, the safer the system is. On the contrary, if $\Qdegsymu$ is negative, then some reachable states violate the safety condition $\dLfmlpost$. If $\Qdegsymu$ is 0, then the system cannot be considered safe. 
We introduce \emph{forward and backward robustness}, $\Frobust{\Pwoattack}{\dLfmlpre}{\dLfmlpost}{\sensorset}{\offsetsym}{\Qdegratio}{}$ and $\Brobust{\Pwoattack}{\dLfmlpre}{\dLfmlpost}{\sensorset}{\offsetsym}{\Qdegratio}{}$ respectively, to quantify the robustness $\Qdegratio$, with $\Qdegratio \leq 1$, for a system $\Pwoattack$ against \emph{bounded sensor attacks\/}, as the ratio between the safety of the attacked system and the degree of safety of the original system;  here, the value of ($1-\Qdegratio$) quantifies the \emph{percentage of safety that is lost} due to the attack. The closer $\Qdegratio$ is to $1$, the more robust the system is.   To reason about the notions of robustness, we introduce two simulation distances, \emph{forward and backward simulation distance}, defined based on the behavior distances between the original system and the compromised system, to characterize upper bounds of the degree of forward and backward safety loss caused by the sensor attacks. 
%
To reason with forward and backward simulation distances, we propose a new \emph{modality} and a \emph{proof system} to reason with the modality. The proposed modality can be used to encode both forward and backward simulation distances in an intuitive and concise way. 
We formally prove the soundness of the encodings and its associated proof system. 
We showcase our formal notions on two non-trivial examples:  an autonomous vehicle that needs to avoid collision and a water tank system.



  \paragraph{Future work} As observed in~\cite{KrCa2013,ifm2018,lanotte2020formal}, \emph{timing} is a critical issue when attacking CPSs. 
We aim at generalizing our threat model to deal with more sophisticated time-sensitive sensor attacks, where the attacker may specify (possibly periodic)  attack windows in which  offsets might be potentially different in each   window, depending on the system state. This might be necessary to implement stealthy attacks working around adaptive  IDSs.
\looseness=-1






\let\oldbibliography\thebibliography
\renewcommand{\thebibliography}[1]{%
	\oldbibliography{#1}%
	\setlength{\itemsep}{2pt}%
}

\bibliographystyle{elsarticle-num}
\bibliography{bib}

\appendix

\vspace{-3mm}
\section{Definitions} \label{appendix:definitions}

We present the formal definition of variable set (i.e., \VAR{\DLfmlsym}), which involve bound variables, must bound variables, and free variables.

\vspace{-2mm}
\begin{definition}[Bound variables] The set $\BV{\DLfmlsym}$ of bound variables of \dL formula $\DLfmlsym$ is defined inductively as:
   \begin{align*}
     \BV{\DLfmlcomp{\DLtermsym}{\DLtermsymp}} & = \emptyset  ~~~~\hostE{\sim} \in \{\hostE{ <, \leq, =, >, \geq} \} \\
     \BV{\DLfmlneg{\DLfmlsym}} & = \BV{\phi} \\
 \BV{\DLfmlconj{\DLfmlsym}{\DLfmlsymp}}    &  = \BV{\DLfmlsym} \cup \BV{\DLfmlsymp} \\
    \BV{\DLfmlUQ{\DLvarsym}{\DLfmlsym}}  &  = \{ \DLvarsym \} \cup \BV{\DLfmlsym} \\
    \BV{\DLfmlModalA{\DLprogsym}{\DLfmlsym}}   &  = \BV{\DLprogsym} \cup \BV{\DLfmlsym}
  \end{align*}
  The set $\BV{\Pwoattack}$ of bound variables of hybrid program $\Pwoattack$, i.e., those may potentially be written to, is defined inductively as:
  \begin{align*}
    \BV{\DLassign{\DLvarsym}{\DLtermsym}} = \BV{\DLassignN{\DLvarsym}}  & = \{ \DLvarsym \} \\
     \BV{\DLtest{\DLfmlsym}} & = \emptyset \\
    \BV{\dLprogODE{\dLODE{\DLvarsym}{\DLtermsym}}{\evolconstraint}}  & =   \{ \DLvarsym , \DLvarsym^\prime\}    \\
    \BV{\DLseq{\DLprogsym}{\DLprogsymp}} = \BV{\DLchoice{\DLprogsym}{\DLprogsymp}} & =  \BV{\DLprogsym} \cup \BV{\DLprogsymp}                    \\
    \BV{\DLloop{\DLprogsym}} &  = \BV{\DLprogsym} 
  \end{align*}
\end{definition}
\vspace*{-5mm}

\begin{definition}[Must-bound variables] The set $\MBV{\Pwoattack}$ $\subseteq$ $\BV{\Pwoattack}$ of most bound variables of hybrid program $\Pwoattack$, i.e., all those that must be written to on all paths of $\Pwoattack$, is defined inductively as:
  \begin{align*}
    \MBV{\DLassign{\DLvarsym}{\DLtermsym}} = \MBV{\DLassignN{\DLvarsym}}  & = \{ \DLvarsym \} \\
     \MBV{\DLtest{\DLfmlsym}} & = \emptyset \\
    \MBV{\dLprogODE{\dLODE{\DLvarsym}{\DLtermsym}}{\evolconstraint}}  & =   \{ \DLvarsym , \DLvarsym^\prime\}     \\
    \MBV{\DLchoice{\DLprogsym}{\DLprogsymp}} & =  \MBV{\DLprogsym} \cap \MBV{\DLprogsymp}                    \\
    \MBV{\DLseq{\DLprogsym}{\DLprogsymp}} & =  \MBV{\DLprogsym} \cup \MBV{\DLprogsymp}                    \\
    \MBV{\DLloop{\DLprogsym}} &  = \emptyset
  \end{align*}  
\end{definition}

\begin{definition}[Free variables]
  The set $\FV{\theta}$ of variables of term $\DLtermsym$ is defined inductively as:
   \begin{align*}
     \FV{\DLtermsym} & = \{ \DLtermsym \} \\
     \FV{\DLconst} & = \emptyset \\
     \FV{\DLtermcomp{\DLtermsym}{\DLtermsymp}} & = \FV{\DLtermsym} \cup \FV{\DLtermsymp}
   \end{align*}
  The set $\FV{\phi}$ of free variables of \dL formula $\phi$ is defined inductively as:
   \begin{align*}
     \FV{\DLfmlcomp{\DLtermsym}{\DLtermsymp}} & = \FV{\DLtermsym} \cup \FV{\DLtermsymp} \\
     \FV{\DLfmlneg{\DLfmlsym}} & = \FV{\DLfmlsym} \\
     \FV{\DLfmlconj{\DLfmlsym}{\DLfmlsymp}}   &  = \FV{\DLfmlsym} \cup \FV{\DLfmlsymp} \\
     \FV{\DLfmlUQ{\DLvarsym}{\DLfmlsym}}  &  = \FV{\DLfmlsym} \setminus \{ \DLvarsym \} \\
     \FV{\DLfmlModalA{\DLprogsym}{\DLfmlsym}}   &  = \FV{\DLprogsym} \cup (\FV{\DLfmlsym} \setminus \MBV{\DLprogsym})
   \end{align*}
   The set $\FV{\Pwoattack}$ of bound variables of hybrid program $\Pwoattack$ is defined inductively as:
   \begin{align*}
     \FV{\DLassign{\DLvarsym}{\DLtermsym}}  & = \FV{\DLtermsym} \\
     \FV{\DLassignN{\DLvarsym}}     & = \emptyset   \\
     \FV{\DLtest{\DLfmlsym}} & = \FV{\DLfmlsym} \\
     \FV{\dLprogODE{\dLODE{\DLvarsym}{\DLtermsym}}{\evolconstraint}}  & =   \{ \DLvarsym \} \cup \FV{\DLtermsym} \cup \FV{\evolconstraint}   \\
     \FV{\DLchoice{\DLprogsym}{\DLprogsymp}} & =  \FV{\DLprogsym} \cup \FV{\DLprogsymp} \\
     \FV{\DLseq{\DLprogsym}{\DLprogsymp}} & =  \FV{\DLprogsym} \cup (\FV{\DLprogsymp} \setminus \MBV{\DLprogsym})                    \\
     \FV{\DLloop{\DLprogsym}} &  = \FV{\DLprogsym} 
   \end{align*}  
 \end{definition}
 \begin{definition}[Variable sets] The set $\VAR{\Pwoattack}$, variables of hybrid program $\Pwoattack$ is $\BV{\Pwoattack} \cup \FV{\Pwoattack}$. The set $\VAR{\DLfmlsym}$, variables of \dL formula $\DLfmlsym$ is $\BV{\DLfmlsym} \cup \FV{\DLfmlsym}$. 
 \end{definition}

 \section{Proofs} \label{sec:appendix-proof}

In this section, we provide the proofs of all results stated in the paper.

\subsection{Proofs of results in Section~\ref{sec:q-safety}}
\label{app:sec:q-safety}
\vspace{3 mm}

\noindent
\begin{proof}[\textbf{Proof of Proposition~\ref{prop:coherency}}]
We prove the two properties separately.

Let us start with first property.
If $\Fsafety{\Pwoattack}{\dLfmlpre}{\dLfmlpost}{\Qdegsymu}$, then, 
by definition, it holds that $\Qdegsymu = \infsym \{ \signdistStateToSet{\statesymp}{\dLfmlsemstate{\dLfmlpost}} ~|~ 
\statesymp \in \dLfmlsemstate{\SP{\Pwoattack}{\dLfmlpre}}$.
Since, by the hypothesis, $\Qdegsymu > 0$, then by definition of $\signdistStateToSet{\_\,}{\_}$ it follows that all states 
$\statesymp \in \dLfmlsemstate{\SP{\Pwoattack}{\dLfmlpre}}$ satisfy $\statesymp \in \dLfmlsemstate{\dLfmlpost}$, thus implying that all states
$\statesym \in \dLfmlsemstate{\dLfmlpre}$ are such that 
all states $\statesymp$ with $(\statesym,\statesymp) \in \dLfmlsemstate{\Pwoattack}$ satisfy $\statesymp \in \dLfmlsemstate{\dLfmlpost}$.
By the definition of the modality of necessity, this implies that all states $\statesym \in \dLfmlsemstate{\dLfmlpre}$ are also in $\dLfmlsemstate{\WP{\Pwoattack}{\dLfmlpost}}$,
thus implying 
$\distStateToSet{\statesym}{\dLfmlsemstate{\WP{\Pwoattack}{\dLfmlpost}}} \ge 0$ and, consequently,
$\infsym \{ \signdistStateToSet{\statesym}{\dLfmlsemstate{\WP{\Pwoattack}{\dLfmlpost}}} ~|~ 
\statesym \in \dLfmlsemstate{\dLfmlpre} \} \ge 0$.
By definition, property
 $\Bsafety{\Pwoattack}{\dLfmlpre}{\dLfmlpost}{\Qdegsymr}$ for some $\Qdegsymr \geq 0$ follows.
  
Let us consider now the second property.  
If $\Bsafety{\Pwoattack}{\dLfmlpre}{\dLfmlpost}{\Qdegsymr}$, then, by definition, it holds that
$\Qdegsymr= \infsym \{ \signdistStateToSet{\statesym}{\dLfmlsemstate{\WP{\Pwoattack}{\dLfmlpost}}} ~|~ 
\statesym \in \dLfmlsemstate{\dLfmlpre}\}$.
Since, by the hypothesis, $\Qdegsymr > 0$, then by definition of $\signdistStateToSet{\_\,}{\_}$ it follows that all states in $\dLfmlsemstate{\dLfmlpre}\}$ are also in $\dLfmlsemstate{\WP{\Pwoattack}{\dLfmlpost}}$, thus implying that
all states 
$\statesym \in \dLfmlsemstate{\dLfmlpre}$ are such that all states $\statesymp$ with $(\statesym,\statesymp) \in \dLfmlsemstate{\Pwoattack}$ satisfy $\statesymp \in \dLfmlsemstate{\dLfmlpost}$.
By the definition of modality $\SP{\Pwoattack}{\dLfmlpre}$,
this implies that all states $\statesymp \in \dLfmlsemstate{\SP{\Pwoattack}{\dLfmlpre}}$ are also in $\dLfmlsemstate{\dLfmlpost}$,
thus implying
$\distStateToSet{\statesymp}{\dLfmlsemstate{\dLfmlpost}} \ge 0$ and, consequently,
$\infsym \{ \signdistStateToSet{\statesymp}{\dLfmlsemstate{\dLfmlpost}} ~|~ 
\statesymp \in \dLfmlsemstate{\SP{\Pwoattack}{\dLfmlpre}} \} \ge 0$.
By definition, property $\Fsafety{\Pwoattack}{\dLfmlpre}{\dLfmlpost}{\Qdegsymu}$ for some $\Qdegsymu \geq 0$ follows.
\end{proof}

\subsection{Proofs of results in Section~\ref{sec:robust-safety}}
\label{app:sec:robust-safety}
\vspace{3 mm}

\noindent
\begin{proof}[\textbf{Proof of Theorem~\ref{theorem:forward-rob-mono}}]
	We prove the inequality $\Qdegsymu_1  \leq   \Qdegsymu$, the inequality 
	$\Qdegsymu_2  \leq \Qdegsymu_1  $ can be proved similarly.  
  The behaviors of the system with compromised sensors subsume the behaviors of the original program, since the sensed values $q_{s}$ can take the correct physical value $q_{p}$. Therefore we know that $\dLfmlsemstate{\SP{\Qattacked{\Pwoattack}{\sensorset}{\offsetsym_1}}{\dLfmlpre}}$ contains all states of $\dLfmlsemstate{\SP{\Pwoattack}{\dLfmlpre}}$. Then, according to the definition of forward safety, 
  $\Qdegsymu = \infsym \{ \signdistStateToSet{\statesymp}{\dLfmlsemstate{\dLfmlpost}} ~|~ 
\statesymp \in \dLfmlsemstate{\SP{\Pwoattack}{\dLfmlpre}} \}$ can only be no smaller than $\Qdegsymu_1 = \infsym \{ \signdistStateToSet{\statesymp}{\dLfmlsemstate{\dLfmlpost}} ~|~ 
\statesymp \in \dLfmlsemstate{\SP{\Qattacked{\Pwoattack}{\sensorset}{\offsetsym_1}}{\dLfmlpre}} \}$.
\end{proof}

\vspace{3mm}

\noindent
\begin{proof}[\textbf{Proof of Theorem~\ref{theorem:backward-rob-mono}}]
	We prove the inequality $\Qdegsymr_1  \leq   \Qdegsymr$, the inequality 
	$\Qdegsymr_2  \leq \Qdegsymr_1  $ can be proved similarly.  
  As observed sbove in the proof of Theorem~\ref{theorem:forward-rob-mono}, the behaviors of the system with compromised sensors subsume the behaviors of the original program. Therefore we know that $\dLfmlsemstate{\WP{\Pwoattack}{\dLfmlpost}}$ contains all states of  $\dLfmlsemstate{\WP{\Qattacked{\Pwoattack}{\sensorset}{\offsetsym_1}}{\dLfmlpost}}$. 
  Then, according to the definition of backward safety, we can conclude that value 
$\Qdegsymr = \infsym \{ \signdistStateToSet{\statesym}{\dLfmlsemstate{\WP{\Pwoattack}{\dLfmlpost}}} ~|~ 
\statesym \in \dLfmlsemstate{\dLfmlpre} \}$
can only be no smaller than value
$\Qdegsymr_1 = \infsym \{ \signdistStateToSet{\statesym}{\dLfmlsemstate{\WP{\Qattacked{\Pwoattack}{\sensorset}{\offsetsym_1}}{\dLfmlpost}}} ~|~ 
\statesym \in \dLfmlsemstate{\dLfmlpre} \}$.
\end{proof}

\subsection{Proofs of results in Section~\ref{sec:h-eq}}
\label{app:sec:h-eq}
\vspace{3mm}

\noindent
 \begin{proof}[\textbf{Proof of Proposition~\ref{lemma:distance-h}}]
We show $\Qdegsymu_1  =   \Qdegsymu$.
Property $\Qdegsymu_2  =   \Qdegsymu$ can be proved analogously.

Property $\Qdegsymu_1  =   \Qdegsymu$ follows if we show that for all states $\statesym \in \DLsem{\DLfmlsymp}$ it holds
\begin{equation}
\label{lemma:distance-h_proof_obligation}
\signdistStateToSet{\statesym}{\dLfmlsemstate{\DLfmlsym}} = 
\HsigndistStateToSet{\statesym}{\dLfmlsemstate{\DLfmlsym}}{\VAR{\DLfmlsym}}
\end{equation}
where,
by definition, we have:
\begin{equation*}
\signdistStateToSet{\statesym}{\dLfmlsemstate{\DLfmlsym}}
=
\begin{cases}
  \infsym\{ \diststate{\statesym}{\statesymp} ~|~ \statesymp \in \DLsem{\DLfmlneg{\DLfmlsym}} \}
 & \text{if } \statesym \in \dLfmlsemstate{\DLfmlsym} 
 \\ 
  -\infsym\{\diststate{\statesym}{\statesymp} ~|~ \statesymp \in \DLsem{\DLfmlsym} \}
& \text{if } \statesym \not\in \dLfmlsemstate{\DLfmlsym} 
\end{cases}
\end{equation*}
and
\begin{equation*}
\HsigndistStateToSet{\statesym}{\dLfmlsemstate{\DLfmlsym}}{\VAR{\DLfmlsym}}
=
\begin{cases}
  \infsym\{ \Hdiststate{\statesym}{\statesymp}{\VAR{\DLfmlsym}} ~|~ \statesymp \in \DLsem{\DLfmlneg{\DLfmlsym}} \}
 & \text{if } \statesym \in \dLfmlsemstate{\DLfmlsym} 
 \\
 - \infsym\{ \Hdiststate{\statesym}{\statesymp}{\VAR{\DLfmlsym}} ~|~ \statesymp \in \DLsem{\DLfmlsym} \}
& \text{if } \statesym \not\in \dLfmlsemstate{\DLfmlsym}.
\end{cases}
\end{equation*}

We prove Equation~\ref{lemma:distance-h_proof_obligation}, by distinguishing two cases, $\statesym \in \dLfmlsemstate{\DLfmlsym}$ and $\statesym \not\in \dLfmlsemstate{\DLfmlsym}$.

\underline{Case $\statesym \in \dLfmlsemstate{\DLfmlsym}$}.
Since $\diststate{\statesym}{\statesymp} \ge \Hdiststate{\statesym}{\statesymp}{\VAR{\DLfmlsym}}$ for all states $\statesymp$,  we infer 
$\signdistStateToSet{\statesym}{\dLfmlsemstate{\DLfmlsym}} \ge \HsigndistStateToSet{\statesym}{\dLfmlsemstate{\DLfmlsym}}{\VAR{\DLfmlsym}}$.
We can prove that also $\signdistStateToSet{\statesym}{\dLfmlsemstate{\DLfmlsym}} \le \HsigndistStateToSet{\statesym}{\dLfmlsemstate{\DLfmlsym}}{\VAR{\DLfmlsym}}$ holds, thus confirming 
Equation~\ref{lemma:distance-h_proof_obligation}.
For an arbitrary state $\statesymp \in \DLsem{\DLfmlneg{\DLfmlsym}}$,  $\Hdiststate{\statesym}{\statesymp}{\VAR{\DLfmlsym}}$ is equal to $\diststate{\statesym}{\statesymvb}$, where $\statesymvb$ is the state such that:
  \[ \statesymvb = 
    \begin{cases}
      x \mapsto \statesymp(x) & \text{if}\ x \in \VAR{\DLfmlsym}\\
      x \mapsto \statesym(x) & \text{otherwise.}
    \end{cases}
  \] 
Notice that $\statesymvb$ belongs to $\DLsem{\DLfmlneg{\DLfmlsym}}$, since $\statesymp \in \DLsem{\DLfmlneg{\DLfmlsym}}$ and states $\statesymvb$ and $\statesymp$ assign the same value to all variables occurring in $\DLfmlsym$.
  By the arbitrariness of $\statesymp$ in $\DLsem{\DLfmlneg{\DLfmlsym}}$,  we get that all values in set $\{ \Hdiststate{\statesym}{\statesymp}{\VAR{\DLfmlsym}} ~|~ \statesymp \in \DLsem{\DLfmlneg{\DLfmlsym}} \}$ are also in set $\{ \diststate{\statesym}{\statesymp} ~|~ \statesymp \in \DLsem{\DLfmlneg{\DLfmlsym}} \}$, thus implying
  $\infsym\{ \Hdiststate{\statesym}{\statesymp}{\VAR{\DLfmlsym}} ~|~ \statesymp \in \DLsem{\DLfmlneg{\DLfmlsym}} \} \ge 
    \infsym\{ \diststate{\statesym}{\statesymp} ~|~ \statesymp \in \DLsem{\DLfmlneg{\DLfmlsym}} \}$,
   which gives 
$\signdistStateToSet{\statesym}{\dLfmlsemstate{\DLfmlsym}} \le \HsigndistStateToSet{\statesym}{\dLfmlsemstate{\DLfmlsym}}{\VAR{\DLfmlsym}}$.
  
 \underline{Case $\statesym \not\in \dLfmlsemstate{\DLfmlsym}$}.
Since $\diststate{\statesym}{\statesymp} \ge \Hdiststate{\statesym}{\statesymp}{\VAR{\DLfmlsym}}$ for all states $\statesymp$,  we infer 
$\signdistStateToSet{\statesym}{\dLfmlsemstate{\DLfmlsym}} \le \HsigndistStateToSet{\statesym}{\dLfmlsemstate{\DLfmlsym}}{\VAR{\DLfmlsym}}$.
We can prove that also $\signdistStateToSet{\statesym}{\dLfmlsemstate{\DLfmlsym}} \ge \HsigndistStateToSet{\statesym}{\dLfmlsemstate{\DLfmlsym}}{\VAR{\DLfmlsym}}$ holds, thus confirming 
Equation~\ref{lemma:distance-h_proof_obligation}.
For an arbitrary state $\statesymp \in \DLsem{\DLfmlsym}$, $\Hdiststate{\statesym}{\statesymp}{\VAR{\DLfmlsym}}$  is equal to $\diststate{\statesym}{\statesymvb}$, where $\statesymvb$ is the state such that:
  \[ \statesymvb = 
    \begin{cases}
      x \mapsto \statesymp(x) & \text{if}\ x \in \VAR{\DLfmlsym}\\
      x \mapsto \statesym(x) & \text{otherwise.}
    \end{cases}
  \] 
Notice that $\statesymvb$ belongs to $\DLsem{\DLfmlsym}$, 
since $\statesymp \in \DLsem{\DLfmlsym}$ and states $\statesymvb$ and $\statesymp$ assign the same value to all variables occurring in $\DLfmlsym$.
  By the arbitrariness of $\statesymp$ in $\DLsem{\DLfmlsym}$,  we get 
  that all values in set $\{ \Hdiststate{\statesym}{\statesymp}{\VAR{\DLfmlsym}} ~|~ \statesymp \in \DLsem{\DLfmlsym} \}$ are also in set $\{ \diststate{\statesym}{\statesymp} ~|~ \statesymp \in \DLsem{\DLfmlsym} \}$,
  thus implying
  $ - \infsym\{ \Hdiststate{\statesym}{\statesymp}{\VAR{\DLfmlsym}} ~|~ \statesymp \in \DLsem{\DLfmlsym} \} \le 
  - \infsym\{ \diststate{\statesym}{\statesymp} ~|~ \statesymp \in \DLsem{\DLfmlsym} \}$,
   which gives 
$\signdistStateToSet{\statesym}{\dLfmlsemstate{\DLfmlsym}} \ge \HsigndistStateToSet{\statesym}{\dLfmlsemstate{\DLfmlsym}}{\VAR{\DLfmlsym}}$.
\end{proof}

\vspace{5 mm}

\noindent
\begin{proof}[\textbf{Proof of Theorem~\ref{thm:forward_rob}}]
  Let $\Hsymbol$ be the set of variables $\VAR{\dLfmlpost}$. We need to prove 
\[\Fsafety{\Qattacked{\Pwoattack}{\sensorset}{\offsetsym}}{\dLfmlpre}{\dLfmlpost}{\Qdegsymup}
 \text{ for }\Qdegsymup \ge \Qdegsymu -\Hdistdegsym \]
where, by definition of forward safety, we know that 
\[
  \Qdegsymup  =  \infsym \{ \signdistStateToSet{\statesymp}{ \dLfmlsemstate{\dLfmlpost} } ~|~ \statesymp \in {\dLfmlsemstate{\SP{\Qattacked{\Pwoattack}{\sensorset}{\offsetsym}}{\dLfmlpre}}}\}
\]
Therefore, the proof obligation is
\begin{equation}
\label{thm:forward_rob:proof_obligation}
\infsym \{ \signdistStateToSet{\statesymp}{ \dLfmlsemstate{\dLfmlpost} } ~|~ \statesymp \in
{\dLfmlsemstate{\SP{\Qattacked{\Pwoattack}{\sensorset}{\offsetsym}}{\dLfmlpre}}}\} \ge 
\Qdegsymu -\Hdistdegsym
\end{equation}

Consider any state $\nu  \in {\dLfmlsemstate{\SP{\Qattacked{\Pwoattack}{\sensorset}{\offsetsym}}{\dLfmlpre}}}$.
The hypothesis $\FHsimulation{\Qattacked{\Pwoattack}{\sensorset}{\offsetsym}}{\Pwoattack}{\dLfmlpre}{\Hsymbol}{\Hdistdegsym}$ ensures that there is some state $\statesymp'  \in \dLfmlsemstate{\SP{\Pwoattack}{\dLfmlpre}}$ with
$\Hdiststate{\statesymp}{\statesymp'}{\Hsymbol} \le d$.
We can show that $\HsigndistStateToSet{\statesymp'}{ \dLfmlsemstate{\dLfmlpost} }{\Hsymbol} \ge \Qdegsymu$.
Indeed, the hypothesis 
$\Fsafety{\Pwoattack}{\dLfmlpre}{\dLfmlpost}{\Qdegsymu}$ coincides, by definition, with property
\[\Qdegsymu  =  \infsym \{ \signdistStateToSet{\statesymp}{ \dLfmlsemstate{\dLfmlpost} } ~|~
\statesymp \in
{\dLfmlsemstate{\SP{\Pwoattack}{\dLfmlpre}}}\}\]
which, by Proposition~\ref{lemma:distance-h}, can be rewritten as 
\[\Qdegsymu  = \infsym \{ \HsigndistStateToSet{\statesymp}{ \dLfmlsemstate{\dLfmlpost} }{\Hsymbol} ~|~
\statesymp \in
{\dLfmlsemstate{\SP{\Pwoattack}{\dLfmlpre}}}\} \]  
from which we infer
$\HsigndistStateToSet{\statesymp'}{ \dLfmlsemstate{\dLfmlpost} }{\Hsymbol} \ge \Qdegsymu$ since $\statesymp' \in \dLfmlsemstate{\SP{\Pwoattack}{\dLfmlpre}}$.

Since $\Hdiststate{\_\,}{\_}{\Hsymbol}$ is a metric,  it is symmetric, thus implying $\Hdiststate{\statesymp}{\statesymp'}{\Hsymbol}=\Hdiststate{\statesymp'}{\statesymp}{\Hsymbol}$,  and satisfies the triangular property.
By the triangular property we infer that for any state $\statesymp'' \not\in  \dLfmlsemstate{\dLfmlpost}$, it holds
\begin{equation}\label{thm:forward_rob_triangular}
\Hdiststate{\statesymp'}{\statesymp''}{\Hsymbol} \le   \Hdiststate{\statesymp'}{\statesymp}{\Hsymbol} + \Hdiststate{\statesymp}{\statesymp''}{\Hsymbol}
\end{equation}
By definition of 
$\HsigndistStateToSet{\statesymp'}{ \dLfmlsemstate{\dLfmlpost} }{\Hsymbol}$ and the properties $\HsigndistStateToSet{\statesymp'}{ \dLfmlsemstate{\dLfmlpost} }{\Hsymbol} \ge \Qdegsymu$ and
$\statesymp'' \not\in  \dLfmlsemstate{\dLfmlpost}$
we infer
$\Hdiststate{\statesymp'}{\statesymp''}{\Hsymbol} \ge \Qdegsymu$. 
From this inequality, property 
$\Hdiststate{\statesymp}{\statesymp'}{\Hsymbol} \le d$ and Equation~\ref{thm:forward_rob_triangular}, we infer
$ \Hdiststate{\statesymp}{\statesymp''}{\Hsymbol} \ge \Qdegsymu - \Hdistdegsym$.
By definition of
$\HsigndistStateToSet{\statesymp}{ \dLfmlsemstate{\dLfmlpost} }{\Hsymbol}$ and the arbitrariness of $\statesymp$ in ${\dLfmlsemstate{\SP{\Qattacked{\Pwoattack}{\sensorset}{\offsetsym}}{\dLfmlpre}}}$,
we infer
\[ \infsym\{ 
\HsigndistStateToSet{\statesymp}{ \dLfmlsemstate{\dLfmlpost} }{\Hsymbol} ~|~  \statesymp  \in {\DLsem{\SP{\Qattacked{\Pwoattack}{\sensorset}{\offsetsym}}{\dLfmlpre}}}\} \ge  \Qdegsymu -\Hdistdegsym \] 
which, by Proposition~\ref{lemma:distance-h}, can be rewritten as: 
\[ \infsym\{ 
\signdistStateToSet{\statesymp}{ \dLfmlsemstate{\dLfmlpost} } ~|~  \statesymp  \in {\DLsem{\SP{\Qattacked{\Pwoattack}{\sensorset}{\offsetsym}}{\dLfmlpre}}}\} \ge  \Qdegsymu -\Hdistdegsym \] 
which coincides with the proof obligation Equation~\ref{thm:forward_rob:proof_obligation}.
This 
completes the proof.
\end{proof}

\vspace{5mm}

\noindent \begin{proof}[\textbf{Proof of Theorem~\ref{thm:backward_rob}}]
Let $\Hsymbol$ be the set of variables $\VAR{\dLfmlpre}$.
  We have to prove 
\[\Bsafety{\Qattacked{\Pwoattack}{\sensorset}{\offsetsym}}{\dLfmlpre}{\dLfmlpost}{\Qdegsymrp}
\text{ for } \Qdegsymrp \ge \Qdegsymr -\Hdistdegsym\]
where, by definition of backward safety, we have 
\[
\Qdegsymrp  =  \infsym \{ \signdistStateToSet{\statesym}{\dLfmlsemstate{\WP{\Qattacked{\Pwoattack}{\sensorset}{\offsetsym}}{\dLfmlpost}}} ~|~ 
\statesym \in \dLfmlsemstate{\dLfmlpre} \}
\]
Therefore, the proof obligation is
\begin{equation*}
\infsym \{ \signdistStateToSet{\statesym}{\dLfmlsemstate{\WP{\Qattacked{\Pwoattack}{\sensorset}{\offsetsym}}{\dLfmlpost}}} ~|~ 
\statesym \in \dLfmlsemstate{\dLfmlpre} \} 
\ge \Qdegsymr -\Hdistdegsym
\end{equation*}
which, by Proposition~\ref{lemma:distance-h}
can be rewritten as 
\begin{equation*}
\infsym \{ \HsigndistStateToSet{\statesym}{\dLfmlsemstate{\WP{\Qattacked{\Pwoattack}{\sensorset}{\offsetsym}}{\dLfmlpost}}}{\Hsymbol} ~|~ 
\statesym \in \dLfmlsemstate{\dLfmlpre} \} 
\ge \Qdegsymr -\Hdistdegsym
\end{equation*}
which can be rewritten as
\begin{equation}
\forall \statesym \in \dLfmlsemstate{\dLfmlpre}. \;
\forall \statesym'  \in \DLsem{\DLfmlModalE{\Qattacked{\Pwoattack}{S_A}{\offsetsym}}{\DLfmlneg{\dLfmlpost}}}. \;
\Hdiststate{\statesym }{\statesym'}{\Hsymbol} \ge \Qdegsymr -\Hdistdegsym
\label{thm:backward_rob_proof_obligation}
\end{equation}


Let us fix any $\statesym \in \dLfmlsemstate{\dLfmlpre}$.
For each $\statesym'  \in \DLsem{\DLfmlModalE{\Qattacked{\Pwoattack}{S_A}{\offsetsym}}{\DLfmlneg{\dLfmlpost}}}$ the property 
$\Hdiststate{\statesym }{\statesym'}{\Hsymbol} \ge \Qdegsymr -\Hdistdegsym$ is 
immediate if $\statesym'  \in \DLsem{\DLfmlModalE{\Pwoattack}{\DLfmlneg{\dLfmlpost}}}$, 
since in that case the hypothesis $\Bsafety{\Pwoattack}{\dLfmlpre}{\dLfmlpost}{\Qdegsymr}$, which coincides with
 $\Qdegsymr$ = $\infsym \{ \signdistStateToSet{\statesym}{\DLsem{\WP{\Pwoattack}{\dLfmlpost}}}  ~|~ 
\statesym \in \DLsem{\dLfmlpre} \}$,
and with 
$\infsym \{ \HsigndistStateToSet{\statesym}{\DLsem{\WP{\Pwoattack}{\dLfmlpost}}}{\Hsymbol}  ~|~ 
\statesym \in \DLsem{\dLfmlpre} \}$
by Proposition~\ref{lemma:distance-h},
ensures that $\Hdiststate{\statesym}{\statesym'}{\Hsymbol} >   \Qdegsymr$.
%
The interesting case is $\statesym' \in \DLsem{\WP{\Pwoattack}{\dLfmlpost}}$.
By the hypothesis $\BHsimulation{\Qattacked{\Pwoattack}{\sensorset}{\offsetsym}}{\Pwoattack}{\dLfmlpost}{\Hsymbol}{\Hdistdegsym}$ there is a state $\statesym'' \in 
 \DLsem{\DLfmlModalE{\Pwoattack}{\DLfmlneg{\dLfmlpost}}}$
 with $\Hdiststate{\statesym'}{\statesym''}{\Hsymbol} \le \Hdistdegsym$.
The hypothesis $\Bsafety{\Pwoattack}{\dLfmlpre}{\dLfmlpost}{\Qdegsymr}$ and Proposition~\ref{lemma:distance-h} 
ensure that 
$\Qdegsymr$ = $\infsym \{ \HsigndistStateToSet{\statesym}{\DLsem{\WP{\Pwoattack}{\dLfmlpost}}}{\Hsymbol}  ~|~ 
\statesym \in \DLsem{\dLfmlpre} \}$.
This inequality together with
$\statesym'' \in  \DLsem{\DLfmlModalE{\Pwoattack}{\DLfmlneg{\dLfmlpost}}}$ give $\Hdiststate{\statesym''}{\statesym}{\Hsymbol} \ge \Qdegsymr$. 
Since $\Hdiststate{\_\,}{\_}{\Hsymbol}$ is a metric,  it is symmetric, thus implying that $\Hdiststate{\statesym}{\statesym''}{\Hsymbol}=\Hdiststate{\statesym''}{\statesym}{\Hsymbol}$.
Moreover $\Hdiststate{\_\,}{\_}{\Hsymbol}$ satisfies the triangular property, which ensures that 
\[
\Hdiststate{\statesym}{\statesym''}{\Hsymbol} \le   \Hdiststate{\statesym}{\statesym'}{\Hsymbol} + \Hdiststate{\statesym'}{\statesym''}{\Hsymbol} 
\]
From this inequality,  
$\Hdiststate{\statesym''}{\statesym}{\Hsymbol} \ge \Qdegsymr$ and
$\Hdiststate{\statesym'}{\statesym''}{\Hsymbol} \le \Hdistdegsym$ we infer 
$
 \Hdiststate{\statesym}{\statesym'}{\Hsymbol} \ge \Qdegsymr -\Hdistdegsym
$.
This  completes the proof.
\end{proof}
$\\$

\subsection{Proofs of results in Section~\ref{sec:proof-calculus}}

\label{app:sec:proof-calculus}
\vspace{3mm}
\noindent
\begin{proof}[\textbf{Proof of Proposition~\ref{lem:newMod}}]
The thesis follows from the definition of the modalities of necessity $\DLfmlModalA{\_}{\_}$ and existency $\DLfmlModalE{\_}{\_}$ and the definition of renaming function $\renamingsym$. 
\end{proof}
$ $\\


\noindent
\begin{proof}[\textbf{Proof of Theorem~\ref{thm:proof-calculus-forward-soundness}}]
We have to prove that for each state $\statesymp_1 \in {\dLfmlsemstate{\SP{\Qattacked{\Pwoattack}{\sensorset}{\offsetsym}}{\dLfmlpre}}}$,
there exists a state $ \statesymp_2 \in 
\dLfmlsemstate{\SP{\Pwoattack}{\dLfmlpre}}$ such that 
$\Hdiststate{\statesymp_1 }{\statesymp_2 }{\Hsymbol} \le d$.
Given an arbitrary state 
$\statesymp_1 \in {\dLfmlsemstate{\SP{\Qattacked{\Pwoattack}{\sensorset}{\offsetsym}}{\dLfmlpre}}}$, there exists a state
$\statesym_1 \in \dLfmlsemstate{\dLfmlpre}$ such that  $(\statesym_1, \statesymp_1  ) \in 
\dLfmlsemstate{\Qattacked{\Pwoattack}{\sensorset}{\offsetsym}}$. 
From $\statesym_1 \in  \dLfmlsemstate{\dLfmlpre}$ and the hypothesis  
 $\dLfmlpre \hostimply
         \exists \renaming{\VAR{\dLfmlpre}}{\renamingsym}. \, (\DLfmlsymp
         \land
    \renaming{\dLfmlpre}{\renamingsym})$
we infer that there is some state $\statesym_2  \in  \dLfmlsemstate{\dLfmlpre} $ such that
$\statesym_1  \uplus
\renaming{\statesym_2 }{\renamingsym} \in  \dLfmlsemstate{\DLfmlsymp}$.
From the hypothesis  
${\DLfmlsymp}\hostimply{         \ModalATT{\Pwoattack}{\sensorset}{\offsetsym}{\renamingsym}{\DLfmlsym} 
     }$   we have that 
$ \statesym_1 \uplus
\renaming{\statesym_2}{\renamingsym} \in 
\dLfmlsemstate{         \ModalATT{\Pwoattack}{\sensorset}{\offsetsym}{\renamingsym}{\DLfmlsym}} $.   
 By the hypothesis  $\DLfmlsym \hostimply \simuHdist{\renamingsym}{\Hsymbol}{\Hdistdegsym}$,
     we derive 
$  \statesym_1 \uplus
\renaming{\statesym_2}{\renamingsym} \in
\dLfmlsemstate{         \ModalATT{\Pwoattack}{\sensorset}{\offsetsym}{\renamingsym}{(\simuHdist{\renamingsym}{\Hsymbol}{\Hdistdegsym}} ) } $.  
From this fact  and Proposition~\ref{lem:newMod},  we derive that there exists  a  state $\statesymp_2$ such that   $(\renaming{\statesym_2}{\renamingsym}, \renaming{\statesymp_2}{\renamingsym}  ) \in 
\dLfmlsemstate{\renaming{\Pwoattack}{\renamingsym}}$ and 
$  \statesymp_1  \uplus
\renaming{\statesymp_2 }{\renamingsym} \in
\dLfmlsemstate{(\simuHdist{\renamingsym}{\Hsymbol}{\Hdistdegsym}\}   } $. 
Now, from
$(\renaming{\statesym_2}{\renamingsym}, \renaming{\statesymp_2}{\renamingsym}  ) \in 
\dLfmlsemstate{\renaming{\Pwoattack}{\renamingsym}}$
we derive that $(\statesym_2, \statesymp_2  ) \in 
\dLfmlsemstate{\Pwoattack}$.
Moreover, since
 $\statesym_2   \in \dLfmlsemstate{\dLfmlpre}$,  we have that  
$\statesymp_2 \in \dLfmlsemstate{\SP{\Pwoattack}{\dLfmlpre}}$.
It remainse to prove that
$\Hdiststate{\statesymp_1}{\statesymp_2}{\Hsymbol} \le d$.
From $  \statesymp_1  \uplus
\renaming{\statesymp_2}{\renamingsym} \in
\dLfmlsemstate{(\simuHdist{\renamingsym}{\Hsymbol}{\Hdistdegsym}\}   } $ we have that
$\sqrt{ \sum_{x \in \Hsymbol} \left(\statesymp_1(x) - (\renaming{\statesymp_2}{\renamingsym})( \renaming{x}{\renamingsym})\right)^2} \leq \Hdistdegsym$.  
Since 
$\sqrt{ \sum_{x \in \Hsymbol} \left(\statesymp_1 (x) -  (\renaming{\statesymp_2}{\renamingsym})( \renaming{x}{\renamingsym})\right)^2} =\sqrt{ \sum_{x \in \Hsymbol} \left(\statesymp_1(x) - \statesymp_2 (x) \right)^2} = \Hdiststate{\statesymp_1 }{\statesymp_2 }{\Hsymbol}$ 
we have that  
$\Hdiststate{\statesymp_1 }{\statesymp_2 }{\Hsymbol} \le d$, which concludes the proof.
\end{proof}
$\\$

\noindent
\begin{proof}[\textbf{Proof of Theorem~\ref{thm:proof-calculus-backward-soundness}}]
We have to prove that for each state $\statesym_1    \in {\dLfmlsemstate{\DLfmlModalE {\Qattacked{\Pwoattack}{\sensorset}{\offsetsym}}{\DLfmlneg{\dLfmlpost}}}}$, 
there is a state $\statesym_2 \in \dLfmlsemstate{\DLfmlModalE{\Pwoattack}{\DLfmlneg{\dLfmlpost}}}$ with $\Hdiststate{\statesym_1}{\statesym_2}{\Hsymbol} \le \Hdistdegsym$.
Given an arbitrary  state
$\statesym_1 \in {\dLfmlsemstate{\DLfmlModalE {\Qattacked{\Pwoattack}{\sensorset}{\offsetsym}}{\DLfmlneg{\dLfmlpost}}}}$, by the hypothesis 
 $\forall~\VAR{\Pwoattack}~\exists \renaming{\VAR{\Pwoattack}}{\renamingsym}.\, \DLfmlsym$,
there is some state $\statesym_2$ such that
$\statesym_1  \uplus
\renaming{\statesym_2 }{\renamingsym} \in  \dLfmlsemstate{\DLfmlsym}$.
From  the hypothesis 
$ 
       \DLfmlsym \hostimply  \ModalATT{\Pwoattack}{\sensorset}{\offsetsym}{\renamingsym}{(\DLfmlneg{\dLfmlpost} \hostimply \renaming{\DLfmlneg{\dLfmlpost}}{\renamingsym})} 
     $,    we have that 
$ \statesym_1 \uplus
\renaming{\statesym_2}{\renamingsym} \in 
\dLfmlsemstate{ 
\ModalATT{\Pwoattack}{\sensorset}{\offsetsym}{\renamingsym}{(\DLfmlneg{\dLfmlpost} \hostimply \renaming{\DLfmlneg{\dLfmlpost}}{\renamingsym})}} 
       $.   
Since $\statesym_1    \in {\dLfmlsemstate{\DLfmlModalE {\Qattacked{\Pwoattack}{\sensorset}{\offsetsym}}{\DLfmlneg{\dLfmlpost}}}}$, then there exists a state   
$ \statesymp_1   \in  \dLfmlsemstate{\DLfmlneg{\dLfmlpost}} $  such that $(\statesym_1, \statesymp_1 ) \in 
\dLfmlsemstate{\Qattacked{\Pwoattack}{\sensorset}{\offsetsym}}$.
From this fact and since   $ \statesym_1 \uplus
\renaming{\statesym_2}{\renamingsym} \in 
\dLfmlsemstate{ 
\ModalATT{\Pwoattack}{\sensorset}{\offsetsym}{\renamingsym}{(\DLfmlneg{\dLfmlpost} \hostimply \renaming{\DLfmlneg{\dLfmlpost}}{\renamingsym})}} 
       $, by Proposition~\ref{lem:newMod},  
there exists  a  state $\statesymp_2$ such that    $(\renaming{\statesym_2}{\renamingsym}, \renaming{\statesymp_2}{\renamingsym} ) \in 
\dLfmlsemstate{\renaming{\Pwoattack}{\renamingsym}}$ and
$\renaming{\statesymp_2}{\renamingsym}  \in  \dLfmlsemstate{\renaming{\DLfmlneg{\dLfmlpost}}{\renamingsym} }$.
Therefore $\renaming{\statesym_2}{\renamingsym} \in \dLfmlsemstate{\DLfmlModalE{\renaming{\Pwoattack}{\renamingsym}}{\renaming{\DLfmlneg{\dLfmlpost}}{\renamingsym}}}$, thus 
implying
$\statesym_2 \in \dLfmlsemstate{\DLfmlModalE{\Pwoattack}{\DLfmlneg{\dLfmlpost}}}$.
It remains to prove that
$\Hdiststate{\statesym_1}{\statesym_2}{\Hsymbol} \le d$. 
By $\statesym_1  \uplus
\renaming{\statesym_2 }{\renamingsym} \in  \dLfmlsemstate{\DLfmlsym}$
and the hypothesis $\DLfmlsym \hostimply \simuHdist{\renamingsym}{\Hsymbol}{\Hdistdegsym} $,  we infer
$\statesym_1 \uplus
\renaming{\statesym_2}{\renamingsym} \in
\dLfmlsemstate{\simuHdist{\renamingsym}{\Hsymbol}{\Hdistdegsym}}$ implying
$\sqrt{ \sum_{x \in \Hsymbol} \left(\statesym_1(x) - (\renaming{\statesym_2}{\renamingsym})( \renaming{x}{\renamingsym})\right)^2} \leq \Hdistdegsym$.  
Since 
$\sqrt{ \sum_{x \in \Hsymbol} \left(\statesym_1 (x) - (\renaming{\statesym_2 }{\renamingsym})( \renaming{x}{\renamingsym})\right)^2} =\sqrt{ \sum_{x \in \Hsymbol} \left(\statesym_1(x) - \statesym_2(x) \right)^2} = \Hdiststate{\statesym_1 }{\statesym_2 }{\Hsymbol}$ 
we have that  
$\Hdiststate{\statesym_1 }{\statesym_2 }{\Hsymbol} \le d$ which concludes the proof.
\end{proof}
$\\$

\medskip

\newcommand{\rulenameFsem}{\textsc{sem}}

\noindent
\begin{proof}[\textbf{Proof of Proposition~\ref{lem:proof-calculus-forward-soundness}}]
We prove the soundness of these proof rules using the default \dL axioms and rules~\cite{platzer2017complete, Platzer18book}. We write \Patt to denote $\Qattacked{\Pwoattack}{\sensorset}{\offsetsym}$, and \Prename to denote $\renaming{\Pwoattack}{\renamingsym}$. In the proof derivation, we ignore certain parts, using the notation $\mydots$, if they are similar to the other parts presented in the same derivation.

\begin{itemize}
    \item  Rule \rulenameFdef is sound as it is defined according to the definition of the modality.  
    \item  The following derivation shows the soundness proof of the rule \rulenameFseq. The key step is the right branch. The implication ${ \DLfmlModalE{\Prename}{ \DLfmlModalA{ \Qatt }{ \DLfmlModalE{ \Qrename }{\DLfmlsym} } }
    } \hostimply
    { 
    \DLfmlModalA{ \Qatt }{ \DLfmlModalE{ \Prename }{ \DLfmlModalE{ \Qrename }  
 {\DLfmlsym} } } 
    }$ holds, intuitively, because the diamond modality of $\Prename$ in the conclusion can always take the execution of $\Prename$ that makes the premise hold.  
\begin{prooftree}
    \AxiomC{ 
       $ \sequent{\Gamma}
        { 
         \ModalATT{\DLprogsym}{\sensorset}{\offsetsym}{\renamingsym}{ 
           \ModalATT{\DLprogsymp}{\sensorset}{\offsetsym}{\renamingsym}{\DLfmlsym} 
         }
       } $
    }
    \LeftLabel{\prooflabel{\rulenameFdef}}  
    \UnaryInfC{ 
       $ \sequent{\Gamma}
        { 
         \DLfmlModalA{ \Patt }{ \DLfmlModalE{  \Prename  }{     \DLfmlModalA{ \Qatt }{ \DLfmlModalE{ \Qrename }{\DLfmlsym}  }     } }
       } $
    }

    \AxiomC{
    $\sequent{}
     { { \DLfmlModalE{\Prename}{ \DLfmlModalA{ \Qatt }{ \DLfmlModalE{ \Qrename }{\DLfmlsym} } }
      } \hostimply
      { \DLfmlModalA{ \Qatt }{ \DLfmlModalE{ \Prename }{ \DLfmlModalE{ \Qrename }  
 {\DLfmlsym} } } 
      }
     }$
    }
    \UnaryInfC{
    $\sequent{ \DLfmlModalE{\Prename}{ \DLfmlModalA{ \Qatt }{ \DLfmlModalE{ \Qrename }{\DLfmlsym} } }
    }
    { 
    \DLfmlModalA{ \Qatt }{ \DLfmlModalE{ \Prename }{ \DLfmlModalE{ \Qrename }  
 {\DLfmlsym} } } 
    }$
    }
    \LeftLabel{\prooflabel{ \rulenameMR }}  
    \BinaryInfC{ 
       $ \sequent{\Gamma}
        { 
         \DLfmlModalA{ \Patt }{ \DLfmlModalA{  \Qatt  }{     \DLfmlModalE{ \Prename }{ \DLfmlModalE{ \Qrename }{\DLfmlsym}  }     } }
       } $
    }
    \LeftLabel{\prooflabel{\axiomComposeb \text{~and} \rulenameFdef}}  
    \UnaryInfC{
        $ \sequent {\Gamma}
        {
            \ModalATT{\DLseq{\DLprogsym}{\DLprogsymp}}{\sensorset}{\offsetsym}{\renamingsym}{\DLfmlsym}
        } $
    }  
\end{prooftree}
\item The following derivation shows the soundness proof of the rule \rulenameFmr. 

\begin{prooftree}
    \AxiomC{ 
      $ \sequent{\Gamma}{ \ModalATT{\Pwoattack}{\sensorset}{\offsetsym}{\renamingsym}{\DLfmlsym} }$
    }
    \LeftLabel{\prooflabel{\rulenameFdef}}  
    \UnaryInfC{ 
       $ \sequent{\Gamma}
        { 
          \DLfmlModalA{ \Patt }{ \DLfmlModalE{ \Prename }{ \DLfmlsym }}
       } $
    }

    \AxiomC{
      $\sequent{ 
        { \DLfmlsym }
      }
      { 
        { \DLfmlsymp }
      }$
      }
    \LeftLabel{\prooflabel{ \rulenameFsem }}
    \UnaryInfC{
    $\sequent{ 
      \DLfmlModalE{ \Prename }{ \DLfmlsym }
    }
    { 
      \DLfmlModalE{ \Prename }{ \DLfmlsymp }
    }$
    }
    \LeftLabel{\prooflabel{ \rulenameMR }}  
    \BinaryInfC{ 
       $ \sequent{\Gamma}
        { 
         \DLfmlModalA{ \Patt }{ \DLfmlModalE{ \Prename }{ \DLfmlsymp }}
       } $
    }
    \LeftLabel{\prooflabel{\rulenameFdef}}  
    \UnaryInfC{
        $ \sequent{\Gamma}{ \ModalATT{\Pwoattack}{\sensorset}{\offsetsym}{\renamingsym}{\DLfmlsymp} }$
    }  
\end{prooftree}

\item The following derivation shows the soundness proof of the rule \rulenameFand. 

\begin{prooftree}
  \AxiomC{ 
    $ \sequent{\Gamma}{ \ModalATT{\Pwoattack}{\sensorset}{\offsetsym}{\renamingsym}{(\DLfmlsym \hostand \DLfmlsymp)} }$
  }
  \AxiomC{ 
    $ \sequent{\DLfmlsym \hostand \DLfmlsymp }{ \DLfmlsym }$
  }
  \LeftLabel{\prooflabel{ \rulenameFmr }}  
  \BinaryInfC{ 
     $ \sequent{\Gamma}
      { 
        \ModalATT{\DLprogsym}{\sensorset}{\offsetsym}{\renamingsym}{\DLfmlsym}
     } $
  }

  \AxiomC{ 
    $ \sequent{\Gamma}{ \ModalATT{\Pwoattack}{\sensorset}{\offsetsym}{\renamingsym}{(\DLfmlsym \hostand \DLfmlsymp)} }$
  }
  \AxiomC{ 
    $ \sequent{\DLfmlsym \hostand \DLfmlsymp }{ \DLfmlsymp }$
  }
  \LeftLabel{\prooflabel{ \rulenameFmr }}  
  \BinaryInfC{ 
     $ \sequent{\Gamma}
      { 
        \ModalATT{\DLprogsym}{\sensorset}{\offsetsym}{\renamingsym}{\DLfmlsymp}
     } $
  }
  \LeftLabel{\prooflabel{\rulenameAndR}}
  \BinaryInfC{
      $
      \sequent{\Gamma}{\ModalATT{\DLprogsym}{\sensorset}{\offsetsym}{\renamingsym}{\DLfmlsym} \hostand 
       \ModalATT{\DLprogsym}{\sensorset}{\offsetsym}{\renamingsym}{\DLfmlsymp} } 
      $
  }  
\end{prooftree}

\item The following derivation shows the soundness proof of the rule \rulenameFor. 

\begin{prooftree}
\prooftreesize
  \AxiomC{$\sequent
      {\Gamma}
      {\ModalATT{\DLprogsym}{\sensorset}{\offsetsym}{\renamingsym}{\DLfmlsym} \hostor \ModalATT{\DLprogsym}  
            {\sensorset}{\offsetsym}{\renamingsym}{\DLfmlsymp}}$}
  \LeftLabel{\prooflabel{ \rulenameFdef }}
   \UnaryInfC{$\sequent
      {\Gamma}
      {\DLfmlModalA{\Patt}{\DLfmlModalE{  \Prename  }{\DLfmlsym}} \hostor 
        \DLfmlModalA{\Patt}{\DLfmlModalE{  \Prename  }{\DLfmlsymp}}
        }
      $}
  
    \AxiomC{
      $
      \sequent{
        {{\DLfmlsym}} 
      }
      {
        { { (\DLfmlsym \hostor \DLfmlsymp) } }
      } 
      $
  } 
  \LeftLabel{\prooflabel{ \rulenameMonoE }}
  \UnaryInfC{
      $
      \sequent{
        {\DLfmlModalE{  \Prename  }{\DLfmlsym}} 
      }
      {
        { \DLfmlModalE{  \Prename  }{ (\DLfmlsym \hostor \DLfmlsymp) } }
      } 
      $
  } 
  \LeftLabel{\prooflabel{ \rulenameMonoA }}
  \UnaryInfC{
      $
      \sequent{
        \DLfmlModalA{\Patt}{\DLfmlModalE{  \Prename  }{\DLfmlsym}} 
      }
      {
        \DLfmlModalA{ \Patt }{ \DLfmlModalE{  \Prename  }{ (\DLfmlsym \hostor \DLfmlsymp) } }
      } 
      $
  } 
  \AxiomC{\mydots}
  \LeftLabel{\prooflabel{ \rulenameOrL }}
  \BinaryInfC{
      $
      \sequent{
        \DLfmlModalA{\Patt}{\DLfmlModalE{  \Prename  }{\DLfmlsym}} 
        \hostor 
        \DLfmlModalA{\Patt}{\DLfmlModalE{  \Prename  }{\DLfmlsymp}}
      }
      {
        \DLfmlModalA{ \Patt }{ \DLfmlModalE{  \Prename  }{ (\DLfmlsym \hostor \DLfmlsymp) } }
       } 
      $
  } 
  \LeftLabel{\prooflabel{ \rulenameCutR }}
  \BinaryInfC{
      $
      \sequent{\Gamma}
      {
        \DLfmlModalA{ \Patt }{ \DLfmlModalE{  \Prename  }{ (\DLfmlsym \hostor \DLfmlsymp) } }
       } 
      $
  }  
  \LeftLabel{\prooflabel{\rulenameFdef}}
  \UnaryInfC{
      $
      \sequent{\Gamma}{\ModalATT{\DLprogsym}{\sensorset}{\offsetsym}{\renamingsym}{(\DLfmlsym \hostor \DLfmlsymp)} 
       } 
      $
  }  
\end{prooftree}
\item The following derivation shows the soundness proof of the rule \rulenameFchoice. The left \mydots proves the obligation of $\sequent{\Gamma}
    {
      \DLfmlModalA{ {\DLprogsymp}_{\hostE{att}} }{
      ( \DLfmlModalE{  {\DLprogsym}_{\hostE{\epsilon}}  }{ \DLfmlsym }
      \hostor
      \DLfmlModalE{  {\DLprogsymp}_{\hostE{\epsilon}}  }{ \DLfmlsym } 
      ) }
     } $ the same way as the proof for $\sequent{\Gamma}
    {
      \DLfmlModalA{ {\DLprogsym}_{\hostE{att}} }{
      ( \DLfmlModalE{  {\DLprogsym}_{\hostE{\epsilon}}  }{ \DLfmlsym }
      \hostor
      \DLfmlModalE{  {\DLprogsymp}_{\hostE{\epsilon}}  }{ \DLfmlsym } 
      ) }
     }$. 
     And the right \mydots proves the obligation $\sequent{\DLfmlModalE{  {\DLprogsymp}_{\hostE{\epsilon}}  }{ \DLfmlsym } }
      { \DLfmlModalE{  (\DLchoice{\DLprogsym}{\DLprogsymp})_{\hostE{\epsilon}}  }{ \DLfmlsym } }$ the same way as the proof for $\sequent{\DLfmlModalE{  {\DLprogsym}_{\hostE{\epsilon}}  }{ \DLfmlsym } }
      { \DLfmlModalE{  (\DLchoice{\DLprogsym}{\DLprogsymp})_{\hostE{\epsilon}}  }{ \DLfmlsym } }$.

\begin{prooftree}
  \prooftreesize
  \AxiomC{$
  \sequent{\Gamma}{\ModalATT{\DLprogsym}{\sensorset}{\offsetsym}{\renamingsym}{\DLfmlsym} 
   } 
  $}
  \LeftLabel{\prooflabel{\rulenameFdef}}
  \UnaryInfC{ 
    $      
    \sequent{\Gamma}
    {
      \DLfmlModalA{ {\DLprogsym}_{\hostE{att}} }{
      \DLfmlModalE{  {\DLprogsym}_{\hostE{\epsilon}}  }{ \DLfmlsym }
      }
     }
    $
  }
  \AxiomC{ 
    $ \sequent{ \DLfmlModalE{  {\DLprogsym}_{\hostE{\epsilon}}  }{ \DLfmlsym } }{ \DLfmlModalE{  {\DLprogsym}_{\hostE{\epsilon}}  }{ \DLfmlsym }
    \hostor
    \DLfmlModalE{  {\DLprogsymp}_{\hostE{\epsilon}}  }{ \DLfmlsym } 
    }$
  }
  \LeftLabel{\prooflabel{ \rulenameFmr }}  
  \BinaryInfC{ 
    $      
    \sequent{\Gamma}
    {
      \DLfmlModalA{ {\DLprogsym}_{\hostE{att}} }{
      ( \DLfmlModalE{  {\DLprogsym}_{\hostE{\epsilon}}  }{ \DLfmlsym }
      \hostor
      \DLfmlModalE{  {\DLprogsymp}_{\hostE{\epsilon}}  }{ \DLfmlsym } 
      ) }
     }
    $
  }

  \AxiomC{\mydots}
  \LeftLabel{\prooflabel{\axiomChoiceb}}
  \BinaryInfC{
    $      
    \sequent{\Gamma}
    {
      \DLfmlModalA{ (\DLchoice{\DLprogsym}{\DLprogsymp})_{\hostE{att}} }{
      ( \DLfmlModalE{  {\DLprogsym}_{\hostE{\epsilon}}  }{ \DLfmlsym }
      \hostor
      \DLfmlModalE{  {\DLprogsymp}_{\hostE{\epsilon}}  }{ \DLfmlsym } 
      ) }
     }
    $
  } 

   \AxiomC{*}
   \UnaryInfC{ 
   $\sequent{\DLfmlModalE{  {\DLprogsym}_{\hostE{\epsilon}}  }{ \DLfmlsym } }
      { \DLfmlModalE{  (\DLchoice{\DLprogsym}{\DLprogsymp})_{\hostE{\epsilon}}  }{ \DLfmlsym } }$
  }
   \AxiomC{ 
    \mydots
  }
  \BinaryInfC{ 
   $\sequent{\DLfmlModalE{  {\DLprogsym}_{\hostE{\epsilon}}  }{ \DLfmlsym }
      \hostor
      \DLfmlModalE{  {\DLprogsymp}_{\hostE{\epsilon}}  }{ \DLfmlsym } }
      { \DLfmlModalE{  (\DLchoice{\DLprogsym}{\DLprogsymp})_{\hostE{\epsilon}}  }{ \DLfmlsym } }$
  }
  \LeftLabel{\prooflabel{\rulenameMR}}
  \BinaryInfC{
      $      
      \sequent{\Gamma}
      {
        \DLfmlModalA{ (\DLchoice{\DLprogsym}{\DLprogsymp})_{\hostE{att}} }{ \DLfmlModalE{  (\DLchoice{\DLprogsym}{\DLprogsymp})_{\hostE{\epsilon}}  }{ \DLfmlsym } }
       }
      $
  }  
  \LeftLabel{\prooflabel{\rulenameFdef}}
  \UnaryInfC{
      $
      \sequent{\Gamma}{ \ModalATT{\DLchoice{\DLprogsym}{\DLprogsymp}}{\sensorset}{\offsetsym}{\renamingsym}{\DLfmlsym} }
      $
  }  

\end{prooftree}

\item The proof for the rule \rulenameFtest is as follows:
\[
      \ModalATT{\DLtest{\DLfmlsymtest}}{\sensorset}{\offsetsym}{\renamingsym}{\DLfmlsym} \hostlrarrow
      (\DLfmlModalA{ \DLtest{\DLfmlsymtest} }{ \DLfmlModalE{  \DLtest{\renaming{\DLfmlsymtest}{\renamingsym}}  }{ \DLfmlsym } }) \hostlrarrow  (\DLfmlModalA{ \DLtest{\DLfmlsymtest} }{ \renaming{\DLfmlsymtest}{\renamingsym} \hostand \DLfmlsym } ) \hostlrarrow ( \DLfmlsymtest \hostimply (\renaming{\DLfmlsymtest}{\renamingsym} \land \DLfmlsym) 
\]

\item The following derivation shows the soundness proof of the rule \rulenameFinv. The key step is to derive $
\sequent{\DLfmlsymInv}{\ModalATT{\DLloop{\DLprogsym}}{\sensorset}{\offsetsym}{\renamingsym}{\DLfmlsymInv}}
$ from $
\sequent{\DLfmlsymInv}{\ModalATT{{\DLprogsym}}{\sensorset}{\offsetsym}{\renamingsym}{\DLfmlsymInv}}
$. This can be done by induction on the number of iteration taken by $\DLloop{\DLprogsym}$, in particular, assume we can derive $
\sequent{\DLfmlsymInv}{\ModalATT{{\DLprogsym}^n}{\sensorset}{\offsetsym}{\renamingsym}{\DLfmlsymInv}}
$, then using the rule \rulenameFseq we can derive $
\sequent{\DLfmlsymInv}{\ModalATT{{\DLprogsym}^{n+1}}{\sensorset}{\offsetsym}{\renamingsym}{\DLfmlsymInv}}. 
$

\begin{prooftree}
  \AxiomC{$
  \sequent{\Gamma}{ \DLfmlsymInv } 
  $}

  \AxiomC{
    $
    \sequent{\DLfmlsymInv}{\ModalATT{{\DLprogsym}}{\sensorset}{\offsetsym}{\renamingsym}{\DLfmlsymInv}}
    $
  } 
  \UnaryInfC{
    $
    \sequent{\DLfmlsymInv}{\ModalATT{\DLloop{\DLprogsym}}{\sensorset}{\offsetsym}{\renamingsym}{\DLfmlsymInv}}
    $
  } 
  \LeftLabel{\prooflabel{\rulenameImplyL}}
  \BinaryInfC{
    $
    \sequent{\Gamma}{\ModalATT{\DLloop{\DLprogsym}}{\sensorset}{\offsetsym}{\renamingsym}{\DLfmlsymInv}}
    $
  }

  \AxiomC{
    $
    \sequent{\DLfmlsymInv}{\DLfmlsymp}
    $
  } 
  \LeftLabel{\prooflabel{\rulenameFmr}}
  \BinaryInfC{
      $
      \sequent{\Gamma}{\ModalATT{\DLloop{\DLprogsym}}{\sensorset}{\offsetsym}{\renamingsym}{\DLfmlsymp}}
      $
  }  
\end{prooftree}

\item The following derivation shows the soundness proof of the rule \rulenameFv. Note that the branch on the right says given $ \sequent{{\DLfmlsymp}}{
      \DLfmlModalE{\Prename}{\DLfmlsymp_1} 
    } $, we have 
    $ \sequent{{(\DLfmlsym_1 \hostand \DLfmlsymp)}}{
      \DLfmlModalE{\Prename}{(\DLfmlsym_1 \hostand \DLfmlsymp_1)} 
    } $. This is sound because program $\Prename$ doesn't refer to the variables in $\DLfmlsym_1$.
   
\begin{prooftree}
  \AxiomC{
    $ \sequent{\DLfmlsym}{
      \DLfmlModalA{\Patt}{\DLfmlsym_1} 
    } $
  }
  \LeftLabel{\prooflabel{\rulenameAndL}}
  \UnaryInfC{
    $ \sequent{(\DLfmlsym \hostand \DLfmlsymp)}{
      \DLfmlModalA{\Patt}{\DLfmlsym_1} 
    } $
  }

  \AxiomC{*}
  \UnaryInfC{
    $ \sequent{(\DLfmlsym \hostand \DLfmlsymp)}{
      {\DLfmlsymp} 
    } $
  }
  \LeftLabel{\prooflabel{\axiomV}}
  \UnaryInfC{
    $ \sequent{(\DLfmlsym \hostand \DLfmlsymp)}{
      \DLfmlModalA{\Patt}{\DLfmlsymp} 
    } $
  }
  \LeftLabel{\prooflabel{\axiomBoxDist}}
  \BinaryInfC{
    $ \sequent{(\DLfmlsym \hostand \DLfmlsymp)}{
      \DLfmlModalA{\Patt}{(\DLfmlsym_1 \hostand \DLfmlsymp)} 
    } $
  }

  \AxiomC{
      $ \sequent{{\DLfmlsymp}}{
      \DLfmlModalE{\Prename}{\DLfmlsymp_1} 
    } $
  }
  \UnaryInfC{
    $ \sequent{{(\DLfmlsym_1 \hostand \DLfmlsymp)}}{
      \DLfmlModalE{\Prename}{(\DLfmlsym_1 \hostand \DLfmlsymp_1)} 
    } $
  }
  \LeftLabel{\prooflabel{\rulenameMR}}
  \BinaryInfC{
    $ \sequent{(\DLfmlsym \hostand \DLfmlsymp)}{
        \DLfmlModalA{\Patt}{\DLfmlModalE{\Prename}{ (\DLfmlsym_1 \hostand \DLfmlsymp_1) }}
    } $
  }
  \LeftLabel{\prooflabel{\rulenameWL}}
  \UnaryInfC{
    $ \sequent{(\DLfmlsym \hostand 
    \DLfmlsymp)}{
      \ModalATT{\DLprogsym}{\sensorset}{\offsetsym}{\renamingsym}{(\DLfmlsym_1 \hostand \DLfmlsymp_1)} 
    } $
  }

\end{prooftree}

\item The rule \rulenameFodeForall is sound because, given $\DLfmlModalA{ \dLODE{\DLvarsym}{\DLtermsym}}{\DLfmlModalA{\renaming{\dLODE{\DLvarsym}{\DLtermsym}}{\renamingsym} }{( \DLfmlsymp(t, \renaming{t}{\renamingsym}) \hostimply \DLfmlsym )}} $, we know that for all solutions of $\dLODE{\DLvarsym}{\DLtermsym}$ and $\renaming{\dLODE{\DLvarsym}{\DLtermsym}}{\renamingsym}$, their reachable states satisfy $\DLfmlsym$ when the condition $\DLfmlsymp$ holds. And since $\DLfmlsymp$ only concerns time, and we have no constraints on the time in the model of dynamics, there must exists states where $\DLfmlsymp$ holds. 

\item The rule \rulenameFodemerge is sound because it is a specialized variant of the rule \rulenameFodeForall. A box modality $\DLfmlModalA{ \hostE{\dLODE{\DLvarsym}{\DLtermsym}}, ~ \renaming{ \dLODE{\DLvarsym}{\DLtermsym} }{\renamingsym} }{\DLfmlsym} $ forces the two dynamics evolve for the same duration. Thus, if $\DLfmlModalA{ \hostE{\dLODE{\DLvarsym}{\DLtermsym}}, ~ \renaming{ \dLODE{\DLvarsym}{\DLtermsym} }{\renamingsym} }{\DLfmlsym}$ holds, then for any reachable state of $\hostE{\dLODE{\DLvarsym}{\DLtermsym}}$, a solution of $\renaming{ \dLODE{\DLvarsym}{\DLtermsym} }{\renamingsym}$ can reach a state so $\DLfmlsym$ holds, by evolving for the same duration as $\hostE{\dLODE{\DLvarsym}{\DLtermsym}}$. The evolution constraint $\hostE{t \le \epsilon}$ won't affect as both dynamics evolve for the same duration.

\end{itemize}


\end{proof}



\section{The proof rules of    the proof calculus \dL}
\label{sec:proof_calculus_Platzer}

\newcommand{\symP}{\DLfmlsym}
\newcommand{\symQ}{\DLfmlsymp}


\begin{figure}[H]
  \centering
    \begin{mathpar} 
      \label{rule:notR} \inferrule*[Left=$\rulenameNotR$]
      {\sequent{\Gamma, \symP}{\Delta}}
      {\sequent{\Gamma}{\neg \symP, \Delta}}
  
      \label{rule:andR} \inferrule*[Left=$\rulenameAndR$ ]
      {
        \sequent{\Gamma}{\symP, \Delta} \\
        \sequent{\Gamma}{\symQ, \Delta}
       }
      {\sequent{\Gamma}{\symP \land \symQ, \Delta}}
  
      \label{rule:orR} \inferrule*[Left=$\rulenameOrR$ ]
      {\sequent{\Gamma}{\Delta, \symP, \symQ}}
      {\sequent{\Gamma}{\symP \lor \symQ, \Delta}}
  
      \label{rule:implyR} \inferrule*[Left=$\rulenameImplyR$]
  {\sequent{\Gamma, \symP}{\Delta, \symQ}
  }
  {\sequent{\Gamma}{\symP \rightarrow \symQ, \Delta}}
  
  \inferrule*[Left=id \label{rule:id}]
  {
  }
  {\sequent{\symP, \Gamma}{\symP, \Delta}}
  
  \\\\
  \label{rule:notL} \inferrule*[Left=$\rulenameNotL$ ]
      {\sequent{\Gamma}{\Delta, \symP}}
      {\sequent{\neg \symP, \Gamma}{\Delta}}
  
      \label{rule:andL}    \inferrule*[Left=$\rulenameAndL$ ]
      {
        \sequent{\Gamma, \symP, \symQ}{\Delta}
       }
      {\sequent{\symP \land \symQ, \Gamma}{\Delta}}
  
      \label{rule:orL}  \inferrule*[Left=$\rulenameOrL$ ]
      {\sequent{\symP, \Gamma}{\Delta}
      \\
      \sequent{\symQ, \Gamma}{\Delta}
      }
      {\sequent{\symP \lor \symQ, \Gamma}{\Delta}}
  
      \label{rule:implyL} \inferrule*[Left=$\rulenameImplyL$ ]
  {\sequent{\Gamma}{\Delta, \symP}
  \\
  \sequent{\symQ, \Gamma}{\Delta}
  }
  {\sequent{\symP \rightarrow \symQ, \Gamma}{\Delta}}

  \\\\
  
  \label{rule:eqR} \inferrule*[Left=$\rulenameEqR$ ]
  {\sequent{\Gamma, \symP}{\Delta, \symQ}
  \\
  \sequent{\Gamma, \symQ}{\Delta, \symP}
  }
  {\sequent{\Gamma}{\symP \leftrightarrow \symQ, \Delta}}
  
  \label{rule:eqL} \inferrule*[Left=$\rulenameEqL$ ]
  {\sequent{\symP \land \symQ, \Gamma}{\Delta}
  \\
  \sequent{\neg \symP \land \neg \symQ, \Gamma}{\Delta}
  }
  {\sequent{\symP \leftrightarrow \symQ, \Gamma}{\Delta}}
  
  \label{rule:cut} \inferrule*[Left=$\rulenameCut$ ]
  {\sequent{\Gamma, C}{\Delta}
  \\
  \sequent{\Gamma}{\Delta, C}
  }
  {\sequent{\Gamma}{\Delta}}

  \\\\
  
  \label{rule:closetrue} \inferrule*[Left=$\top$R ]
  {
  }
  {\sequent{\Gamma}{true, \Delta}}
  
  \label{rule:closefalse} \inferrule*[Left=$\bot$L ]
  {
  }
  {\sequent{false, \Gamma}{\Delta}}
  
  \label{rule:wr} \inferrule*[Left=\rulenameWR]
  {\sequent{\Gamma}{\Delta}
  }
  {\sequent{\Gamma}{\symP, \Delta}}
  
  \label{rule:wl} \inferrule*[Left=\rulenameWL]
  {\sequent{\Gamma}{\Delta}
  }
  {\sequent{\symP, \Gamma}{\Delta}}
  
  \label{rule:pr} \inferrule*[Left=PR ]
  {\sequent{\Gamma}{\symQ, \symP, \Delta}
  }
  {\sequent{\Gamma}{\symP, \symQ, \Delta}}
  
  \label{rule:pl} \inferrule*[Left=PL ]
  {\sequent{\symQ, \symP, \Gamma}{\Delta}
  }
  {\sequent{\symP, \symQ, \Gamma}{\Delta}}
  
    \end{mathpar}
    \caption{\label{psc}Standard propositional sequent calculus proof rules with cut rule \cite{Platzer18book}} 
  \end{figure}

\begin{figure}[H]
  \begin{mathpar}
   \label{axiom:assignb} \inferrule*[Left=\axiomAssignb]
  {}
  {\DLfmlModalA{\DLassign{x}{e}}{\symP(x)}\leftrightarrow \symP(e)}
  
  \label{axiom:randomb} \inferrule*[Left=\axiomRandomb]
  {}
  {\DLfmlModalA{\DLassignN{x}}{\symP(x)}\leftrightarrow \forall x, \symP(x)}
  
  \label{axiom:testb} \inferrule*[Left=\axiomTestb]
  {}
  {\DLfmlModalA{\DLtest{\symQ}}{\symP} \leftrightarrow (\symQ \rightarrow \symP)}
  \\\\

  \label{axiom:solve} \inferrule*[Left=\axiomSolve]
  {}
  {\DLfmlModalA{\dLprogODE{ \dLODE{x}{f(x)}}{q(x)} }{\symP(x)} \leftrightarrow \forall t \ge 0 ~((\forall 0\leq s \leq t, q(x(s))) \rightarrow \DLfmlModalA{\DLassign{x}{x(t)}}{\symP(x)})}  \quad  (\text{if~} x'(t) = f(x(t)))
  \\\\
  
  \label{axiom:choiceb} \inferrule*[Left=\axiomChoiceb]
  {}
  {\DLfmlModalA{\DLchoice{\DLprogsym}{\DLprogsymp}}{\symP} \leftrightarrow (\DLfmlModalA{\DLprogsym}{\symP} \land \DLfmlModalA{\DLprogsymp}{\symP})}
  
  \label{axiom:composeb} \inferrule*[Left=\axiomComposeb]
  {}
  {\DLfmlModalA{\DLseq{\DLprogsym}{\DLprogsymp}}{\symP} \leftrightarrow \DLfmlModalA{\DLprogsym}{\DLfmlModalA{\DLprogsymp}{\symP}}}
  
  \label{axiom:iterateb} \inferrule*[Left=\axiomIterateb]
  {}
  {\DLfmlModalA{\DLloop{\DLprogsym}}{\symP} \leftrightarrow \symP \land \DLfmlModalA{\DLprogsym}{\DLfmlModalA{\DLloop{\DLprogsym}}{\symP}}}
  
  \label{axiom:diamond} \inferrule*[Left=\axiomDiamond]
  {}
  {\DLfmlneg{ \DLfmlModalA{\DLprogsym}{\symP} }\leftrightarrow  
  \DLfmlModalE{\DLprogsym}{\DLfmlneg{\symP}}}
  \\\\

  \label{axiom:k} \inferrule*[Left=\axiomK]
  {}
  { \DLfmlModalA{\DLprogsym}{\symP \rightarrow \symQ} \rightarrow  
  (\DLfmlModalA{\DLprogsym}{\symP} \rightarrow \DLfmlModalA{\DLprogsym}{\symQ}) }

  \label{axiom:i} \inferrule*[Left=\axiomI]
  {}
  { \DLfmlModalA{\DLloop{\DLprogsym}}{\symP} \leftrightarrow  
  \symP \land \DLfmlModalA{\DLloop{\DLprogsym}}{(\symP \rightarrow \DLfmlModalA{\DLprogsym}{\symP})}}
  
  \label{axiom:v} \inferrule*[Left=\axiomV]
  {}
  { \symP \rightarrow \DLfmlModalA{\DLprogsym}{\symP}   \quad (FV(\DLprogsym) \cap BV(\DLprogsym) = \emptyset)}
  \\\\
  
  
  \label{axiom:boxdist} \inferrule*[Left=\axiomBoxDist]
  {}
  {
  \DLfmlModalA{\DLprogsym}{(\symP \hostand \symQ)} \hostlrarrow (\DLfmlModalA{\DLprogsym}{\symP} \hostand \DLfmlModalA{\DLprogsym}{\symQ})
  }

  \label{axiom:loop} \inferrule*[Left=\rulenameLoop]
  {\sequent{\Gamma}{J, \Delta} \\
  \sequent{J}{\symP} \\
  \sequent{J}{\DLfmlModalA{\DLprogsym}{J}}
  }
  {\sequent{\Gamma}{\DLfmlModalA{\DLloop{\DLprogsym}}{\symP}, \Delta} }

  \\ 

  \label{rule:cutr} \inferrule*[Left=\rulenameCutR]
  {\sequent{\Gamma}{\symQ, \Delta} \\
  \sequent{\Gamma}{\symQ \rightarrow \symP, \Delta} 
  }
  {\sequent{\Gamma}{\symP, \Delta} }
  
  \label{rule:cutl} \inferrule*[Left=\rulenameCutL]
  {\sequent{\symQ, \Gamma}{\Delta} \\
  \sequent{\Gamma}{\Delta, \symP \rightarrow \symQ} 
  }
  {\sequent{\symP, \Gamma}{\Delta} }

  \label{rule:mr} \inferrule*[Left=\rulenameMR]
  {
  \sequent{\Gamma}{\DLfmlModalA{\DLprogsym}{\DLfmlsym}, \Delta} \\
  \sequent{\DLfmlsym}{\DLfmlsymp} 
  }
  {  \sequent{\Gamma}{\DLfmlModalA{\DLprogsym}{\DLfmlsymp}, \Delta}   }

    \label{rule:monoA} \inferrule*[Left=\rulenameMonoA]
  {\sequent{\Gamma}{{\symP} \hostimply {\symQ}, \Delta}
  }
  {\sequent{\Gamma}{\DLfmlModalA{\DLprogsym}{\symP} \hostimply \DLfmlModalA{\DLprogsym}{\symQ}, \Delta} }
   
  \label{rule:monoE} \inferrule*[Left=\rulenameMonoE]
  {\sequent{\Gamma}{{\symP} \hostimply {\symQ}, \Delta}
  }
  {\sequent{\Gamma}{\DLfmlModalE{\DLprogsym}{\symP} \hostimply \DLfmlModalE{\DLprogsym}{\symQ}, \Delta} }

  \end{mathpar}
    \caption{\label{other-rules}Other key axioms and proof rules of \dL proof calculus \cite{platzer2017complete,Platzer18book}} 
  \end{figure}



\end{document}
\endinput
